\documentclass[useAMS,usenatbib]{mn2e}
\usepackage{graphicx}
\usepackage{latexsym}
\usepackage{xspace}
\usepackage{rotating}
\usepackage{lscape}
\usepackage[percent]{overpic}
\usepackage{color}

\title[]{On the nature of star-forming filaments: I. Filament morphologies}
\author[Smith et~al.]{Rowan J. Smith\thanks{Email: rowan@uni-heidelberg.de}, Simon C. O. Glover, Ralf. S. Klessen \\
Universit\"at Heidelberg, Zentrum f\"ur Astronomie, Institut f\"ur Theoretische Astrophysik, Albert-Ueberle-Str. 2, 69120 Heidelberg, Germany \\
 }
 
\begin{document}

\pagerange{\pageref{firstpage}--\pageref{lastpage}} \pubyear{2010}

\maketitle

\label{firstpage}

\def\mnras{MNRAS}
\def\apj{ApJ}
\def\aap{A\&A}
\def\apjl{ApJL}
\def\apjs{ApJS}
\def\bain{BAIN}
\def\pasp{PASP}
\def\araa{ARA\&A}
\def\ga{\sim}
\def\nat{Nature}
\def\aj{AJ}
\def\pasj{PASJ}


\newcommand{\eq}{Equation }
\newcommand{\fig}{Figure }
\newcommand{\msun}{\,M$_{\odot}$ }
\newcommand{\gcmc}{\,g\,cm$^{-3}$}
\newcommand{\cmc}{\,cm$^{-3}$}
\newcommand{\cms}{\,cm$^{-2}$}
\newcommand{\kms}{\,kms$^{-1}$}
\newcommand{\tab}{Table }
\newcommand{\gcms}{\,g\,cm$^{-2}$\xspace}
\newcommand{\E}{\times 10}

\newcommand{\arepo}{\textsc{arepo}\xspace}
\newcommand{\sph}{\textsc{sph}\xspace}
\newcommand{\treecol}{\textsc{treecol}\xspace}
\newcommand{\healpix}{\textsc{healpix}\xspace}
\newcommand{\gadget}{\textsc{gadget} 2\xspace}
\newcommand{\disperse}{DisPerSE\xspace}
\newcommand{\mpfit}{\textsc{mpfit}\xspace}
\newcommand{\herschel}{\textit{Herschel}\xspace}

\begin{abstract}
We use a suite of high resolution molecular cloud simulations carried out with the moving mesh code \arepo to explore the nature of star-forming filaments. The simulated filaments are identified and categorised from column density maps in the same manner as for recent \herschel observations. When fit with a Plummer-like profile the filaments are in excellent agreement with observations, and have shallow power-law profiles of $p\sim2.2$ without the need for magnetic support. When data within 1 pc of the filament centre is fitted with a Gaussian function, the average FWHM is $\sim0.3$ pc, in agreement with predictions for accreting filaments. However, if the fit is constructed using only the inner regions, as in \herschel observations, the resulting FWHM is only $\sim 0.2$ pc. This value is larger than that measured in IC 5146 and Taurus, but is similar to that found in the Planck Galactic Cold Cores and in Cygnus X. The simulated filaments have a range of widths rather than a constant value. When the column density maps are compared to the 3D gas densities, the filaments seen in column density do not belong to a single structure. Instead, they are made up of a network of short ribbon-like sub-filaments reminiscent of those seen in Taurus. The sub-filaments are pre-existing within the simulated clouds, have radii similar to their Jeans radius, and are not primarily formed through fragmentation of the larger filament seen in column density. Instead, small filamentary clumps are swept together into a single column density structure by the large-scale collapse of the cloud.
\end{abstract}


\section{Introduction}\label{sec-intro}
Stars do not form in isolation but are instead intrinsically linked to the molecular clouds in which they are born through a network of filaments. The filamentary nature of molecular clouds has been known for some time. \citet{Schneider79} presented a catalogue of dense filaments containing embedded cores spaced at intervals of a few times the filament diameter. Recent observations using dust continuum emission have further emphasised the importance of filaments in molecular cloud structure (e.g.\ \citealt{Nutter08,Menshchikov10,Malinen12}; \citeauthor{Juvela12a}~2012a;
\citealt{Schneider12,Peretto12,Zernickel13,Kirk13}).

Filaments and star formation are closely interlinked as dense star-forming cores thread filaments like beads on a string \citep{Andre10,Hacar11}. The picture seems to be that the dense gas within molecular clouds is first assembled into dense filaments that then fragment into dense star-forming cores. Filaments are, therefore, a crucial intermediate stage in star formation.

One particularly striking observation is that of \citet{Arzoumanian11} who find a constant width of $\sim 0.1$ pc for the central dense sections of filaments in the \herschel Gould Belt survey by fitting a Gaussian to the inner section of the filament. \citet{Arzoumanian11} (hereafter A11) also describe their filaments using a Plummer-like function (see also \citealt{Nutter08}) which is characterised by its central density, a flattening radius, and a power law fall off in density beyond that radius. They typically find that the Gould belt filaments have a shallow power law slope of about $p=2$.

This shallow power law profile is a common feature of filament observations. \citet{Hacar11} found that three out of four filaments in L1517 had a shallower power law profile than $p=4$. \citeauthor{Juvela12a}~(2012a; hereafter J12a) found profiles of around $p=2$ in \herschel observations of filaments in cold clouds identified by \textit{Planck}. \citet{Palmeirim13} found a $p=2$ profile in the Taurus B211/3 filament. Shallow power law profiles seem to be a key characteristic of filaments in molecular clouds. That there is a constant filament width is less clear. Like A11, \citet{Palmeirim13} found a central filament width of around $0.1$ pc. However other authors have found wider filament full width half maxima (FWHM). For example, \citet{Hennemann12} found widths between 0.26 pc and 0.34 pc for the DR21 ridge and filaments in Cygnus X. Similarly, J12a found FWHM of around 0.32 pc for filaments within the Planck Galactic cold cores. One difficulty with making these comparisons is that different things are fitted in different studies, with some authors preferring to find the FWHM of the entire filament rather than just the flat density core.

Analytical studies have investigated the evolution of filamentary gas. \citet{Ostriker64} showed that infinite, isothermal self-gravitating cylinders have density profiles with a flat central core and then a steep power-law fall off, equating to $p=4$ in Equation \ref{Plum-dens}, which is steeper than that seen in observations. When the filaments are magnetised, the power law becomes $p=2$ \citep{Tilley03,Hennebelle03}, in closer agreement with observations. More recent theoretical work has studied the effects of external pressure and gravitational collapse \citep{Fischera12,Heitsch13a} upon such filament profiles. \citet{Heitsch13a} showed that filaments should have a constant width for a large part of their evolution. However, he found a FWHM of about 0.3 pc, which is three times larger than the observationally-derived value quoted in A11. Analytically, filaments have also been shown to be prone to fragmentation, highlighting their importance for star formation. \citet{Inutsuka92,Inutsuka97} showed that self-gravitating filaments are unstable to axisymmetric perturbations above a critical line mass $M_{\rm crit}=2 c_{\rm s}^2 /G$. The resulting fragments are spaced at separations of about four times the filament diameter.

In addition to analytical studies, filaments are also ubiquitous in molecular cloud simulations (e.g.\ \citealt{Klessen00,Ballesteros-Paredes02,Bate05,Padoan06,Vazquez-Semadeni06,Heitsch08,Smith09b,Federrath10b,Krumholz11}; \citeauthor{GC12b}~2012b; \citealt{Bonnell13,Hennebelle13a}). In contrast to analytical studies which often adopt the assumption of hydrostatic equilibrium, the filaments seen in simulations are typically dynamically evolving structures. As in observations, dense star-forming cores form within the filaments. For example, \citet{Smith11a} found that $75\%$ of cores in a giant molecular cloud simulation were embedded within filaments and that the cores grew in mass via filamentary accretion flows, which were crucial for building up enough mass to form massive stars. Similarly, \citet{Myers09a,Myers09b} highlighted the importance of filaments in setting stellar masses.

Given the crucial role of filaments in star formation it is important to test how similar the filaments seen in numerical simulations actually are to observed filaments. In this paper we make column density maps of filaments from a suite of high-resolution simulations of turbulent molecular clouds with active star formation. We identify filaments and fit profiles to them in a similar manner to that done observationally to allow us to answer the following key questions about filament morphologies: 1) Do filaments in non-magnetised molecular cloud simulations agree with observations? 2) How well are the filament profiles determined and do they change with time? 3) Is there a constant filament width? 4) How well do the filaments seen in column density represent the true morphology of the filament in three dimensions?

\section{Method}\label{sec-method}
\subsection{Numerical model} \label{sec-hydro}
We perform our simulations using the moving mesh code \arepo \citep{Springel10}. This is a quasi-Lagrangian code that aims to utilise the strengths of both smoothed particle hydrodynamics (SPH) and grid-based adaptive mesh refinement (AMR) codes. The fluid is represented by a series of irregular mesh cells that attempt to move with the flow and that are analogous to SPH particles. However, as the mesh is not completely Lagrangian, there is generally some residual flux of mass, momentum and energy into or out of the cells. These fluxes are computed using a Riemann solver, thereby avoiding the need to introduce artificial viscosity and allowing sharp discontinuities in the flow, such as shock fronts, to be modelled with a thickness of only 1--2 mesh cells.  A problem with previous attempts to use Lagrangian grids to represent turbulent fluid flow is that the grid would become highly tangled as the simulation evolved, significantly compromising its accuracy. \arepo avoids this problem by continuously remaking the grid using the method of Voronoi and Delaunay tessellation (see \citealt{Springel10} for more details). The resulting mesh is adaptable and can be refined to give improved resolution in regions of interest. This allows the study of problems with an extreme dynamic range, that are discontinuous, and that involve fluid instabilities, all while imparting no preferred geometry on the problem.

The chemical evolution of the gas in our simulations is modelled using the hydrogen chemistry of \citet{Glover07a,Glover07b}, together with the highly simplified treatment of CO formation and destruction introduced in \citet{Nelson97}. Full details of the combined network are given in \citeauthor{Glover12a}~(2012a): the network used here is the same the NL97 model in that paper. We assume that the strength and spectral shape of the ultraviolet portion of the interstellar radiation field (ISRF) are the same as the values for the solar neighbourhood derived by \citet{Draine78}; note that this corresponds to a field strength of 1.7~ \citet{Habing68} units. We also include the effects of cosmic rays and adopt a rate $\zeta_{\rm H} = 3 \times 10^{-17} \: {\rm s^{-1}}$ for atomic hydrogen, and a rate twice the size of this for molecular hydrogen.

To treat the attenuation of the ISRF due to H$_{2}$ self-shielding, CO self-shielding, the shielding of CO by H$_{2}$, and by dust absorption, we use the \treecol algorithm developed by \citet{Clark12b}. This algorithm computes a $4\pi$ steradian map of the dust extinction and H$_{2}$ and CO column densities surrounding each \arepo cell, using information from the same oct-tree structure that \arepo uses to evaluate gravitational interactions between cells. The resulting column density map is discretised onto $N_{\rm pix}$ equal-area pixels using the {\sc healpix} pixelation algorithm \citep{healpix}. In the simulations presented here, we set $N_{\rm pix} = 48$. To convert from H$_{2}$ and CO column densities into the corresponding shielding factors, we use shielding functions taken from \citet{Draine96} and \citet{Lee96}, respectively. We assume that the radiation field is uniform and enters through the sides of the box. In this paper, we will deal mainly with densities and projected column densities, and so the main benefit of including the chemistry is to calculate the radiative heating and cooling of the gas self-consistently within the simulation. It is important to model this accurately, as we expect filament formation to be significantly easier when the effective equation of state of the gas is sub-isothermal \citep{Larson85, Peters12b} and so the use of a simple isothermal or polytropic equation of state may lead to unrepresentative results. In future work, we plan to make further use of the chemical information in our simulations by focussing on the observed molecular emission from species such as C$^{18}$O, and exploring what this tells us about the velocity structure of the filaments.

\subsection{Simulations}\label{sec-sim}

\begin{table}
	\begin{center}
	\caption{Summary of simulation properties. \label{sims}}
		\begin{tabular}{l c c c}
   	         \hline
	         \hline
	          ID & Turbulence Type & Initial H$_2$ fraction & Refinement\\
	         \hline
	         S1  & natural mix  & 0.0 & Yes (16)\\
	         S2  & natural mix  & 0.0 & Yes (16)\\
	         S3 & solenoidal   & 0.84 & Yes (16)\\
	         S4 & compressive  & 0.84 & Yes (16)\\
	         S5  & natural mix  & 0.0 & No \\
	         S6 & natural mix & 0.0 & Yes (32) \\
   	         \hline
	         \hline	         
		\end{tabular}
\end{center}
Simulations 1 and 2 have a natural mix of solenoidal and compressive turbulence but different initial turbulent seeds. Simulations 3 and 4 explore the effects of different types of turbulence. Simulations 5 and 6 are the same as Simulation 1, but are at different resolutions.
\end{table}

In this paper, we consider six simulations of $10^4$\msun solar metallicity molecular clouds. The simulations will be referred to as S1 to S6 and Table \ref{sims} summarises their properties. Four of the simulations are initially fully atomic and two are 84\% molecular. This value was chosen to match the mean molecular composition of molecular clouds in \citet{Smith14a} and is necessary for the case with compressive turbulence because high densities are reached before the gas has been able reach an evolved chemical state. The initial condition is that of a uniform sphere of gas with an initial number density of $n \sim 100$ \cmc~ and a radius of 7~pc. This is embedded in a larger 65 pc periodic box containing a tenuous warm medium with a temperature of several thousand Kelvin. The size of this larger box is such that the dense cloud at its centre never encounters the box edges. This setup is analogous to an isolated molecular cloud in the ISM.

To the central spherical cloud we apply a turbulent velocity field with a $P(k)\propto k^{-4}$ power spectrum such that the velocity field obeys Larson's scaling laws \citep{Larson81}. In order to test the role of turbulence in generating the filaments we apply three different turbulent fields to the gas which have 1) solenoidal turbulence, 2) compressive turbulence and 3) a natural mix of both (i.e.\ two-thirds solenoidal, one-third compressive) using the approach of \citet{Girichidis11}. The magnitude of the root mean square turbulent velocity is normalised such that the clouds have an equal amount of kinetic and gravitational potential energy at the start of the simulations. Since turbulence does not provide an isotropic pressure force the cloud does not remain spherical once the field is applied to the gas: regions with outward velocities expand outward into the surrounding medium, and regions with inward velocities move towards the cloud centre. Nevertheless the turbulence in the simulations is decaying and so the clouds will ultimately collapse under their own self-gravity.

Our base resolution is $10^{-2}$ \msun per cell, but in five of the simulations we add an additional refinement criteria to ensure that the Truelove criteria \citep{Truelove97} is satisfied even at the highest densities. This ensures that there is no artificial fragmentation in the gas, and that the filament profiles are well resolved at their centres where closely spaced density measurements are needed to constrain the best fit profiles. Specifically, in S1 to S4 we require the Jeans length to be resolved by a minimum of 16 cells at all densities. If this is not the case, the grid is refined until the condition is satisfied. We calculate this Jeans length assuming a fixed temperature of 10 K for the gas. Very little of the dense gas is colder than this, and most is at least a few K warmer, and so this gives us a conservative estimate for the size of the Jeans length. Although we expect this to be sufficient resolution to properly represent the structure of the filaments, we have nevertheless investigated whether our results are resolution dependent by performing two additional simulations, S5 and S6. In simulation S5, we switched off Jeans refinement and ran the entire simulation at our base resolution of $10^{-2}$ \msun, corresponding to a spatial resolution of around 0.02~pc for gas at a number density $n \sim 10^{4} \: {\rm cm^{-3}}$. In simulation S6, on the other hand, we continued to use Jeans refinement, but now required the Jeans length to be resolved by a minimum of 32 cells at all densities.

\begin{figure}
\begin{center}
\includegraphics[width=3in]{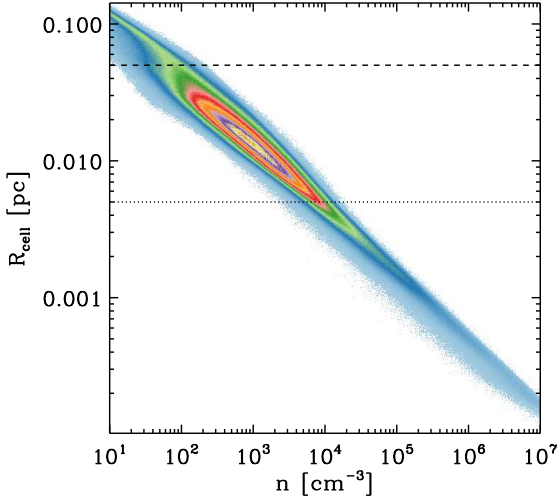}
\caption{Spatial resolution as a function of density in our Jeans refinement scheme. The dashed line shows a cell radius of 0.05 pc, which is equivalent to the filament centre width found observationally. The dotted line shows a radius ten times smaller. Gas at the centre of the simulated filaments has a number density of $10^4 \: {\rm cm^{-3}}$ or higher, meaning that the filament core is always well resolved.}
\label{resolution}
\end{center}
\end{figure}

In \arepo, the cells do not have a constant radius but instead a value given approximately by $r_{\rm cell}=(3V/4\pi)^{1/3}$ for a volume set by the cell mass and density. This means that in higher density regions the gas is resolved at a higher spatial resolution. Figure~\ref{resolution} shows the spatial resolution as a function of density for a simulation with our standard Jeans refinement. The centres of filaments have densities in excess of $n=10^{4}$\cmc ~ which equates to a spatial cell radius of $r_{\rm cell}=5.0\E^{-3}$ pc with our standard  refinement scheme. Based on their observations, A11 find a filament width of around 0.1 pc, and we see from Figure~\ref{resolution} that in gas with $n=10^{4}$\cmc, we resolve this length scale with at least ten cells. In practice, this is an under-estimate of the actual resolution, since the densities that we find at the centres of our filaments are often considerably larger than $n=10^{4} \: {\rm cm^{-3}}$, and the cell radii are correspondingly smaller. We will discuss issues of resolution when studying filaments in more detail in Section \ref{results-res}.

Star formation is modelled in the simulation using sink particles \citep{Bate95}. These were first introduced into \arepo by \citet{Greif11} and we use a slightly modified version of this routine here. Above number densities of $10^7$\cmc, we check whether the densest cell in the simulation and its neighbours satisfy the following three conditions: 1) the cells are gravitationally bound, 2) they are collapsing and 3) the divergence of the accelerations is less that zero, so the particles will not re-expand (see also \citealt{Federrath10a}). If all these conditions are satisfied the cell and its neighbours are replaced with a sink particle, which interacts with the gas cells purely through gravitational forces. Additional material can be accreted by the sink particles if it is within an accretion radius of $r_{\rm acc}= 0.01$ pc and is bound to the sink.  We use relatively large sink particles in this study in order to focus on the geometry of the filaments without interference from very dense collapsing cores which would distort the average of the filament profile. Consequently the sinks should not be thought of as `stars', but as collapsing cores. 

\subsection{Filament Identification}\label{sec-ident}
\begin{figure*}
\begin{center}
\begin{tabular}{c c}
\includegraphics[width=3.5in]{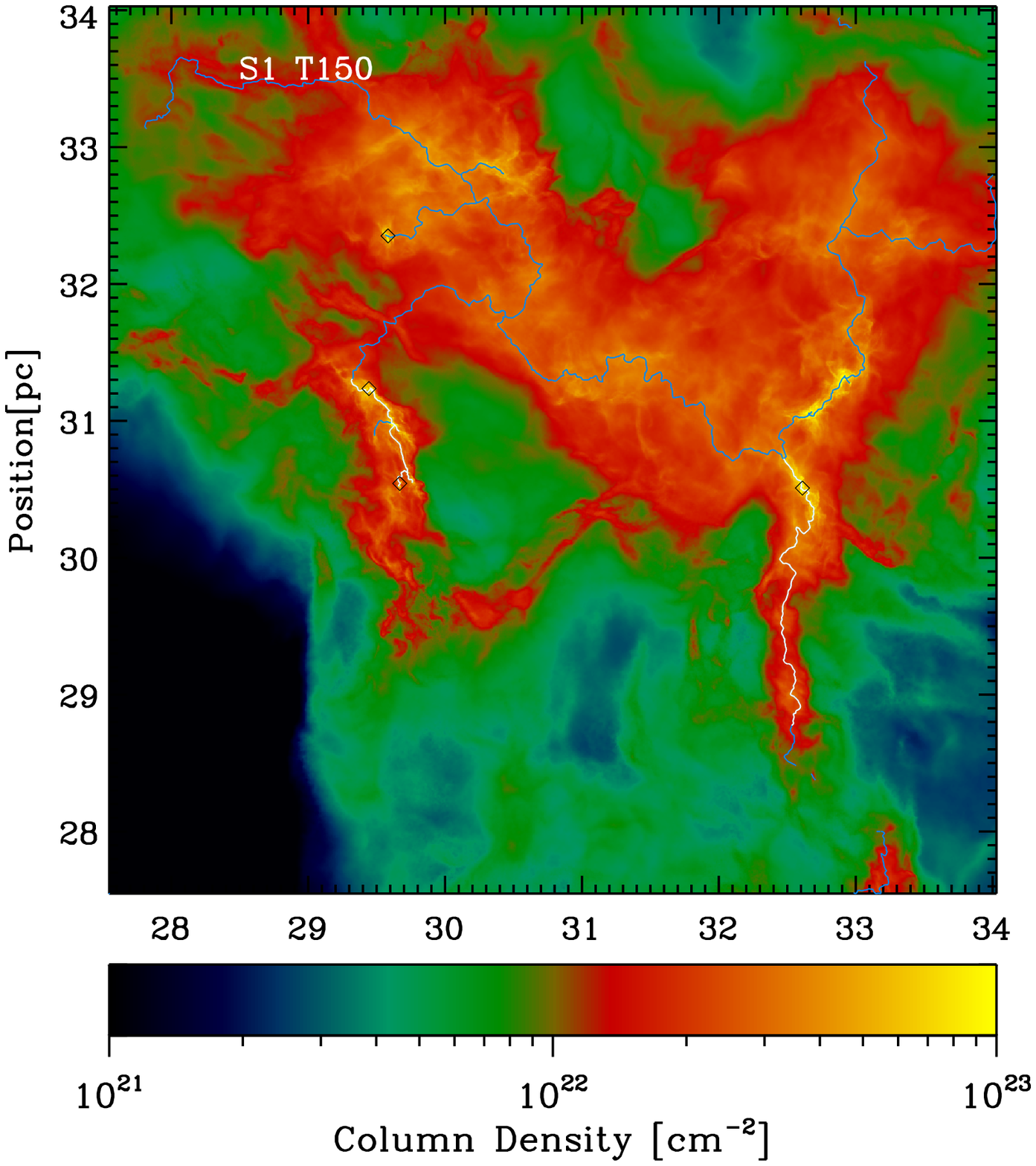}
\includegraphics[width=3.5in]{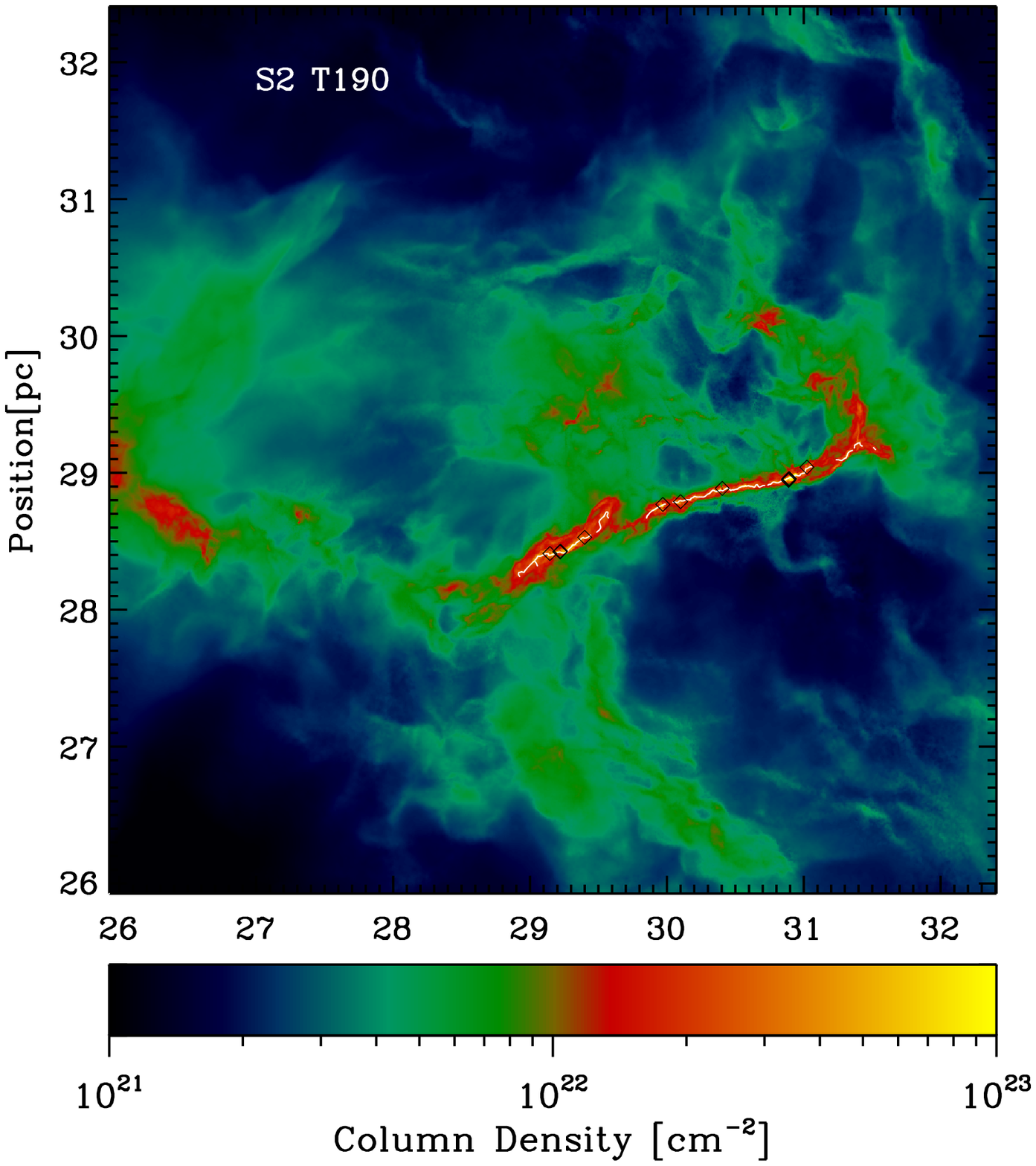}\\
\includegraphics[width=3.5in]{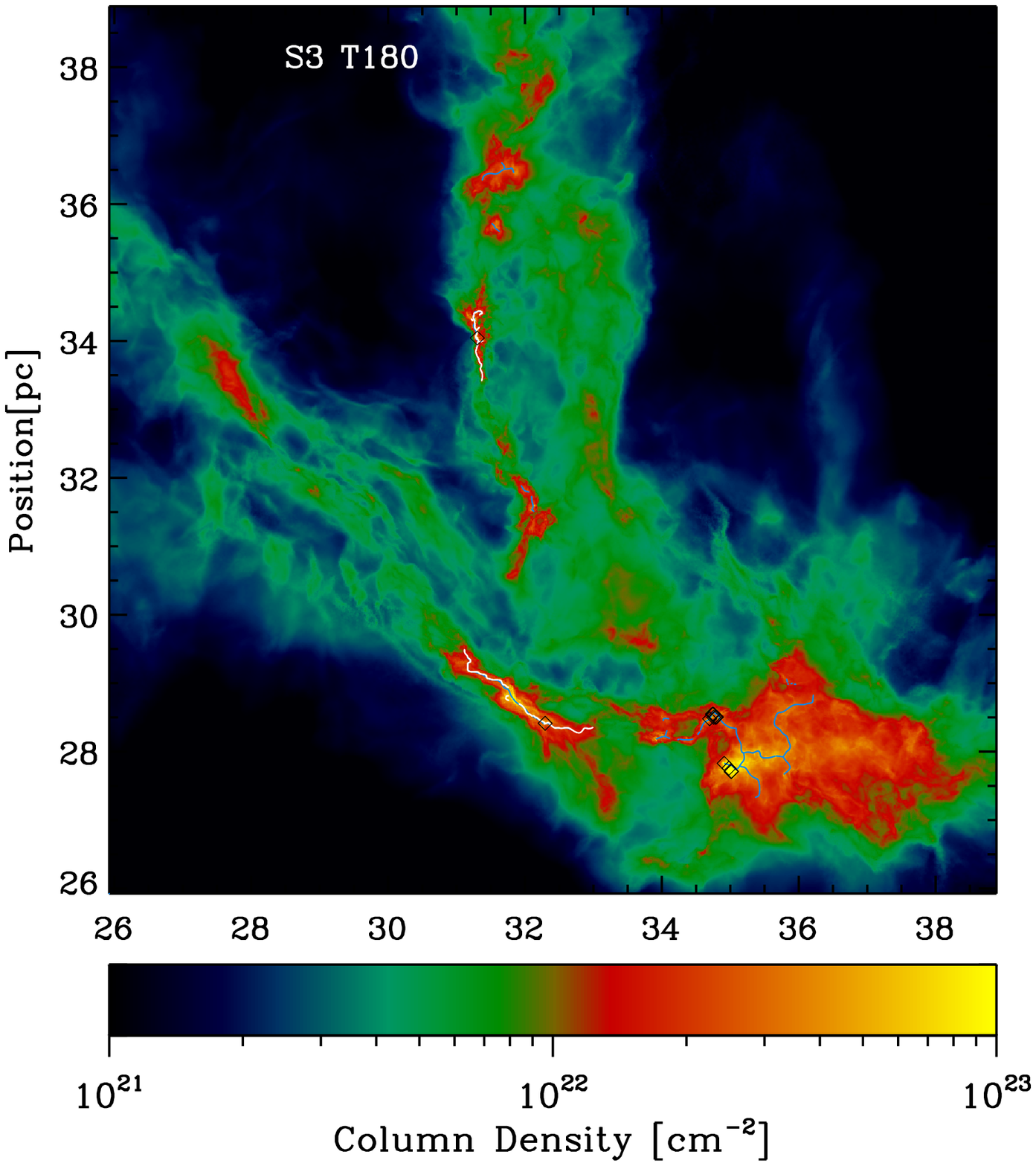}
\includegraphics[width=3.5in]{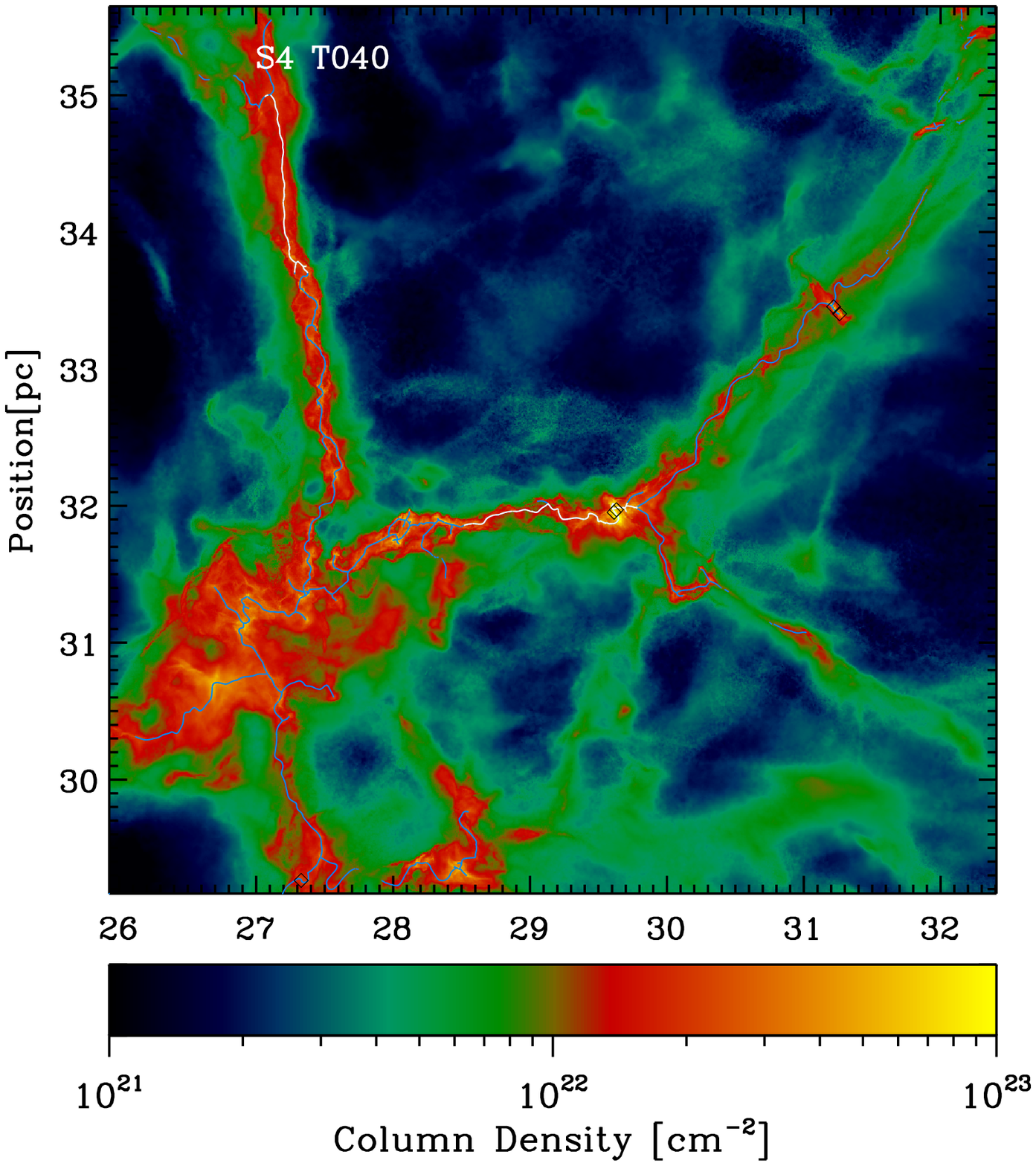}\\
\end{tabular}
\caption{The four simulations at the end of the analysed period. The filament skeleton identified using \disperse is shown in blue, the filaments analysed are shown in white, and star-forming cores are shown by black diamonds.}
\label{s1234}
\end{center}
\end{figure*}

The filaments are identified using the \disperse (DIScrete PERsistent Structures Extractor) algorithm \citep{Sousbie13}. This algorithm constructs a Morse-Smale complex from an input density distribution and identifies the critical points where the density gradient is zero. These can then be connected to delineate structures in the data. Filaments are found by connecting the points such that maxima are connected to saddle-points along Morse field lines. As noise in the data may lead to artificial structures being identified, a density threshold is applied to the data below which filaments are not identified. It is important to note that as the method works by connecting maxima to saddle-points, cores are found to be embedded in filaments almost by definition.

As the motivation of our study is to understand the properties of star-forming filaments like those seen in nearby molecular cloud observations (e.g.\ A11) we apply \disperse to \textit{column density} projections of our simulations. We calculate the column densities by projecting the simulation results onto a 1000 pixel by 1000 pixel grid with side length 6.5~pc, centred on the cloud. The size of each pixel in this grid is $6.5 \times 10^{-3}$~pc, much smaller than the 0.1~pc width measured by A11. 

When finding filaments in our column density images, we apply a cut at $2\E^{22} \: {\rm cm^{-2}}$ so that only dense filaments are identified. This simplifies the network to the filaments which will undergo future star formation. Using \disperse we generate a list of filament segments describing the local orientation of the filament on the column density grid. Figure \ref{s1234} shows an example of the filaments found by \disperse when applied to our simulations. We find the column density profile perpendicular to the filament vector in each segment. Following the approach of A11 these are then averaged to give the mean filament column density profiles and their standard deviation. In Section \ref{results-3D} we will then compare the filaments profiles identified in column density with those found by applying \disperse to the 3D density distribution.

\subsection{Filament profiles}\label{sec-prof}

We now have the filament profiles as a function of position with error bars equivalent to the standard deviation of the average at each point. We fit the commonly used Plummer-like function to describe the number density $n(r)$, 
\begin{equation}\label{Plum-dens}
n(r)=\frac{n_{\rm c}}{[1+(r/R_{\rm flat})^2]^{p/2}} + B_{\rm 3D} [{\rm cm}^{-3}]
\end{equation}
where $n_{\rm c}$ is the central density, $R_{\rm flat}$ is a radius within which the profile is flat, $p$ sets the slope of the power law fall off beyond this radius, and $B_{\rm 3D}$ represents the background density. When projected into 2D this becomes a surface density profile of
\begin{equation}\label{Plum-surface}
\Sigma(r)=  \frac{ A_p n_{\rm c} R_{\rm flat}}{[1+(r/R_{\rm flat})^2]^{(p-1)/2}} + B_{\rm 2D} [{\rm cm}^{-2}]
\end{equation}
where $A_p$ is a finite constant factor resulting from the normalisation during the projection (in this work we hold $A_p$ constant at a value of $\pi/2$ as in \citealt{Ostriker64}). An inspection of this second profile immediately reveals a possible degeneracy between $R_{\rm flat}$, $n_{\rm c}$ and $p$. It is equally possible to gain a good fit to the column density by increasing $n_{\rm c}$ and decreasing $R_{\rm flat}$ or doing the converse. Whichever of these is chosen will then have a knock-on effect on the value of $p$ due to the change in $(r/R_{\rm flat})$. We will discuss this degeneracy further in Section \ref{robust}.

In addition to the Plummer-like profile, we also fit a Gaussian function to get an estimate of the filament width
\begin{equation}\label{gaussian}
\Sigma(r)= A_{22} \exp \left( -\frac{r^2}{2\sigma^2} \right) + B_{\rm G}
\end{equation}
In Equation~\ref{gaussian}, $A_{22}$ gives the height of the Gaussian (usually of the order of $10^{22}$\cms), $r$ is the radius in pc, $\sigma$ is the standard deviation, and $B_{\rm G}$ is the background surface density. 
The key finding of A11 was that filaments in nearby star-forming regions had a characteristic width of 0.1 pc. It is important to emphasise that this is the width of only the inner section of the filament profile: it is not derived for the filament as a whole. Typically, radii up to $0.3-0.4$ pc are included in the Gaussian fits (Arzoumanian, private communication), although the full distribution is used for the Plummer fit. In our work we will calculate the filament FWHM in three different ways: 1) fitting a Gaussian with a background component to only the data within $0.35$ pc of the filament spine for comparison to A11, 2) fitting a Gaussian with a background component to all the data within $1$ pc of the filament spine, and 3) finding the FWHM of the raw column density distribution without fitting.

We find the best fits to the Gaussian and Plummer-like profiles using the \mpfit non-linear least-squares fitting programme \citep{Markwardt09}. This allows us to use the standard deviation in the average profile as a measure of the uncertainty in the profile and propagate this through to estimate the uncertainties in our derived parameters.  

\section{Column Density Profiles}\label{results-col}

\subsection{Cloud Morphologies}\label{results-morph}

\begin{figure*}
\begin{center}
\begin{tabular}{c c c}
\begin{overpic}[scale=.3]{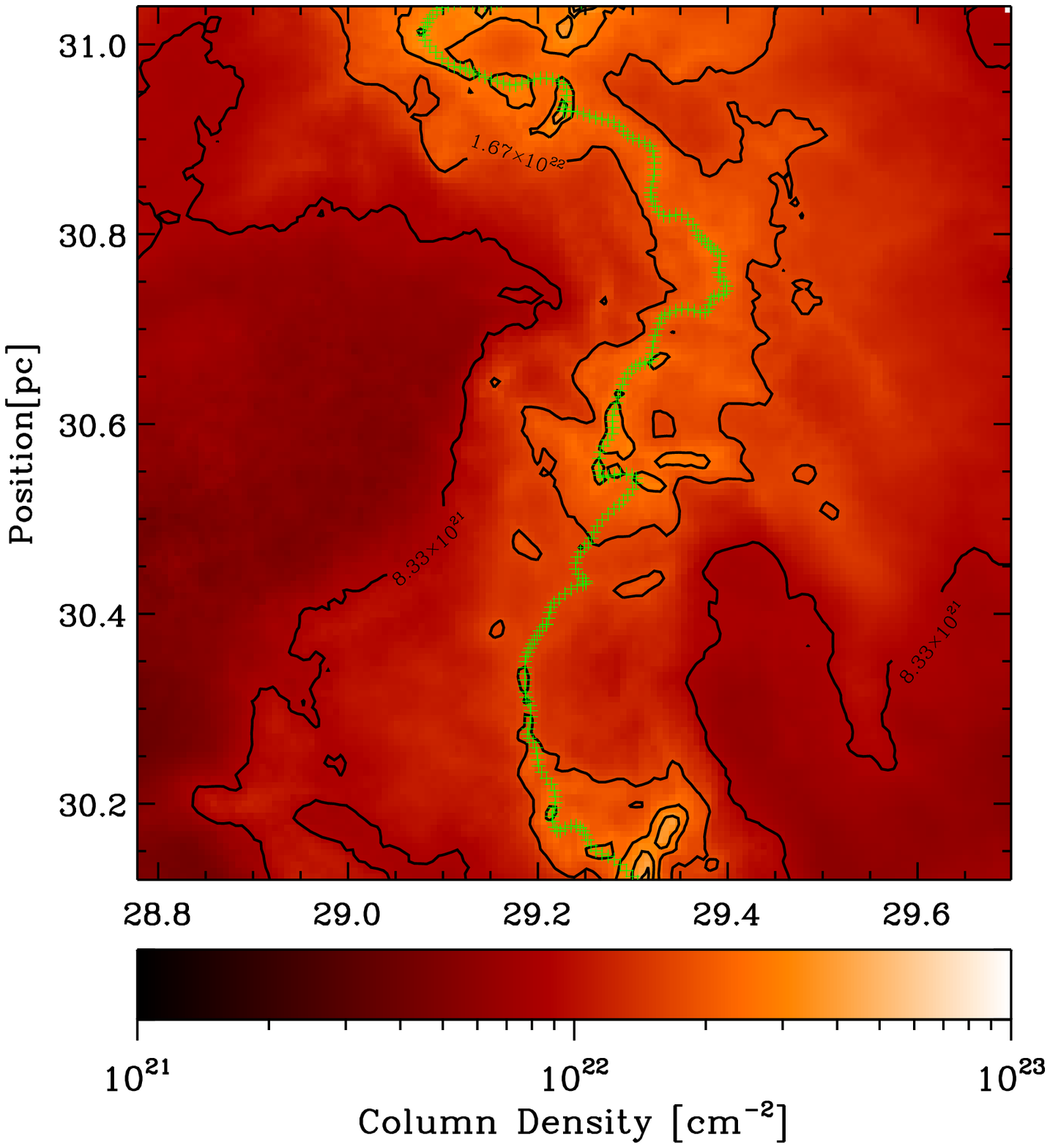}
\put (75,90) {\makebox(0,0){{\color{white}S1-F1-T130}}}
\end{overpic}
\begin{overpic}[scale=.3]{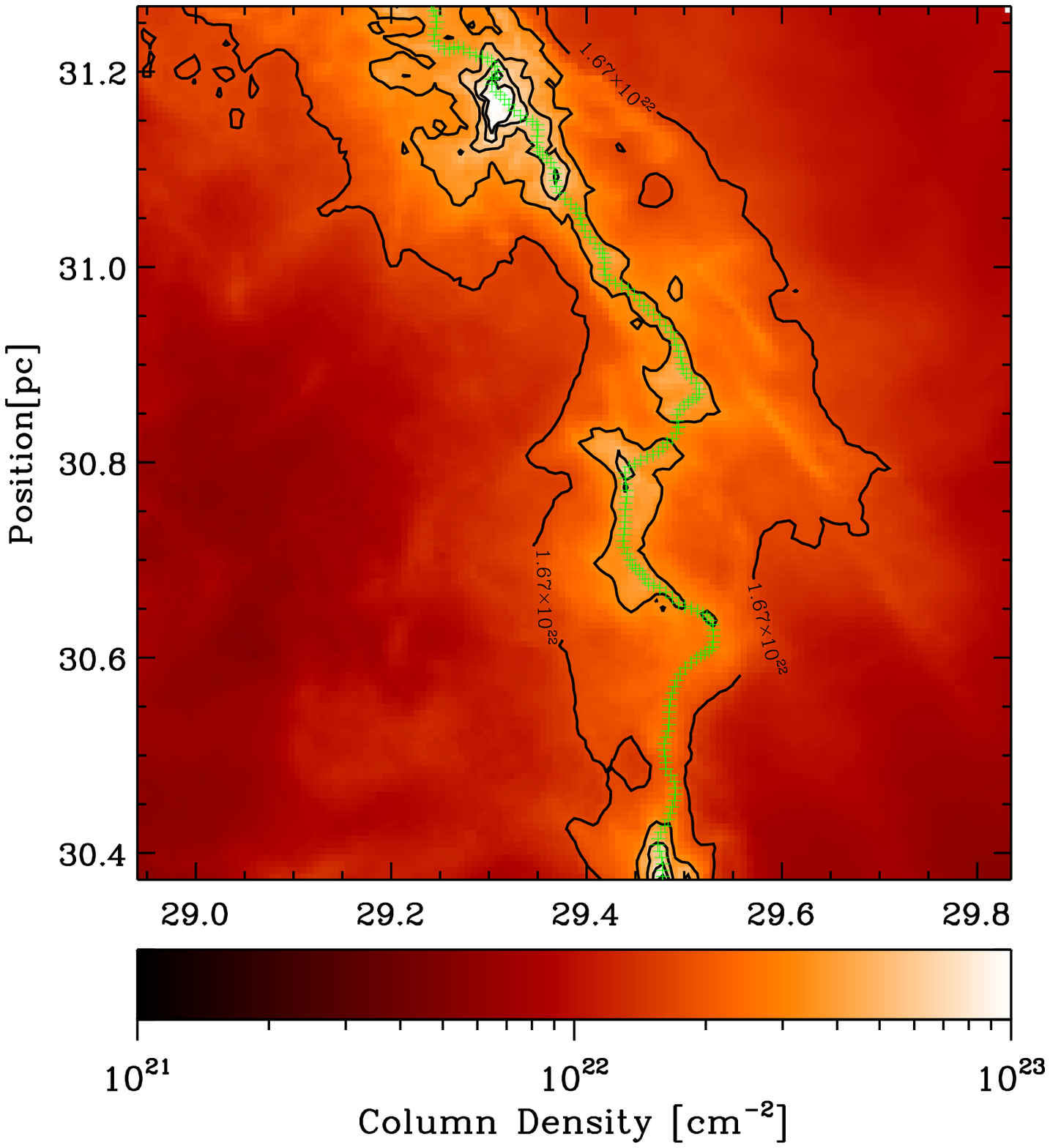}
\put (75,90) {\makebox(0,0){{\color{white}S1-F1-T140}}}
\end{overpic}
\begin{overpic}[scale=.3]{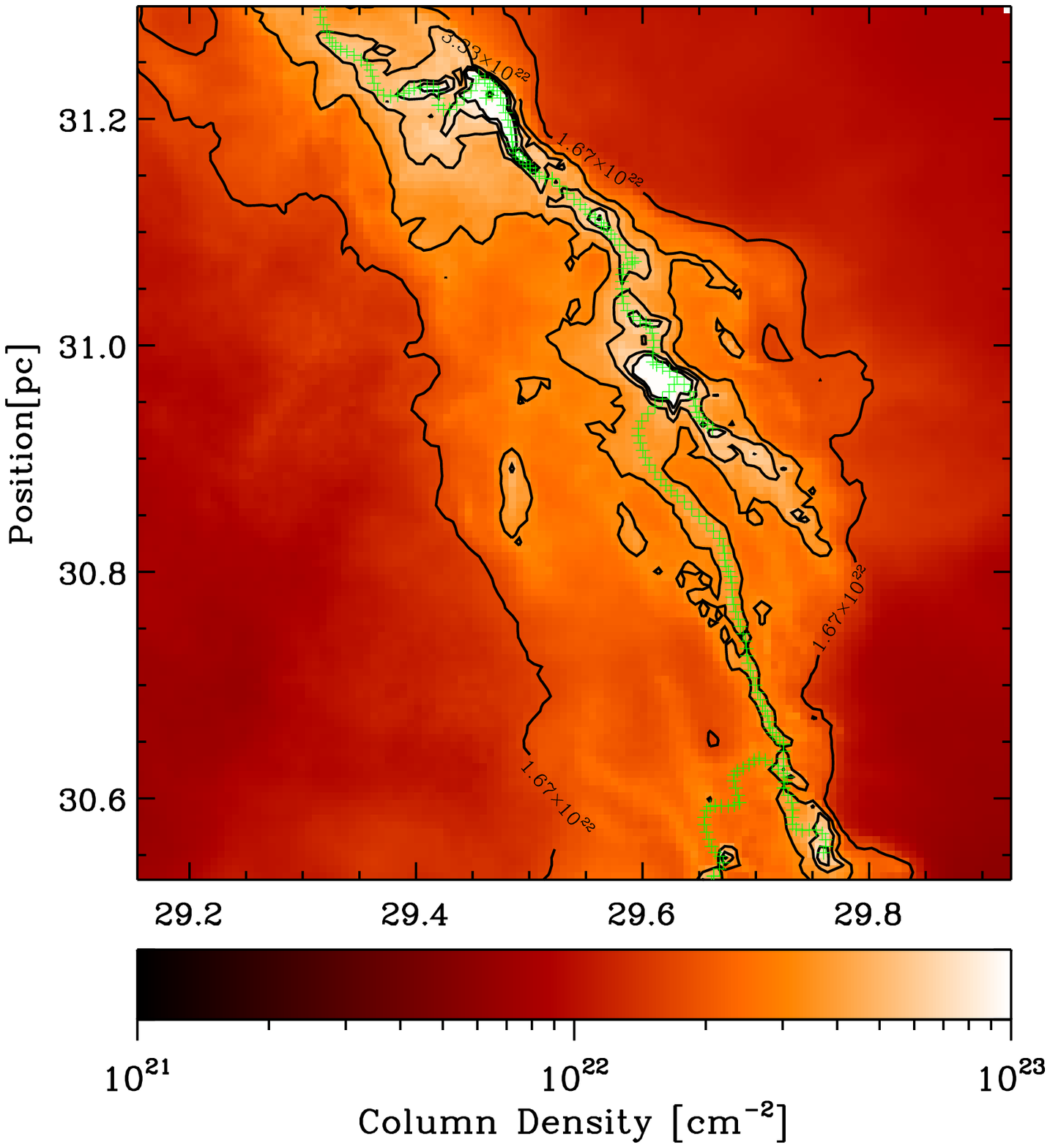}
\put (75,90) {\makebox(0,0){{\color{white}S1-F1-T150}}}
\end{overpic}
\\
\begin{overpic}[scale=.38]{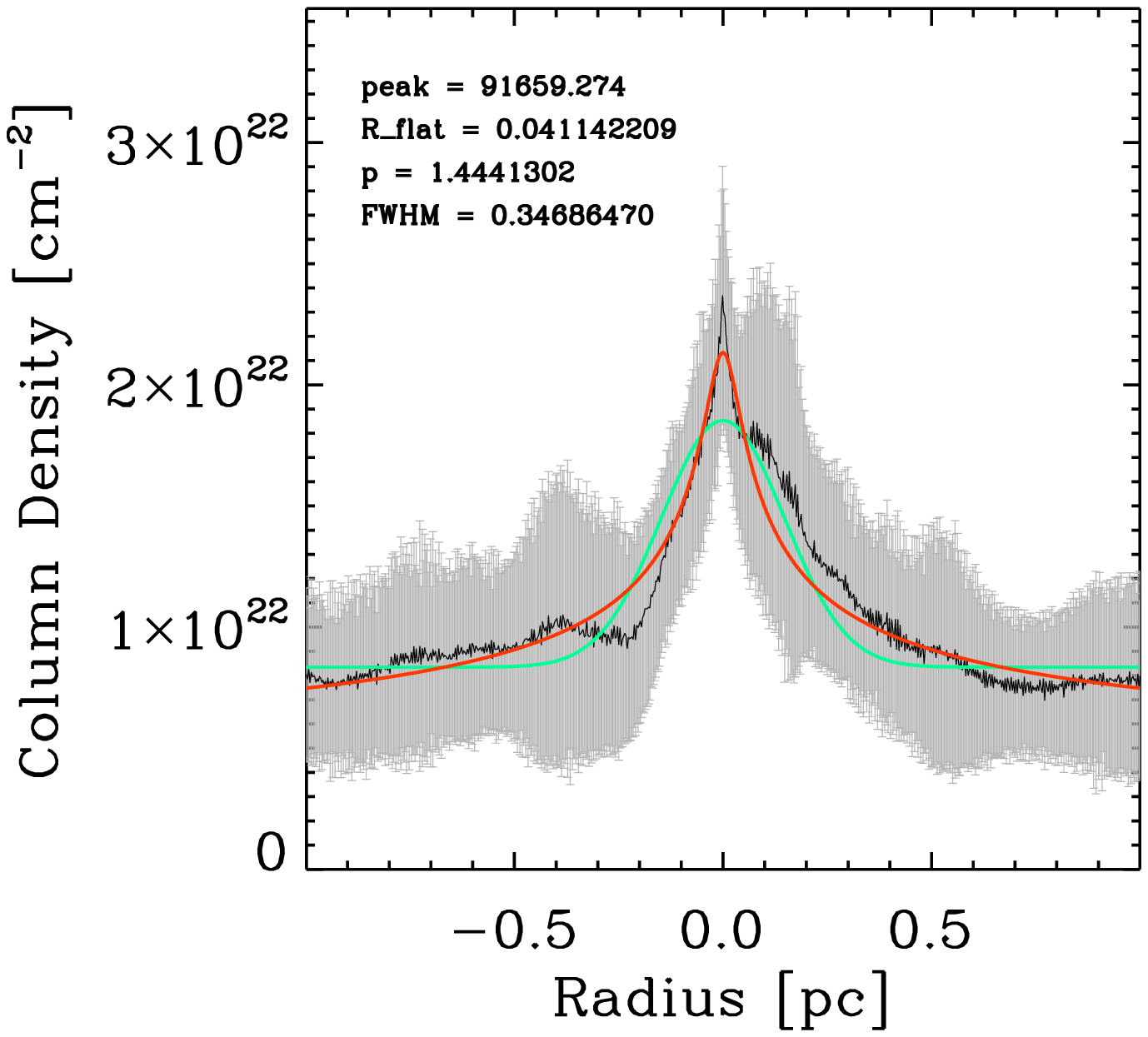}
\put (75,80) {\makebox(0,0){{\color{white}S1-F1-T130}}}
\end{overpic}
\begin{overpic}[scale=.38]{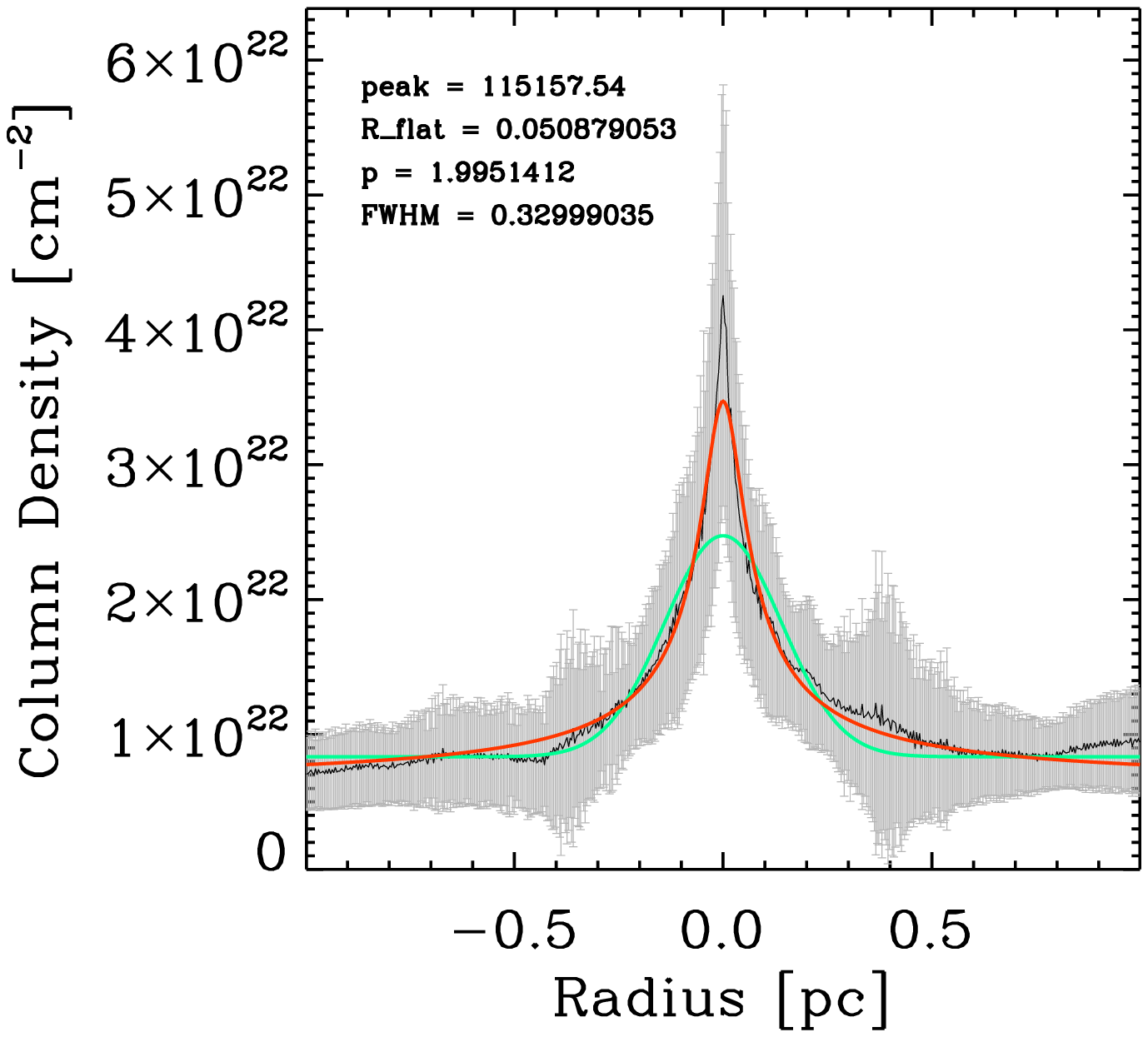}
\put (75,80) {\makebox(0,0){{\color{white}S1-F1-T140}}}
\end{overpic}
\begin{overpic}[scale=.38]{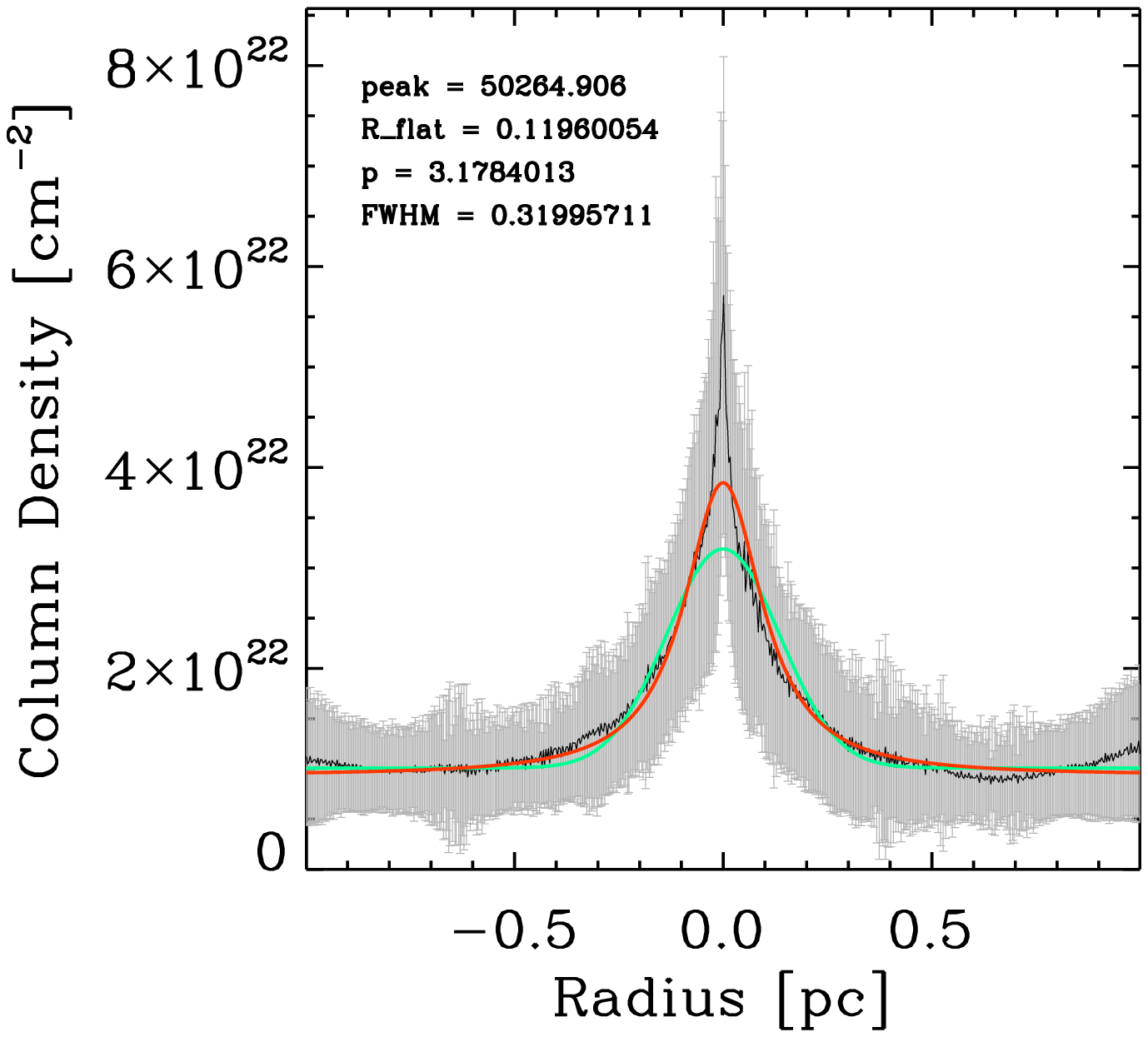}
\put (75,80) {\makebox(0,0){{\color{white}S1-F1-T150}}}
\end{overpic}
\end{tabular}
\caption{\textit{Top}: Evolution of the column density in Filament 1 in S1. The green points show the 2D spine of the filament. \textit{Bottom}: The black lines shows the average column density profile, the grey line shows the standard deviation of the column density average, the red line shows the best Plummer-like fit, and the green line the best Gaussian fit to the entire distribution.}
\label{S1F1}
\end{center}
\end{figure*}

Figure \ref{s1234} shows the morphology of the simulations using our standard refinement scheme at the end of the analysed period. The filament skeleton identified using \disperse is shown in grey, and the filaments that we focus our analysis on are shown in white. S1 contains a natural mix of compressive and solenoidal turbulent modes and forms a compact clouds with a number of filaments. We select two filaments for analysis which extend from the cloud centre out into the surrounding medium and can therefore be identified without confusion from any other surrounding filaments. S2 also has a mixed turbulent field but a different turbulent seed. In this case the dense gas has become concentrated into a single long filament in which most of the star formation occurs. S3 contains only solenoidal modes and so the gas distribution is initially not very filamentary (since filaments are mainly formed by the compressive modes of turbulence). However, as the cloud begins to collapse under its own self-gravity, filaments form. S4 contains only compressive modes and rapidly forms many filaments due to turbulent compression of the gas before gravity has had much of a chance to act on the gas. The fact that simulations S3 and S4 both exhibit strong filamentary structure demonstrates that both gravity and turbulence are effective at generating
filamentary structure.

We begin our analysis of the filaments when the first sink particle (i.e.\ collapsing core) is formed in each simulation, and then take a second and third snapshot of the filament properties at intervals of $8.7\E^5$ yr, which corresponds to a time difference of 10 in code units. This enables us to investigate how the filament profiles evolve in time. We use the following naming scheme to label our filaments; the first part of the filament ID (S) denotes the simulation, the second part (F) shows the filament number, and the last part (T) describes the time in the simulation in code units.
 
Figure~\ref{S1F1} shows the column density evolution of Filament 1 in S1. The gas becomes more concentrated at the centre of the filament as it evolves, as shown by the increasing central density of the filament average profile. The filament is not smooth, but has multiple clumpy structures along its length.

\begin{figure*}
\begin{center}
\begin{tabular}{c c c}
\begin{overpic}[scale=.3]{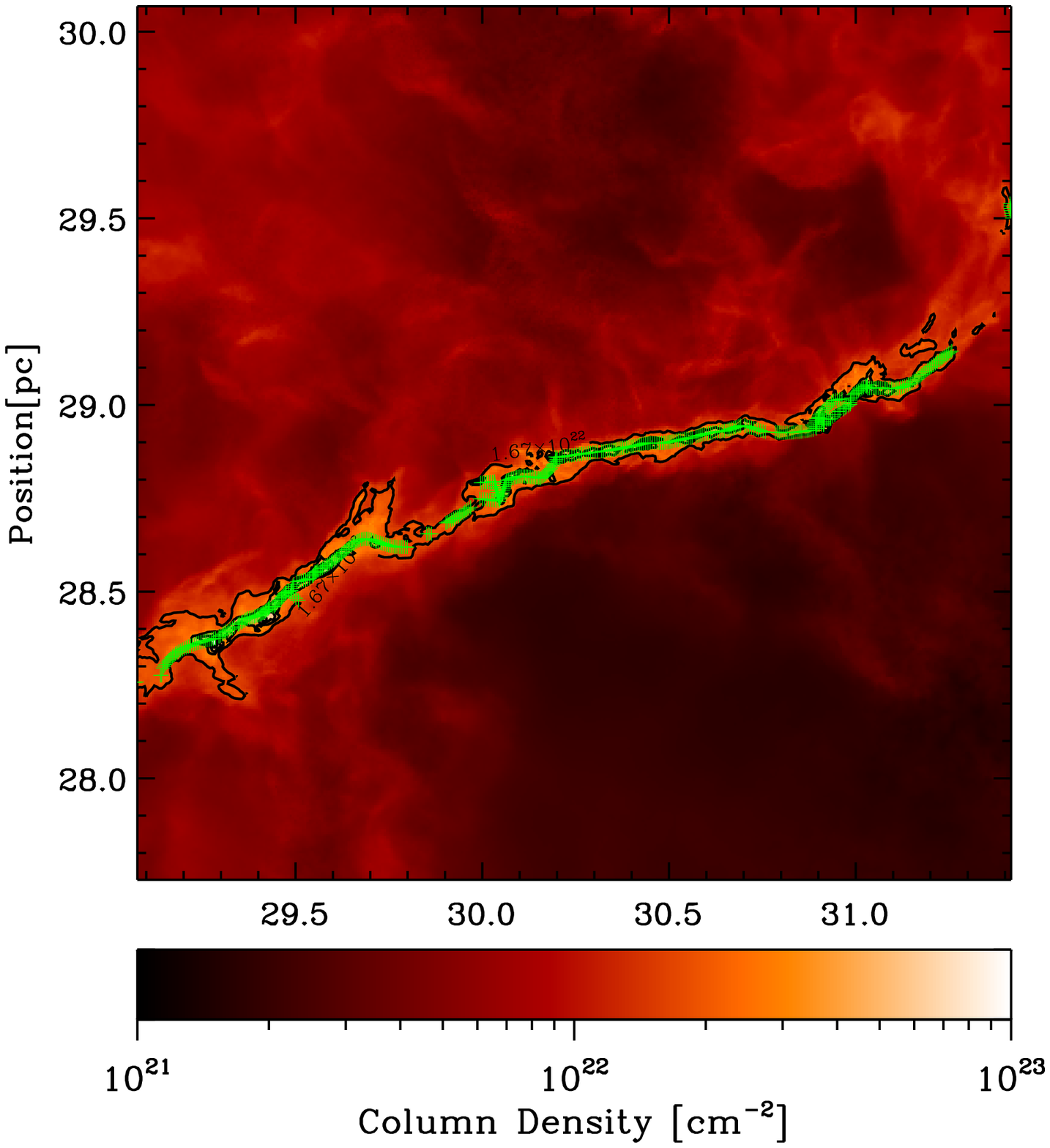}
\put (75,90) {\makebox(0,0){{\color{white}S2-F1-T180}}}
\end{overpic}
\begin{overpic}[scale=.3]{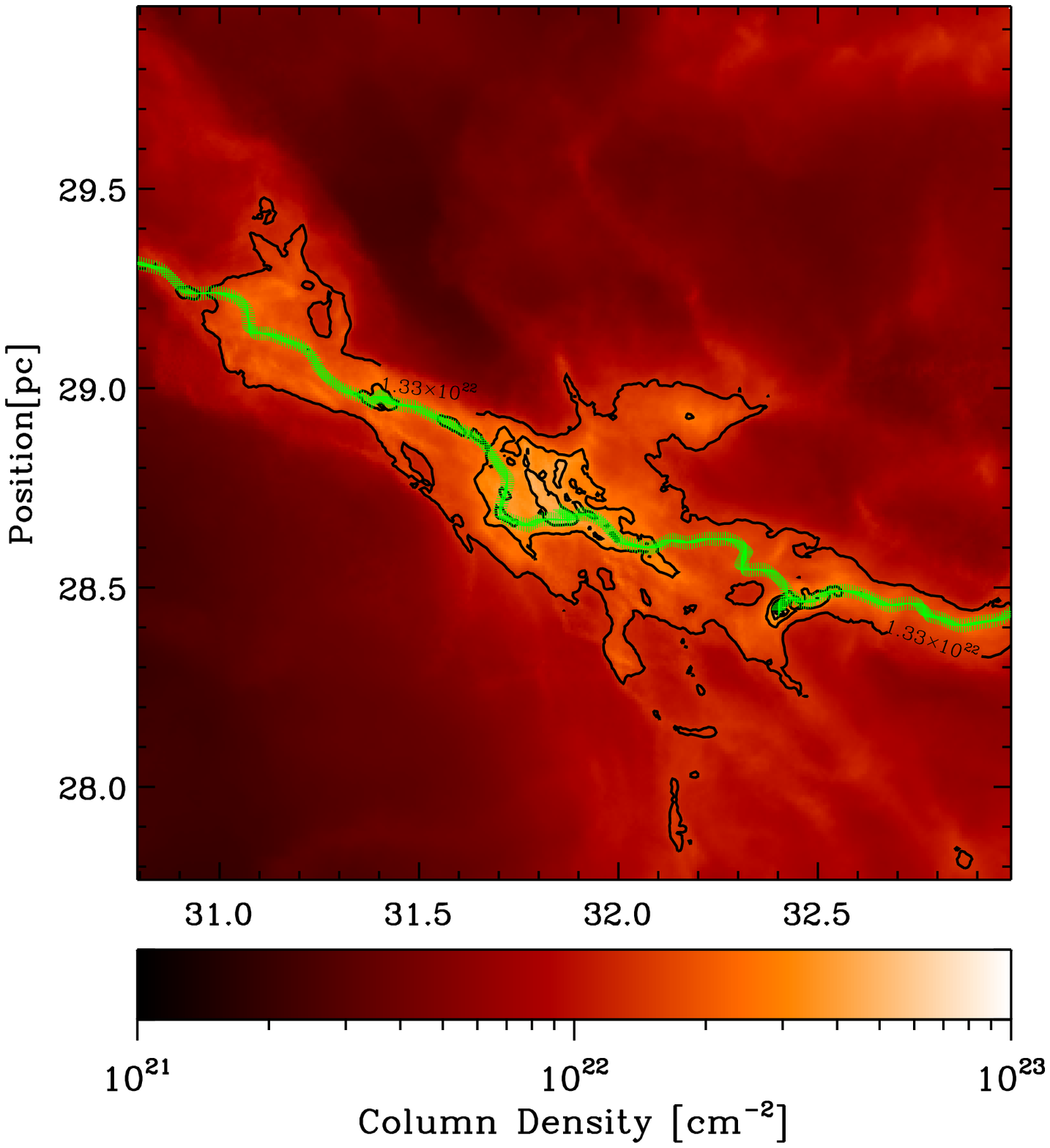}
\put (75,90) {\makebox(0,0){{\color{white}S3-F1-T170}}}
\end{overpic}
\begin{overpic}[scale=.3]{./final_figs/Fg4_contour_F1_S180}
\put (75,90) {\makebox(0,0){{\color{white}S4-F1-T030}}}
\end{overpic}
\\
\begin{overpic}[scale=.38]{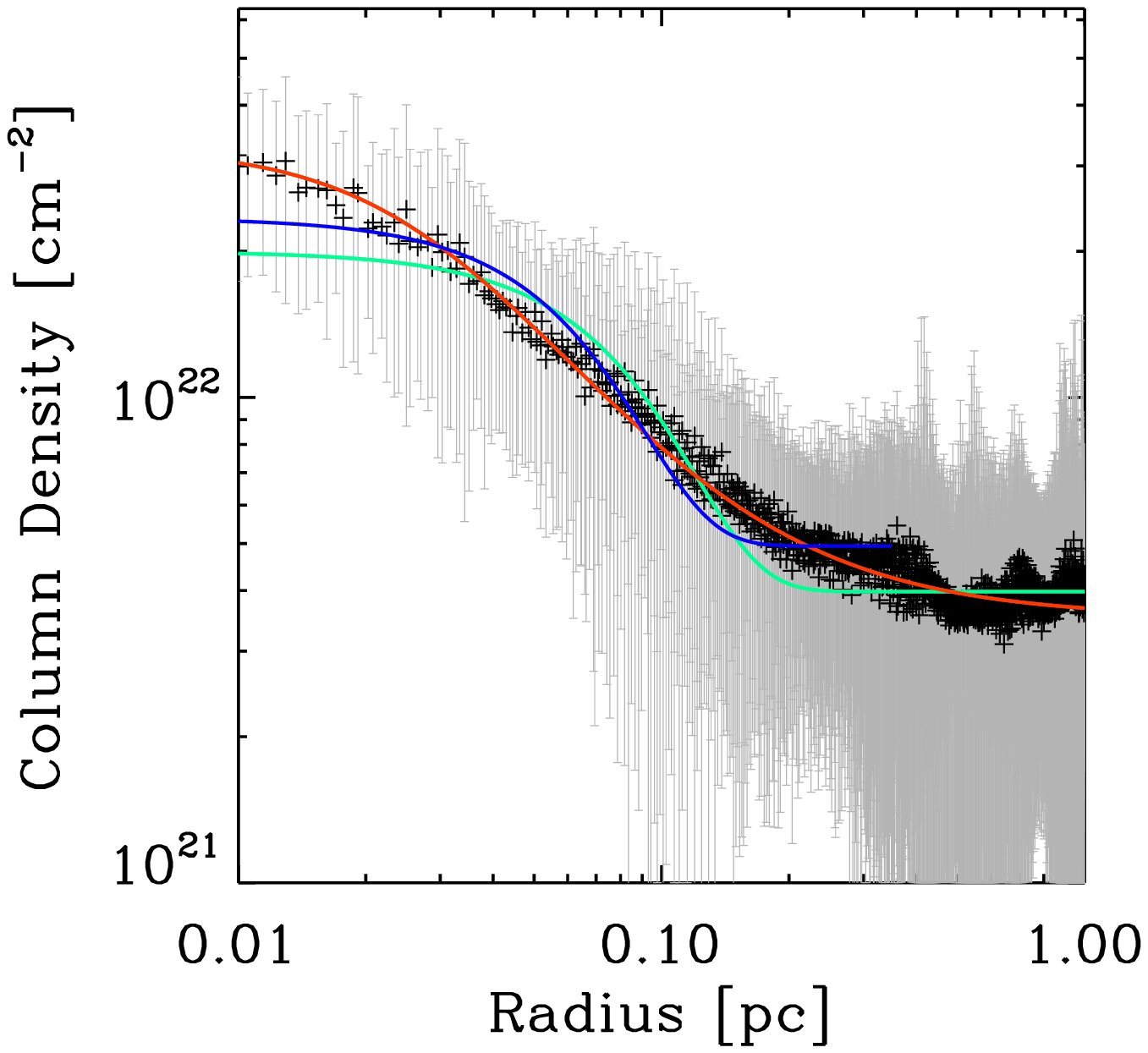}
\put (75,80) {\makebox(0,0){S2-F1-T180}}
\end{overpic}
\begin{overpic}[scale=.38]{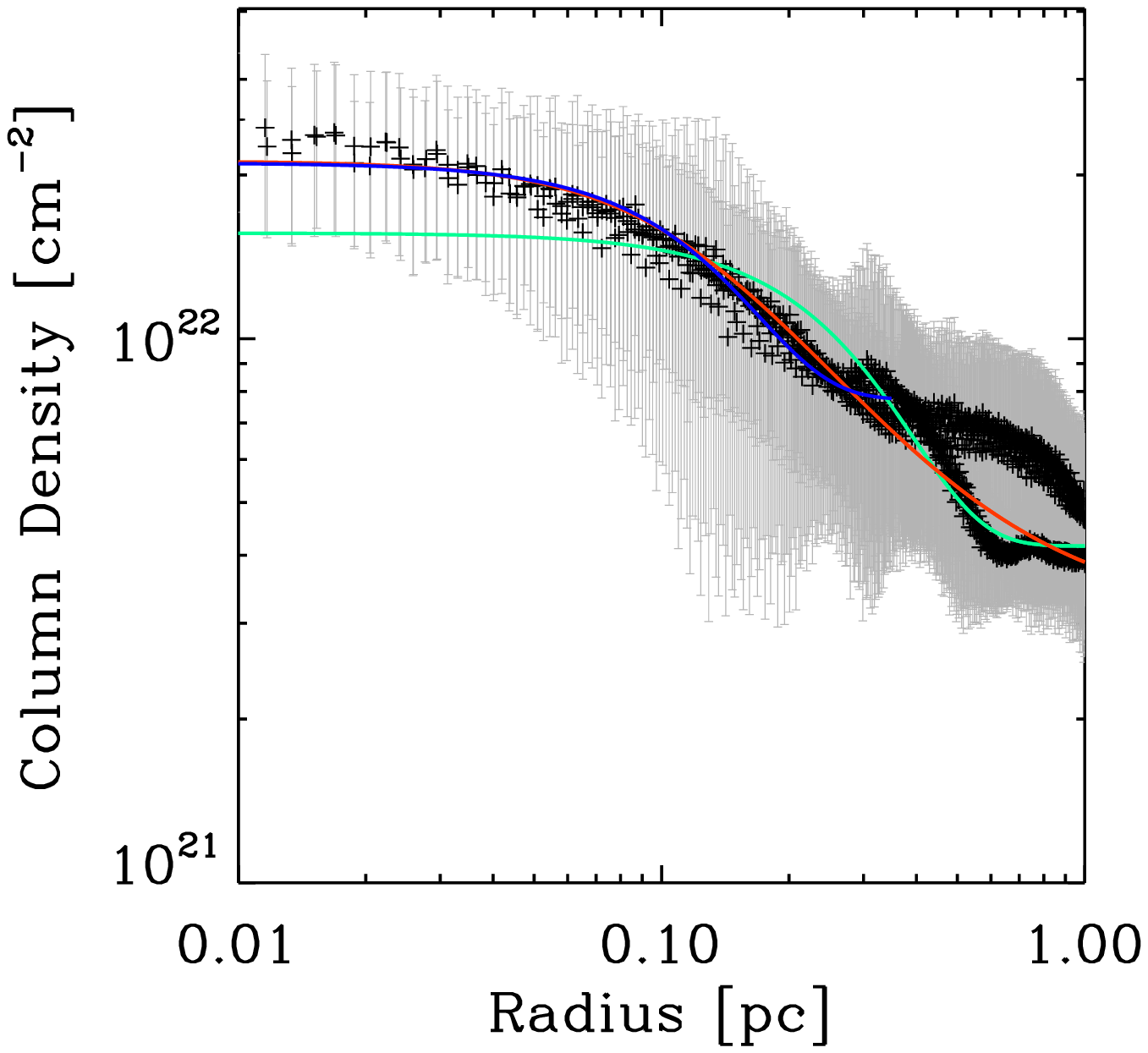}
\put (75,80) {\makebox(0,0){S3-F1-T170}}
\end{overpic}
\begin{overpic}[scale=.38]{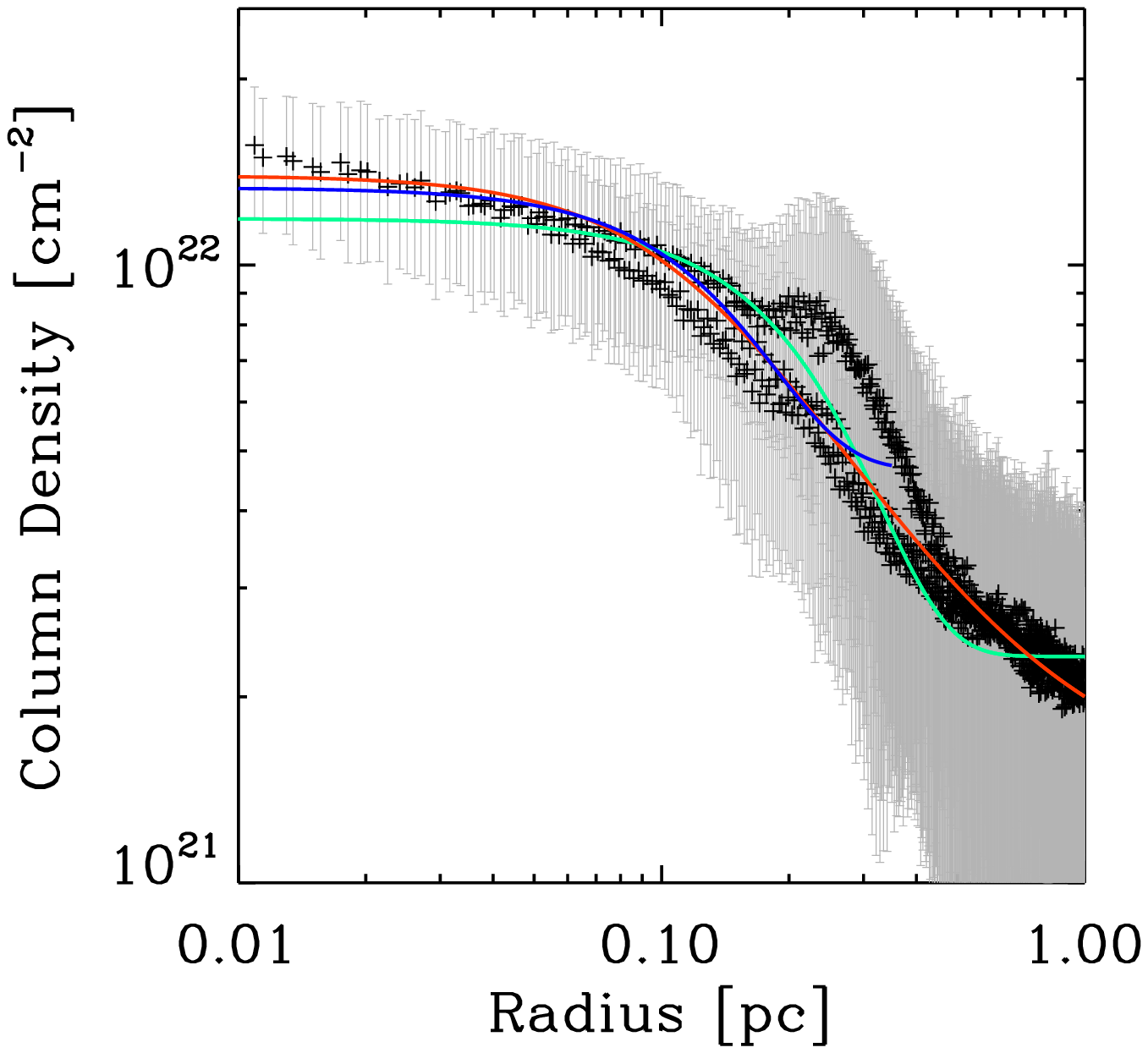}
\put (75,80) {\makebox(0,0){S4-F1-T030}}
\end{overpic}
\end{tabular}
\caption{\textit{Top}: The filaments S2F1T180, S3F1T170 and S4F1T030 from left to right. The green points show the 2D spine of the filament. \textit{Bottom}: The filament column density profiles in log space. The black lines show the average column density profiles, the grey error bars show the standard deviation of the average, the red line shows the best Plummer-like fit, the green line shows the best Gaussian fit to data within 1~pc of the filament spine, and the blue line shows the best Gaussian fit to the inner 0.35 pc of the filament.}
\label{S234F1}
\end{center}
\end{figure*}

Figure \ref{S234F1} shows the column density distribution and radial column density profile of a representative filament in each of simulations S2, S3 and S4. The filaments from S2 and S3 both show the same clumpy structure as in S1, suggesting incipient fragmentation along the length of the filament. However, in S4 the column density contours align more closely with the 2D spine of the filament. Despite this, the averaged column density profiles do not look substantially different in the three cases, which suggests that the effect of clumping within the filament is averaged out.

\subsection{Best Fits}\label{results-fits}

\begin{table*}
	\centering
		\begin{tabular}{l l l l l l l}
   	         \hline
	         \hline
		ID  & length & mass & $n_{\rm c}$ & $p$ & $R_{\rm flat}$ & FWHM \\
		 & [pc] & [\msun] & [$10^4$ \cmc] & & [$10^{-2}$ pc] & [pc] \\
	         \hline 
	         S1-F1-T130 & 1.46 & 84.1 & $9.17 \pm 77\%$ & $1.44 \pm 18\%$ & $4.11 \pm 52\%$ & $0.347 \pm 7\%$ \\
		 S1-F1-T140 & 1.15 & 98.1 & $11.5 \pm 55\%$ & $2.00 \pm 16\%$ & $5.09 \pm 42\%$ & $0.330 \pm 6\%$ \\
		 S1-F1-T150 & 1.36 & 111.9 &$5.03 \pm 47\%$ & $3.18 \pm 28\%$ & $11.9 \pm 37\%$ & $0.320 \pm 7\%$ \\
		\hline
	          S1-F2-T130 & 2.54 & 123.8 & $2.17\pm 35\%$ & $2.69 \pm 22\%$ & $12.3 \pm 29\%$ &$0.338 \pm 5\%$ \\
	          S1-F2-T140 & 2.45 & 161.8 & $3.67\pm 41\%$ & $2.66 \pm 22\%$ & $10.3 \pm 33\%$ &$0.307 \pm 6\%$ \\
	          S1-F2-T150 & 2.53 & 194.1 & $6.80\pm 56\%$ & $2.03 \pm 21\%$ & $7.78 \pm 42\%$ &$0.353 \pm 7\%$ \\   
	         \hline
	          S2-F1-T170 & 2.14 & 102.4 & $8.77\pm 35\%$ & $2.27 \pm 10\%$ & $4.53 \pm 26\%$ &$0.268 \pm 5\%$ \\
	          S2-F1-T180 & 2.65 & 122.5 & $21.3\pm 37\%$ & $2.47 \pm 9\%$   & $2.84 \pm 26\%$ &$0.154 \pm 6\%$ \\
	          S2-F1-T190 & 2.55 & 120.3 & $34.4\pm 49\%$ & $2.07 \pm 8\%$   & $1.75 \pm 34\%$ &$0.177 \pm 6\%$ \\
	          \hline
	          S3-F1-T160 & 3.54 & 144.9 &$3.02 \pm 30\%$ &$2.31 \pm 15\%$ & $8.00 \pm 24\%$ & $0.301 \pm 5\%$ \\
	          S3-F1-T170 & 2.32 & 152.5 & $2.89 \pm 26\%$&$2.50 \pm 12\%$ & $13.0 \pm 20\%$ & $0.525 \pm 3\%$ \\
	          S3-F1-T180 & 1.92 & 195.16&$3.59 \pm 27\%$&$2.82 \pm 12\%$ & $14.0 \pm 21\%$ & $0.503 \pm 3\%$ \\
	          \hline
	          S3-F2-T160 & 1.54 & 57.5 & $ 3.90 \pm 18\%$ & $1.39 \pm 5\%$   & $6.31 \pm 16\%$ &$0.506 \pm 5\%$ \\
	          S3-F2-T170 & 1.87 & 68.3 & $ 2.01 \pm 41\%$ & $2.50 \pm 22\%$ & $11.1 \pm 33\%$ &$0.353 \pm 6\%$ \\
	          S3-F2-T180 & 1.48 & 55.8 & $ 11.8 \pm 63\%$ & $1.94 \pm 14\%$ & $3.86 \pm 47\%$ &$0.342 \pm 8\%$ \\
   	         \hline
	          S4-F1-T020 & 0.73 & 50.6 & $9.42 \pm 42 \%$ &$1.29 \pm 2\%$   & $2.75 \pm 40\%$ & $0.270 \pm 9\%$ \\
	          S4-F1-T030 & 1.88 & 75.6 & $2.08 \pm 27\%$ & $2.44 \pm 14\%$ & $12.5 \pm 22\%$ & $0.421 \pm 4\%$ \\
	          S4-F1-T040 & 1.52 & 77.2 & $2.96 \pm 21\%$ & $2.85 \pm 11\%$ & $12.1 \pm 18\%$ & $0.358 \pm 3\%$ \\
	         \hline
	          S4-F2-T160 & 1.38 & 41.9 & $54.2 \pm 69\%$ & $1.29 \pm 6\%$ & $0.581 \pm 54\%$ & $0.360 \pm 6\%$ \\
	          S4-F2-T170 & 1.51 & 70.5 & $19.8 \pm 41\%$ & $1.50 \pm 2\%$ & $2.19 \pm 31\%$   & $0.413 \pm 7\%$ \\
	          S4-F2-T180 & 1.75 & 88.1 & $5.73 \pm 54\%$ &$2.46 \pm 21\%$& $9.09 \pm 42\%$   & $0.371 \pm 6\%$ \\
	         \hline
	         \hline
	         Mean & 1.92 & 104.6 & 10.7 & 2.20 & 7.43 & 0.348 \\
	         St. Deviation & 0.63 & 44.0 & 12.5 & 0.54 & 4.25 & 0.091\\
	         \hline
	         \hline
		\end{tabular}
	\caption{Best-fit filament profiles found from fitting the simulated column density distributions out to a radius of 1~pc from the central filament spine using \disperse. The best fit values of $n_{\rm c}$, $p$, and $R_{\rm flat}$ come from fitting a Plummer-like function (Eq.~\ref{Plum-surface}), and the FWHM comes from fitting a Gaussian (Eq.~\ref{gaussian}). \label{master}}
\end{table*}

Table \ref{master} summarises the lengths, masses and best fits obtained from all the filaments (examples of which can be seen in Figures \ref{S1F1} and \ref{S234F1}). In this section we concentrate on the best fits to Plummer-like and Gaussian models using the full data range. In Section \ref{width}, we will discuss the Gaussian FWHM found using a restricted range as in A11. The filaments have lengths of a few pc and total masses of around one hundred solar masses (calculated from the column density within 0.3 pc of the 2D spine of the filament). All of our studied filaments are super-critical \citep{Inutsuka97} and are either forming stars or will do so in the near future. The filaments have an average peak density $n_{\rm c} = 1.07\E^5$ \cmc, an average power-law index of $p = 2.20$, and a mean flattening radius $R_{\rm flat} = 0.074$~pc. Table \ref{lit} compares the average fitted parameters to observational estimates. The observational datasets also include some longer and lower density filaments that are not necessarily star-forming. Nevertheless, filaments in molecular clouds are found to have fairly universal properties in A11 and so our filament properties should match the observations if the simulations are an accurate depiction of reality.

\begin{table}
	\centering
		\begin{tabular}{l l l l l}
		\hline
	         \hline
	         Dataset & $p$ &$\Delta p$ & $R_{\rm flat}$ [pc] & $\Delta R_{\rm flat}$ [pc]   \\
	         \hline
	         This work & 2.20 & 0.54 & 0.074 & 0.042 \\
	         A11 & 1.68 & 0.27 & 0.046 & 0.026 \\
	         J12a & 2.28 & 1.71 & 0.078 & 0.099 \\
	         \citet{Palmeirim13} & 2.0 & - & 0.035 & - \\
	         \citet{Malinen12} & 2.27 & - & 0.043 & - \\
	         \hline
	         \end{tabular}
	
	\caption{Comparison to observations for the Plummer-like models. A11 represents \citet{Arzoumanian11} and J12a represents \citeauthor{Juvela12a}~(2012a). A11 and J12a are averages from large datasets, and the $\Delta$ shows the standard deviation of the average. The remaining papers are measurements from specific filaments. The power-law index $p$ found in the simulations is in excellent agreement with the observations. The flattening radius from the simulations is slightly higher than in some studies, but is very similar to that found in J12a.  
	\label{lit}}
\end{table}

In agreement with the observations, we typically find that the best fitting Plummer-like profiles have a relatively shallow power law index $p \sim 2$. The prediction for an isothermal cylinder in hydrostatic equilibrium is $p=4$ \citep{Ostriker64}. Clearly both our theoretical models and the observations disagree with this value. One potential explanation proposed for this disparity is support by magnetic fields, as magnetised filaments typically exhibit shallower profiles \citep{Fiege00}. However, our simulations do not include magnetic fields, and so magnetic support cannot be the explanation for the shallow profiles that we find for our simulated filaments. Another potential explanation  comes from \citet{Fischera12} who found that while filaments that were highly over-pressured with respect to their environment tended towards $\rho(r) \propto r^{-4}$ at large radii, in agreement with the classical isothermal result, a smaller over-pressure of 6-12 times the background level results in profiles more consistent with $\rho(r) \propto r^{-2}$ at large radii. Ongoing accretion in a direction perpendicular to the filament may also flatten the density profile. Both of these effects are operative in our simulated filaments.

The average flattening radius in the simulations is in excellent agreement with the findings of J12a, but is slightly higher than in the other studies. \citet{Palmeirim13} studied the Taurus B211/3 filament and found $R_{\rm flat}=0.035$~pc, and \citet{Malinen12} studied a pair of filaments in TMC-1 and found values of $p = 2.27$ and $R_{\rm flat} = 0.043$~pc. These are long dense filaments which are among the most visible structure within their parent molecular cloud. Within the simulated dataset the most comparable filament is S2-F1 which has some of the narrowest flattening radii within our distribution and so there may be some selection effects. Overall, the properties of the simulated filaments are broadly consistent with the observations.

When we compare the derived properties for S3, which has initially only solenoidal turbulent modes, with S4, which has initially only compressive modes, we see that there is little difference between the best fit profiles in the two cases. This implies that regardless of how the filament was initially formed, the same physical processes determine the gas distribution in both cases. Pressure confinement and accretion from the surrounding medium are present in all cases and so we suggest that the models of \citet{Fischera12} and \citet{Heitsch13a,Heitsch13b} may be a way to understand filament evolution in an idealised analytical manner. However, while these models are a good way to understand the physical principles behind filament formation, they lack the complexity seen in the simulations. One major difference is that analytical models typically adopt the assumption of hydrostatic equilibrium, but the filaments in our simulations are dynamically evolving density features (as we will discuss in Section \ref{form}). The simulations also show considerable spatial variability with the filaments not being perfectly smooth continuous objects.

\subsection{Robustness of Plummer-like fit}\label{robust}

\begin{figure*}
\begin{center}
\begin{tabular}{c c}
\includegraphics[width=2.7in]{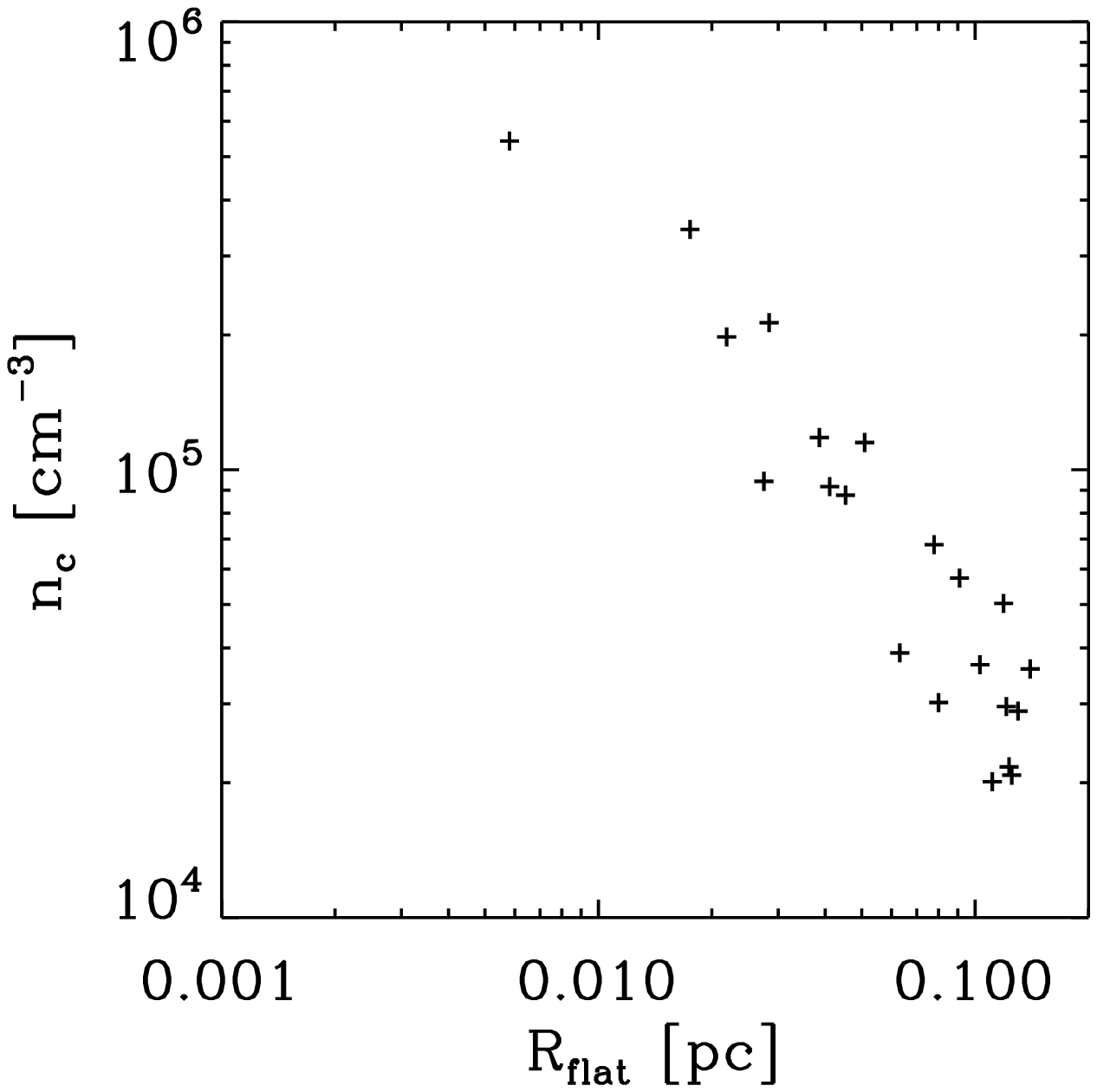}
\includegraphics[width=2.7in]{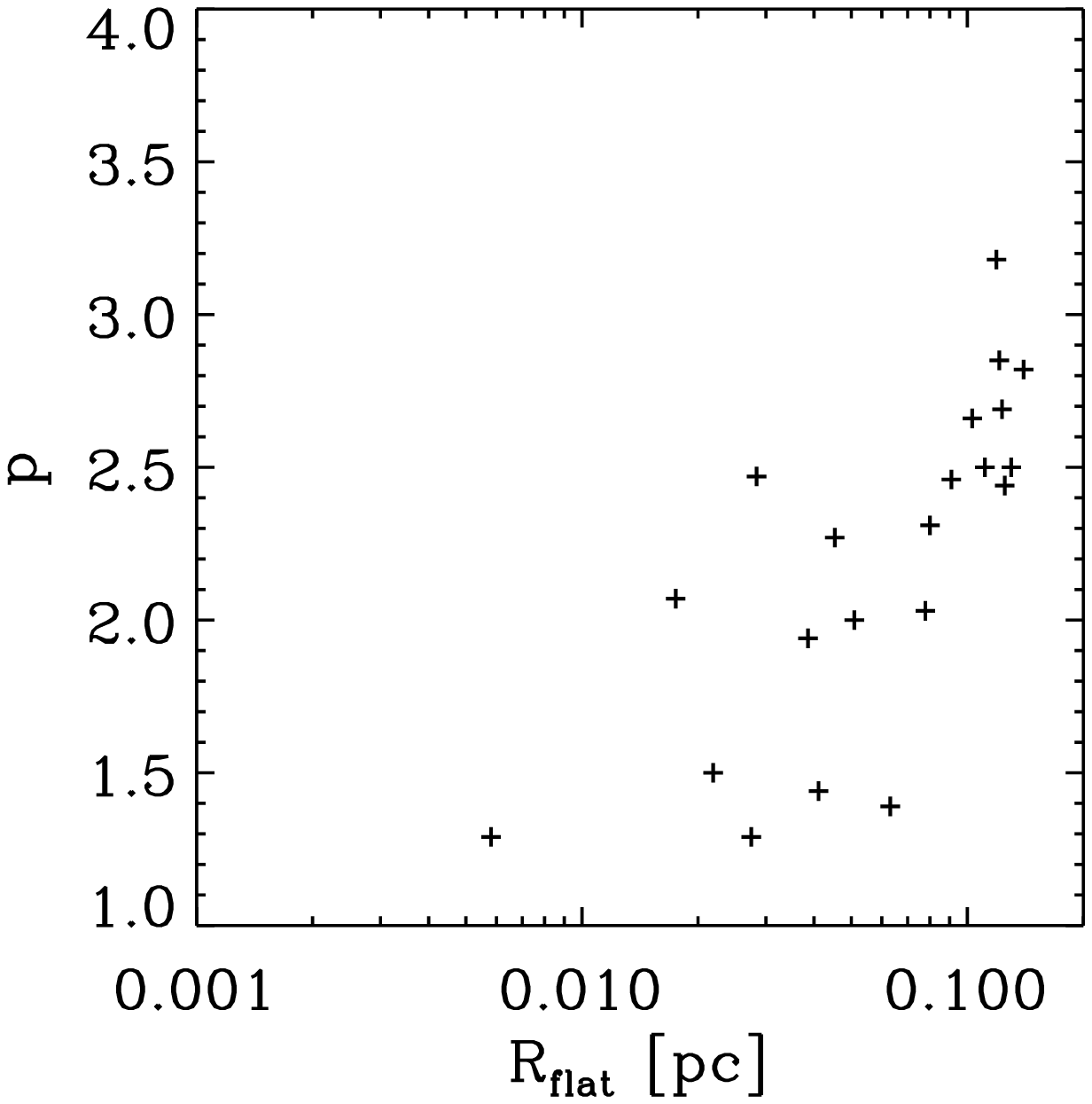} \\
\end{tabular}
\caption{Dependence of the flattening radius $R_{\rm flat}$, central density $n_{\rm c}$ and power law index $p$  upon each other in the Plummer-like model best fits.}
\label{fil_eg}
\end{center}
\end{figure*}

Now that we have established that the simulated filaments are in good agreement with observations, we can examine how robust our determination of the fitted parameters is. In Table \ref{master} we show the estimated uncertainties in the fitted parameters by propagating our errors during the $\chi$-squared minimisation using \mpfit. The Plummer-like model parameters have large uncertainties on their best fit values, particularly in the value of the flattening radius $R_{\rm flat}$ and the central density $n_{\rm c}$, which can have uncertainties of up to 77\%. As mentioned in Section \ref{sec-prof}, this is likely due to the fact that the numerator of the column density profile contains the product of $n_{\rm c}$ and $R_{\rm flat}$ leading to a possible degeneracy between them in the best fit. 

In Figure \ref{fil_eg} we show how the three key Plummer model parameters depend on one another. The central density $n_{\rm c}$ is anti-correlated with the flattening radius $R_{\rm flat}$, while the power $p$ is correlated with $R_{\rm flat}$. If we hold $R_{\rm flat}$ constant during the fit at a value $R_{\rm flat}=0.04$ pc, the uncertainties in $n_{\rm c}$ and $p$ drop to around 5\% (see Table~\ref{rfixed}), demonstrating that there is little degeneracy between these two parameters. When fitting the profiles with a fixed $R_{\rm flat}$, we obtain a higher mean central density, $n_{\rm c} = 1.65 \times 10^{5} \: {\rm cm^{-3}}$, and a shallower power law index, $p = 1.67$, than when we allow all three parameters to vary. It is perhaps interesting to note at this juncture that A11 found a mean $R_{\rm flat}$ of around 0.04 pc and a power law index $p=1.68$, very similar to the value we find for the same $R_{\rm flat}$. It is also important to acknowledge that the degeneracy involved in fitting Plummer-like models to filaments was noted already by \citeauthor{Juvela12b}~(2012b), who produced synthetic dust emission maps of filaments generated in high-resolution isothermal magnetohydrodynamical simulations, and found that the best fit profiles that they obtained were sensitive to the noise added to the data.

\subsection{Evolution of the filament profiles}

\begin{figure}
\begin{center}
\includegraphics[width=2.5in]{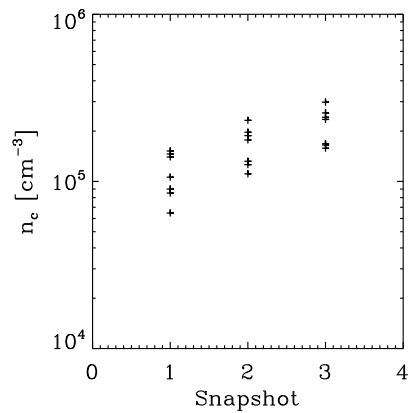} 
\caption{Evolution in central density in successive snapshots when the central radius is held constant at 0.04 pc.}
\label{fil_ev}
\end{center}
\end{figure}

The filaments from our simulations are extracted at three different times to see how they evolve. We find that in general the best fits in Table \ref{master} show no clear evolutionary trend for the filaments. While this is initially quite surprising it is consistent with the findings of \citet{Heitsch13a} that the filament widths remain quite similar for much of their lifetime. However, it is inconsistent with the finding by the same author that the central density of the filament should increase with time. It should be noted that \citet{Heitsch13a} uses an assumption of hydrostatic equilibrium which is not truly valid for our dynamically evolving filaments. However, we still compare to this model since it is one of the few analytical works in the literature that attempts to unite filaments and accretion. Figure \ref{fil_eg} shows that a smaller flattening radius becomes a better fit for higher central densities. To test whether the change in flattening radius hides the density evolution we plot in Figure \ref{fil_ev} the evolution of the central density when $R_{\rm flat}$ is held constant at 0.04 pc. In this case a clear trend of increasing central density with time is seen in the data.

\subsection{A constant filament width?}\label{width}

\begin{table}
	\centering
		\begin{tabular}{l c c c}
   	         \hline
	         \hline
		ID & FWHM$_{\rm 1pc}$ & FWHM$_{\rm 0.35pc}$ & $3 R_{\rm flat}$ \\
	         \hline
	          S1-F1-T130 & 0.347 & 0.204 & 0.123\\ 
		 S1-F1-T140 & 0.330 & 0.207 & 0.153\\
		 S1-F1-T150 & 0.320 & 0.221 & 0.357\\
		\hline
	          S1-F2-T130 & 0.338 & 0.262 & 0.369 \\
	          S1-F2-T140 & 0.307 & 0.229 & 0.309 \\
	          S1-F2-T150 & 0.353 & 0.257 & 0.233\\   
	          \hline 
	          S2-F1-T170 & 0.268 & 0.183 & 0.136\\ 
	          S2-F1-T180 & 0.154 & 0.118 & 0.390 \\
	          S2-F1-T190 & 0.177 & 0.133 & 0.052 \\
	          \hline
	          S3-F1-T160 & 0.301 & 0.222 & 0.024\\ 
	          S3-F1-T170 & 0.525 & 0.238 & 0.039\\ 
	          S3-F1-T180 & 0.503 & 0.259 & 0.042\\ 
	          \hline
	          S3-F2-T160 & 0.506 & 0.183 & 0.189\\ 
	          S3-F2-T170 & 0.353 & 0.244 & 0.333\\ 
	          S3-F2-T180 & 0.314 & 0.172 & 0.116\\ 
	          \hline
	          S4-F1-T020 & 0.271 & 0.143 & 0.082 \\ 
	          S4-F1-T030 & 0.421 & 0.260 & 0.375\\ 
	          S4-F1-T040 & 0.357 & 0.228 & 0.363\\ 
	          \hline
	          S4-F2-T020 & 0.360 & 0.076 & 0.017 \\ 
	          S4-F2-T030 & 0.413 & 0.153 & 0.066 \\ 
	          S4-F2-T040 & 0.371 & 0.251 & 0.273\\
   	         \hline
	         \hline
	         Mean & 0.348 & 0.202 & 0.192\\
	         St. Deviation & 0.091 & 0.052 & 0.132\\
	         \hline	  
	         \hline       
		\end{tabular}
		\caption{Derived filament widths using three different methods. 1) The FWHM of a Gaussian best fit to column densities within 1 pc of the filament spine, 2) the FWHM of a Gaussian best fit to column densities within 0.35 pc of the filament spine, and 3) three times the Plummer best fit flattening radius. All three methods give different results for the individual filaments, suggesting that the filament width is not a well defined concept.}
	\label{widths}
\end{table}

One of the most important findings of A11 was that their observed filaments had a constant width of 0.1 pc. If borne out by further studies this would represent an important constraint on the physics of molecular clouds. One problem with testing the filament widths is that this is not a well defined quantity since filaments merge smoothly into their environment. A11 find their filament width by fitting a Gaussian function with a background contribution to the column densities within 0.3-0.4~pc of the main filament spine (Arzoumanian, private communication). J12a similarly fit the filament using only data out to a radius of 0.4~pc. However, there is no physical reason why this data range should be preferred, rather than using the full range. Figure \ref{S234F1} shows with green and blue lines the best Gaussian fit for data within 1 pc and 0.35 pc of the filament spine. The fits are different and neither reproduces the data as well as the Plummer-like model. A11 propose that a rough conversion between their Gaussian FWHM and the Plummer-like flattening radius is FWHM $ \sim 1.5 \times (2R_{\rm flat})$. 

In this section we test the filaments widths we obtain using the following three methods: 1) The FWHM of a Gaussian best fit to column densities within 1 pc of the filament spine, 2) the FWHM of a Gaussian best fit to column densities within 0.35 pc of the filament spine, and 3) three times the Plummer best fit flattening radius. All our Gaussian fits include a contribution from a constant background. In Table \ref{widths} we show the resulting filament widths using the different methods. The three methods all yield very different answers for the width of each filament, clearly showing that the filament width is not a well defined observational quantity and should be treated with caution.

Figure \ref{S234F1} shows that the Gaussian distribution is typically more flat at small radii than the measured column densities. This is particularly true of the fit out to 1 pc, but often applies to the 0.35 pc filament too. When comparing our best Gaussian fits to those in Figure 4b of A11, we see that their Gaussian best fit does not follow the column density distribution and appears to have a FWHM greater than 0.1 pc. This is because they are not actually plotting their best fit, but rather a new Gaussian with the same width as their best Gaussian fit convolved with their beam normalised to the peak column density without a background (Arzoumanian, private communication). In general, we find that the width of the best Gaussian fit depends sensitively on the value assumed for the background column density, and hence on the properties of the material surrounding the filaments, rather than the properties of the filaments themselves. The average best fit width obtained using radii out to 1 pc is 1.5 times larger than that using radii up to 0.35 pc in the simulations, which shows that the calculated width is range dependent. An inspection of Table \ref{widths} also shows that the width can also vary significantly with time, suggesting that the best Gaussian fit is very sensitive to small changes in the column density profile. Unfortunately, using the flattening radius from the Plummer-like model also has difficulties due to the degeneracies in the model described in the previous section. Generally we would recommend that when using a Gaussian fit authors always clearly state the range to which they are fitting their data, how they are accounting for the background, and plot an example of their best fit. Otherwise it is very hard to interpret and compare filament widths between studies.

Regardless of our methods, the mean filament widths shown in Table \ref{widths} are clearly inconsistent with the constant 0.1 pc filament width proposed by A11. This could be considered a failure of the model, but it should be noted that not all observations are consistent with this value either. In agreement with A11, \citet{Palmeirim13} find widths of $ 0.09 \pm 0.02$ pc in Taurus B211/3, and \citet{Malinen12} find widths of $\sim 0.1$ pc using a range of different methods fitting out to a distance of 0.3 pc for two filaments in Taurus TMC1. However, in the Planck Galactic Cold Cores, J12a find a mean filament width of 0.336 pc with a dispersion of 0.186 pc fitting out to a distance of 0.4 pc in a similar way to A11. \citet{Hennemann12} find weighted mean values of the filament central widths between 0.26 and 0.34 pc for filaments in Cygnus X using \herschel.

\begin{figure}
\begin{center}
\includegraphics[width=3in]{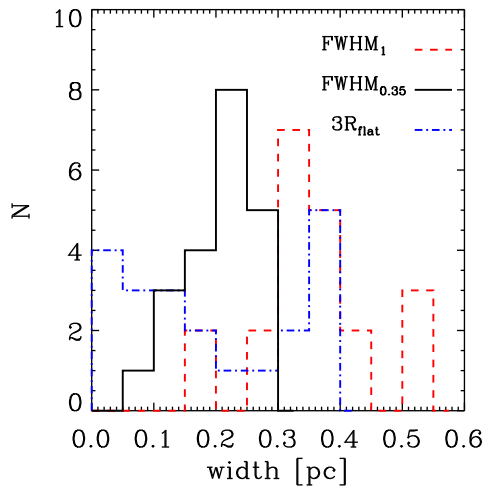}
\caption{Histograms of the filament widths calculated using three different methods. 1) The FWHM of a Gaussian best fit to column densities within 1 pc of the filament spine, 2) the FWHM of a Gaussian best fit to column densities within 0.35 pc of the filament spine, and 2) three times the Plummer best fit flattening radius. All of the methods produce a range of widths rather than a constant value.}
\label{FWHM_hist}
\end{center}
\end{figure}

In Figure \ref{FWHM_hist} we plot histograms of the filament widths derived using the different methods. As well as the mean value disagreeing with the 0.1 pc measurement of A11, the distributions are not narrow, but have a large scatter. This finding is also inconsistent with there being a constant filament width in our simulations. It is intriguing to note that some of the simulations do have narrow 0.1 pc widths, for example the long filament in S2. However the basic initial condition for this simulation was the same as S1, with the only difference being a different choice of turbulent seed. As a final point we highlight that the mean FWHM found for the simulations using the entire distribution is in excellent agreement with the prediction of FWHM $\sim 0.3$~pc for accreting bound filaments by \citet{Heitsch13a}.

\subsection{Resolution Dependence} \label{results-res}

Our fiducial simulations use an extremely high resolution such that the central 0.1~pc of the filament is resolved by a minimum of ten elements (and typically many more) everywhere along the filament length as discussed in Section \ref{sec-sim} and illustrated in Figure~\ref{resolution}. The issue of resolution was highlighted by \citet{Hennebelle13a}, who found that in higher resolution simulations the filaments had narrower widths. \citet{Hennebelle13a} carried out non-isothermal MHD simulations of filament formation using the RAMSES AMR code, with a maximum effective resolution of $2048^3$ zones. The size of his simulation volume was 50 pc, and so this numerical resolution corresponds to a physical cell size of 0.025~pc. This is only a factor of a few smaller than the values of $R_{\rm flat}$ that we derive for many of our filaments, and it is therefore quite plausible that simulations of filaments with this resolution or less do not produce converged results. Our simulations, on the other hand, have a spatial resolution that is a factor of five higher at a number density $n = 2 \times 10^{4} \: {\rm cm^{-3}}$, typical of the cores of filaments, and that is even smaller in higher density regions. This makes it more likely that our results are converged, but nevertheless it is important to verify this.

\begin{figure*}
\begin{center}
\begin{tabular}{c c c}
\begin{overpic}[scale=.3]{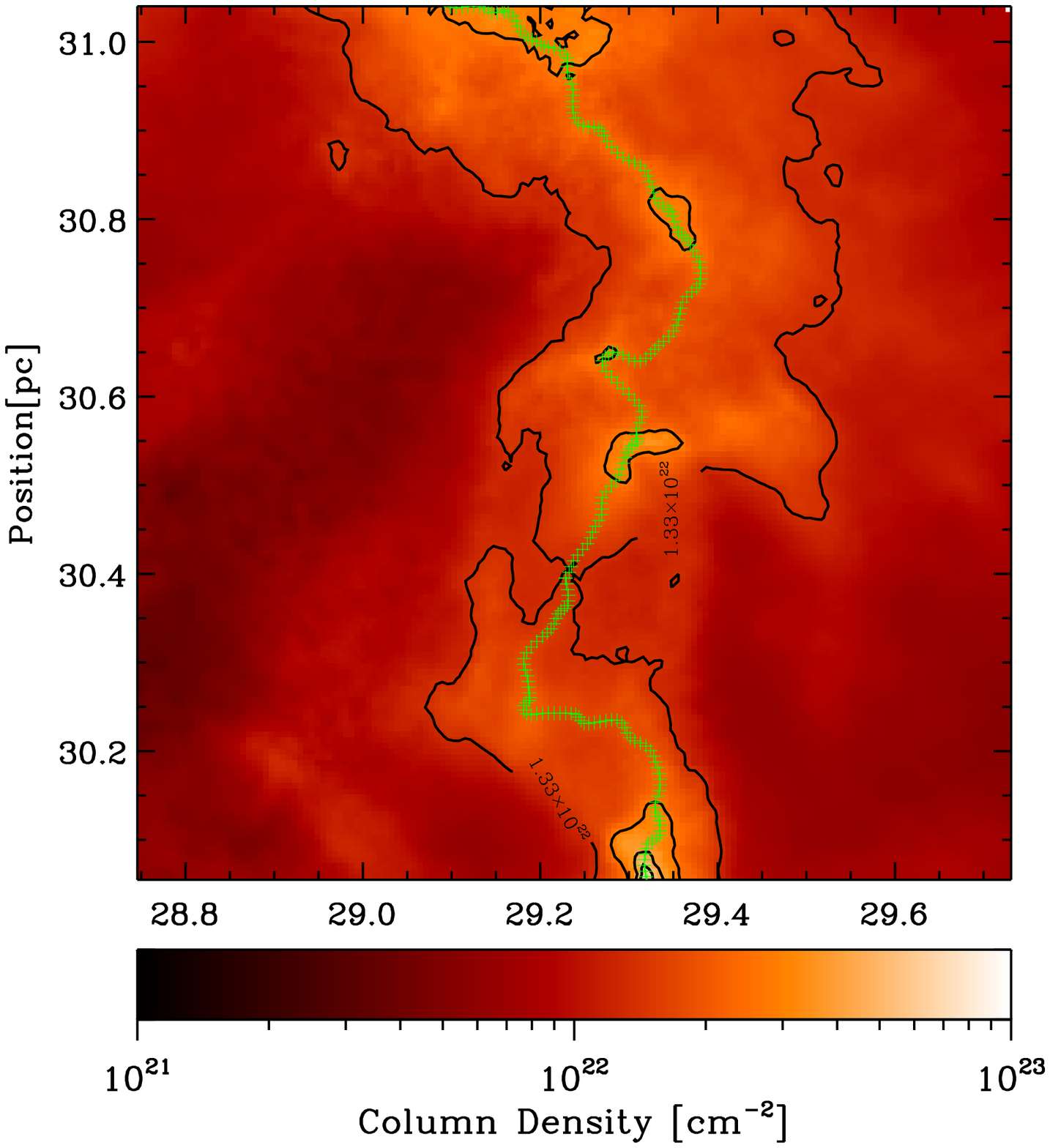}
\put (75,90) {\makebox(0,0){{\color{white}S5-F1-T130}}}
\put (35,90) {\makebox(0,0){{\color{white}low res.}}}
\end{overpic}
\begin{overpic}[scale=.3]{./final_figs/Fg3_S1_contour_F1_S130}
\put (75,90) {\makebox(0,0){{\color{white}S1-F1-T130}}}
\put (35,90) {\makebox(0,0){{\color{white}standard}}}
\end{overpic}
\begin{overpic}[scale=.3]{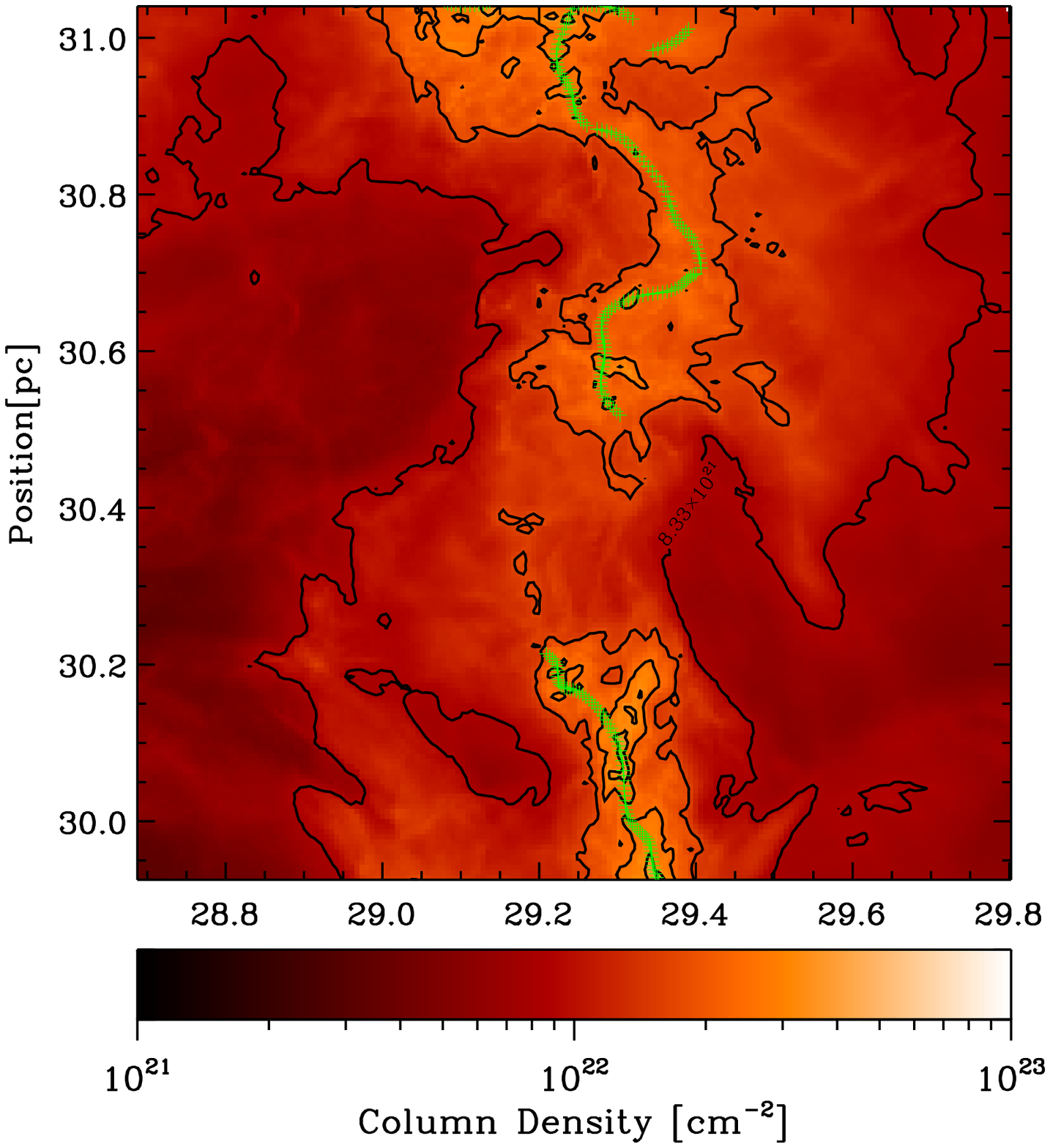}
\put (75,90) {\makebox(0,0){{\color{white}S6-F1-T130}}}
\put (35,90) {\makebox(0,0){{\color{white}high res.}}}
\end{overpic}
\end{tabular}
\caption{The same filament at increasing resolution from left to right. The contours show the column density morphology in the region of the filament. The overall shape of the filament does not change significantly but there is much more substructure in the higher resolution simulations. The filament found by \disperse, shown in green, also varies due to the higher density material being better resolved.}
\label{S156res}
\end{center}
\end{figure*}

To investigate how our results depend upon resolution we ran two additional simulations. Simulation 5 had a mass resolution of 0.0025 \msun and no additional Jeans refinement. Simulation 6 had the same base resolution but used a Jeans refinement scheme two times as strict as in our fiducial case, i.e. it required the Jeans length to be resolved everywhere by 32 elements rather than our standard 16. The initial turbulent field was the same as in Simulation 1 in both cases, and so the same structures develop. Figure \ref{S156res} shows the same region as Simulation 1 Filament 1 (S1-F1-T130) in all three cases. While the large-scale structure of the filament does not change much with resolution there is a far greater degree of substructure in the higher resolution filaments. Additionally, the shape of the filament found by \disperse varies due to the maximum density at the centre of the filament being higher in the higher resolution simulations.

\begin{table*}
	\centering
		\begin{tabular}{l c c c c c c}
   	         \hline
	         \hline
		ID & resolution & $n_{\rm c}$ & $p$ & $R_{\rm flat}$ & FWHM$_{\rm 1pc}$  \\
		 & & [\cmc] & & [pc] & [pc]\\
	         \hline
	          S5-F1-T130 & base & $6.33\E^4$ & 1.65 & 0.0537 & 0.370   \\
	          S1-F1-T130 & 16 per L$_{\rm J}$ &$9.17\E^4$ & 1.44 & 0.0411 & 0.347 \\ 
	          S6-F1-T130 & 32 per L$_{\rm J}$& $11.3\E^4$ & 1.39 & 0.0420 & 0.468 \\
	         \hline	  
	         \hline       
		\end{tabular}
		\caption{The properties of one of the filaments at various resolutions. The difference in the fit obtained between the fiducial and high resolution runs is smaller than the uncertainty in the fiducial fit for the Plummer-like model.}
	\label{resfits}
\end{table*}

Table \ref{resfits} describes the best fit properties of the above filament at the three different resolutions. For the Plummer-like fits there is a general trend of greater central density with increasing resolution. The power law index $p$ also decreases with increasing resolution. However, the difference in $p$ and $R_{\rm flat}$ between the high resolution and fiducial cases are very small, and in all cases the change in the best fit value is smaller than the uncertainty in the calculated fiducial value. For this reasons we conclude that our simulations are sufficiently well-resolved to realistically categorise the filaments. 


The best fit Gaussian FWHM do not appear to be so well converged as the Plummer-like fits. As noted above in Section~\ref{width}, this is probably because the Gaussian profile is extremely sensitive to small changes in the column density profile. The filament selected by \disperse has a different geometry in the highest density simulation which leads to a different sub-set of the data being included in the fit. When the same skeleton is used as found in our fiducial case the best Gaussian fit becomes FWHM$_{\rm 1pc} =0.363$~pc, in better agreement with the other simulations.

Another consideration is whether the results from our observational comparison are affected by the finite beam size of the observations. To address this, we investigated what happens when we convolve our column density maps with a beam of finite thickness and include the effects of the beam in our model of the filament profile when doing the $\chi^2$ fitting, as is done observationally. In all cases, we found that the best fits recovered with and without the beam convolution were nearly identical. We therefore conclude that the beam does not affect the observational determination of the filament profiles (although it may affect where a core would be identified by eye). Note that this is again consistent with the results of \citeauthor{Juvela12b}~(2012b), who found that their derived filament widths were relatively insensitive to the assumed size of the beam. 

Finally, we tested that the choice of grid-size for our column density projection was not affecting the results. Using the filament spine vector for S1-F1-T130 at our standard 1000$\times$1000 resolution we calculated what the average column density profile would be using a 500$\times$500 projection. There was a small change in the central density ($n_{\rm c}=8.42\E^4$~\cmc~ for the $500^2$ grid, and $n_{\rm c}=9.16\E^4$~\cmc~for the $1000^2$ grid). However, the change in the flattening radius and power-law index was very small ($R_{\rm flat}=0.044$, $p=1.47$ for the $500^2$ grid, and $R_{\rm flat}=0.041$, $p=1.44$ for the $1000^2$ grid). We therefore conclude that our choice of grid projection is not significantly affecting the results.

\section{Density profiles in 3D}\label{results-3D}

\subsection{Are the column density features true 3D structures?}

Having determined that our simulated filaments are consistent with observations, we now investigate how well the features that are apparent in the column density projections corresponds with actual physical structures in three-dimensional space. In this section we concentrate our \mbox{analysis} on just the first filament from the middle snapshot in S1-4 to allow us to study the 3D morphology in more detail.

\begin{figure*}
\begin{center}
\begin{tabular}{c c c}
\begin{overpic}[scale=.3]{./final_figs/Fg3_S1_contour_F1_S130}
\put (75,90) {\makebox(0,0){{\color{white}S1-F1-T130}}}
\end{overpic}
\begin{overpic}[scale=.3]{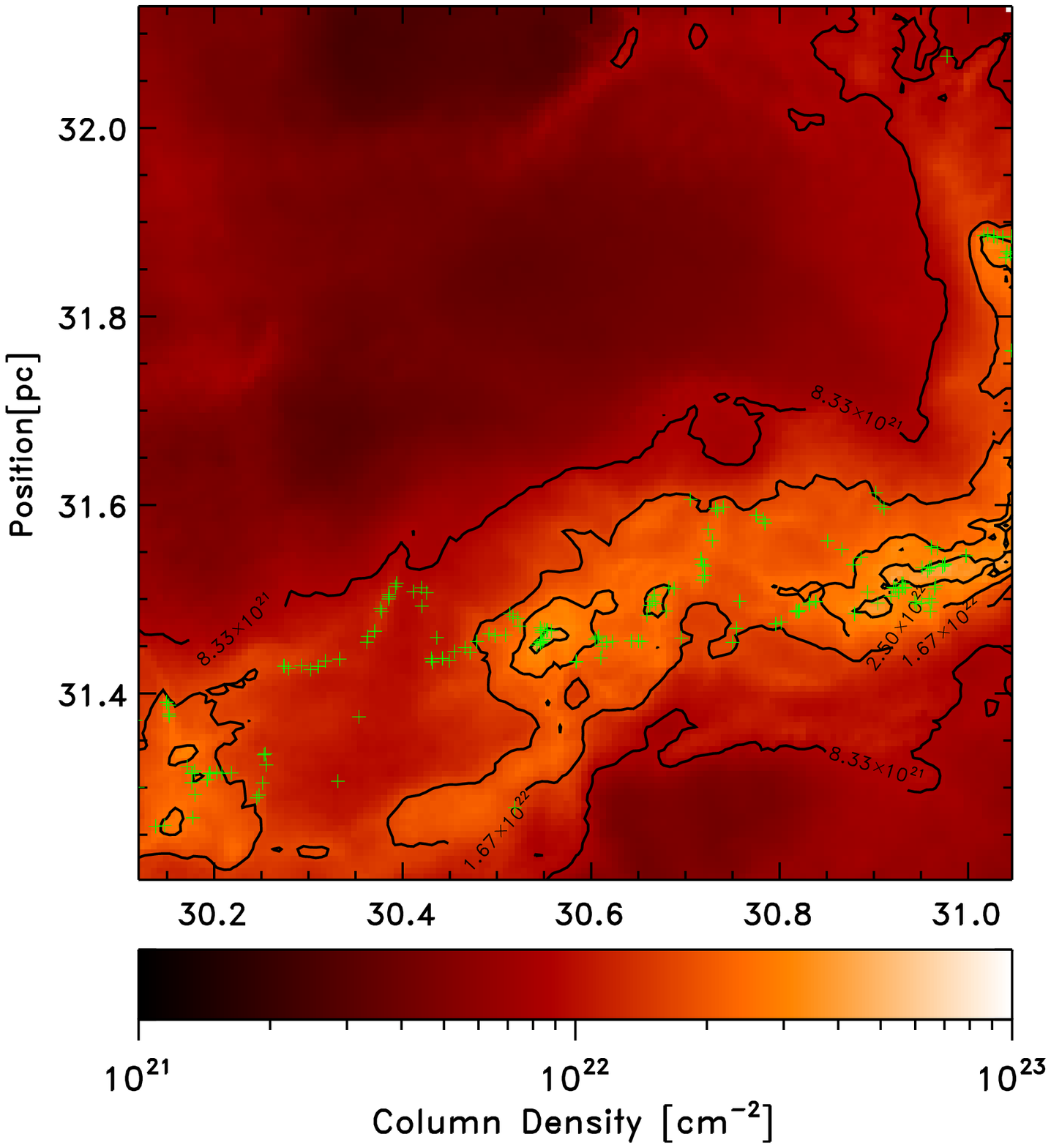}
\put (35,90) {\makebox(0,0){{\color{white}yz plane}}}
\end{overpic}
\begin{overpic}[scale=.3]{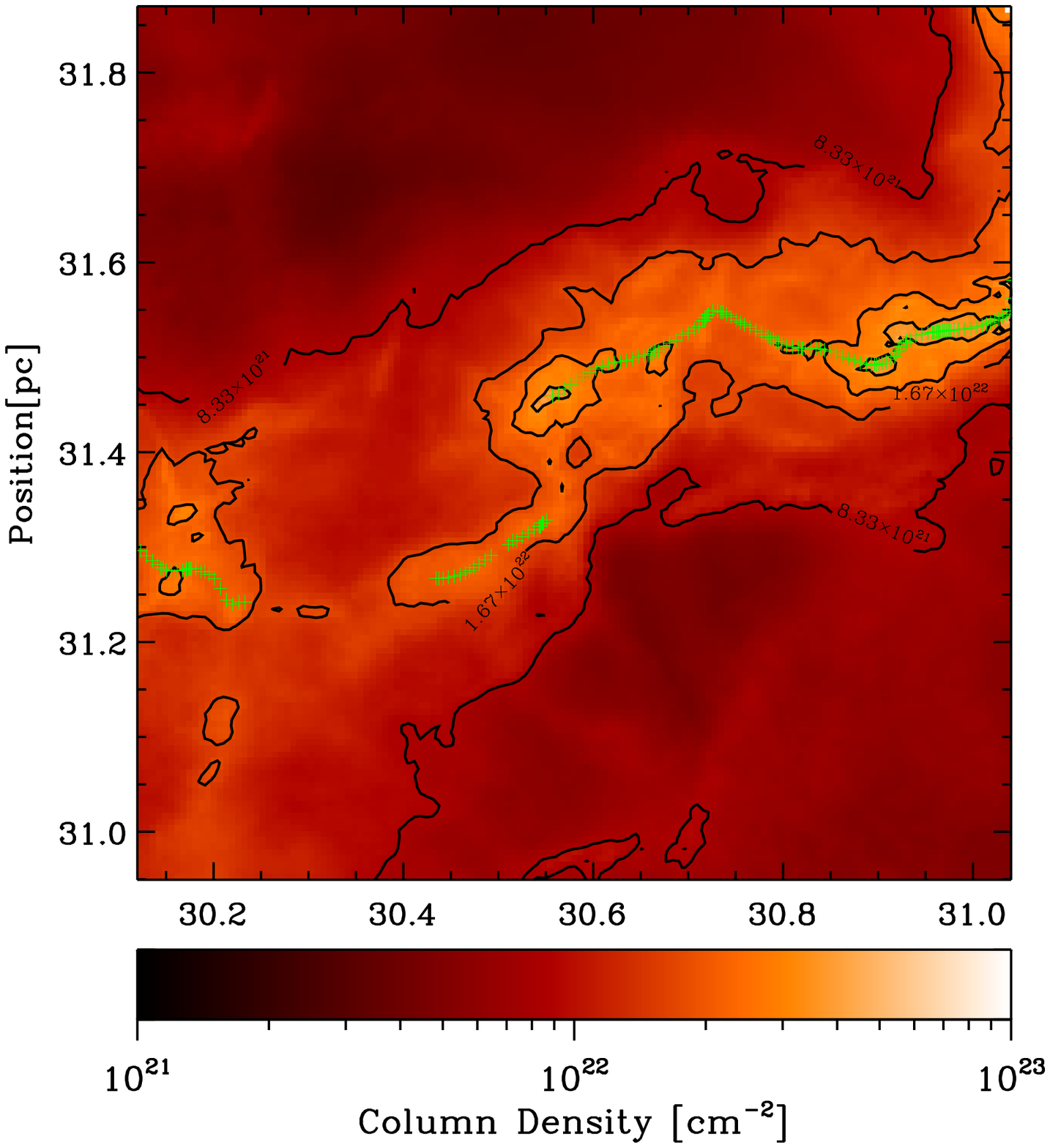}
\put (35,90) {\makebox(0,0){{\color{white} yz plane}}}
\end{overpic}
\\
\begin{overpic}[scale=.3]{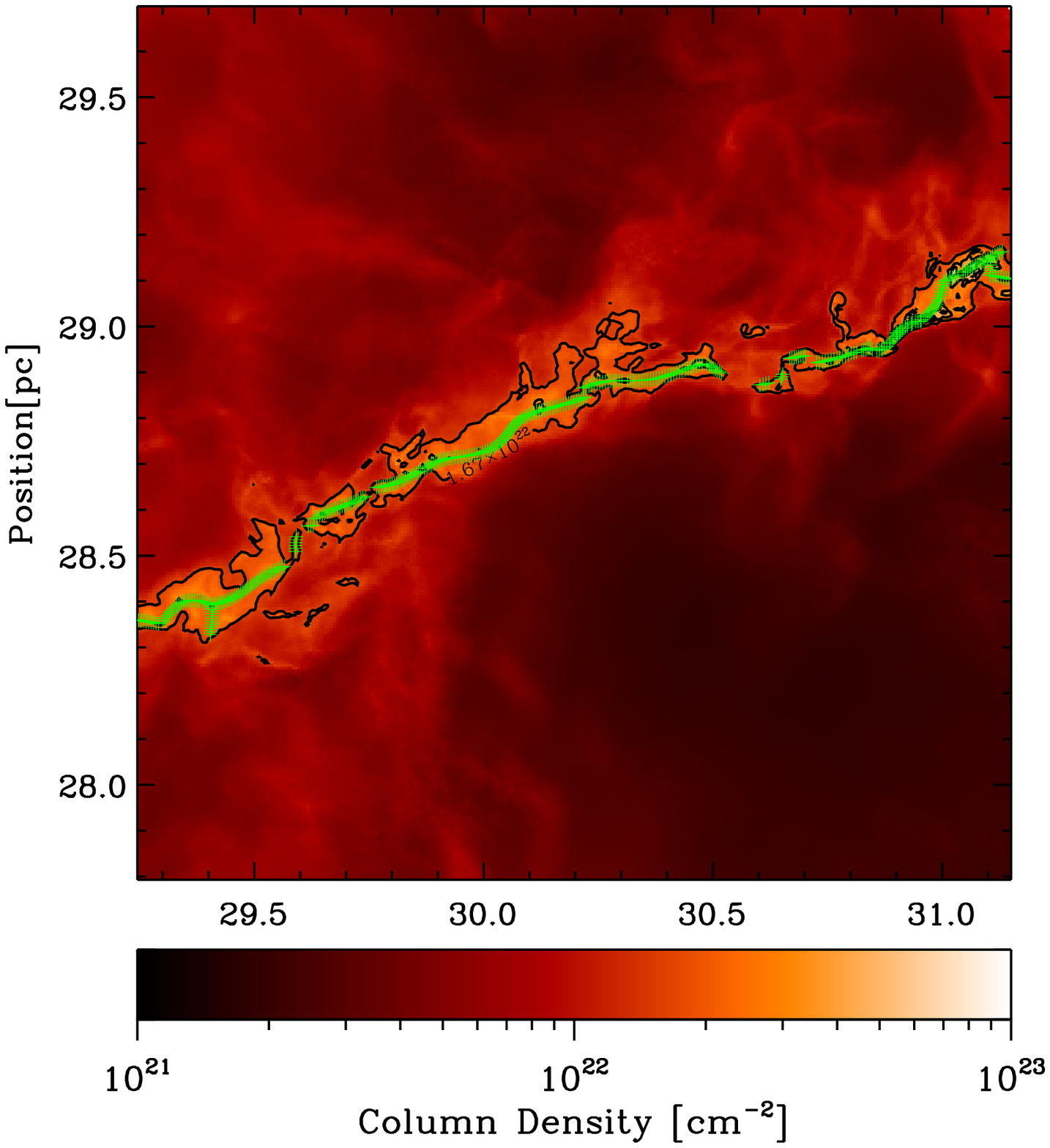}
\put (75,90) {\makebox(0,0){{\color{white}S2-F1-T170}}}
\end{overpic}
\begin{overpic}[scale=.3]{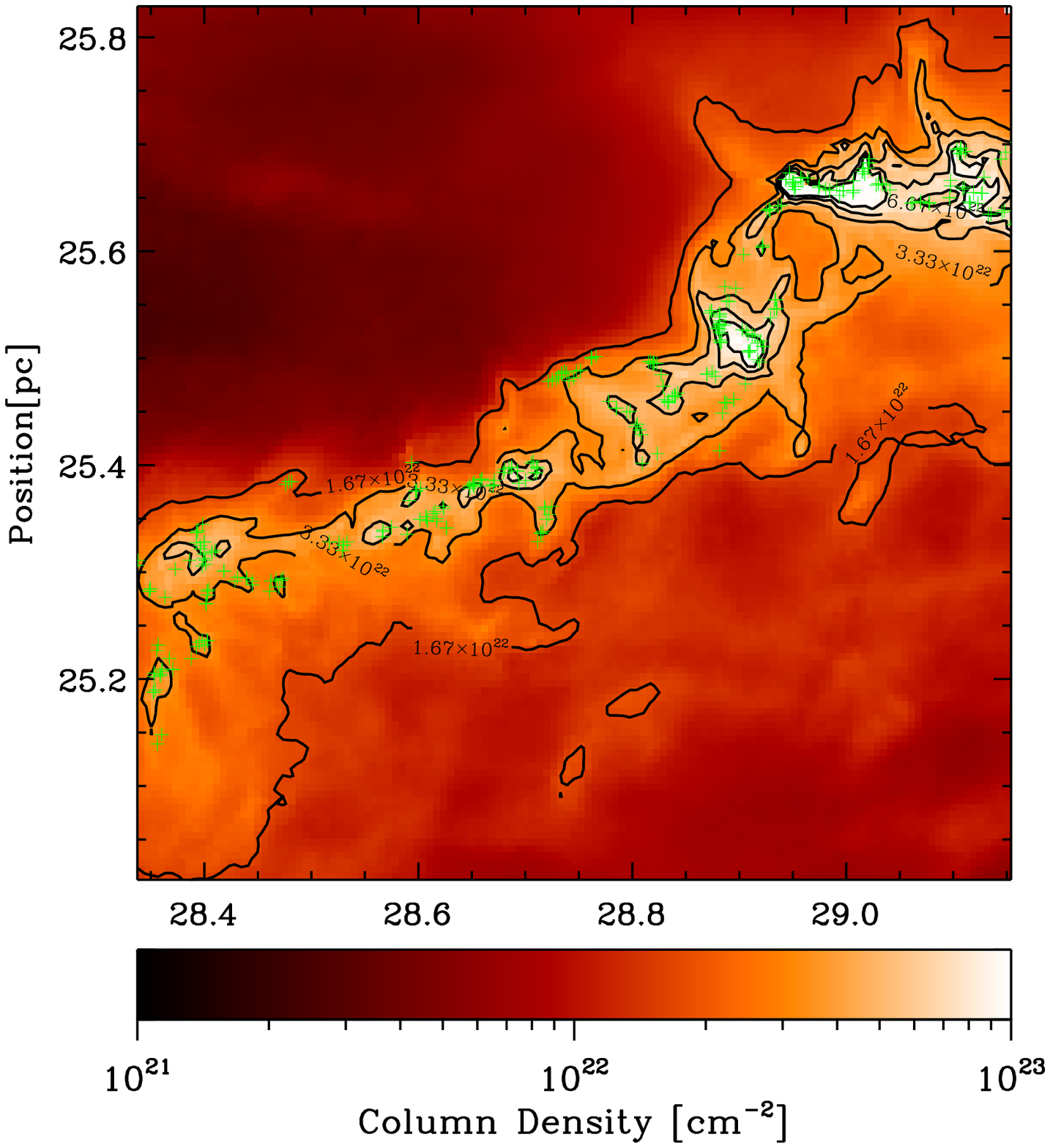}
\put (35,90) {\makebox(0,0){{\color{white} yz plane}}}
\end{overpic}
\begin{overpic}[scale=.3]{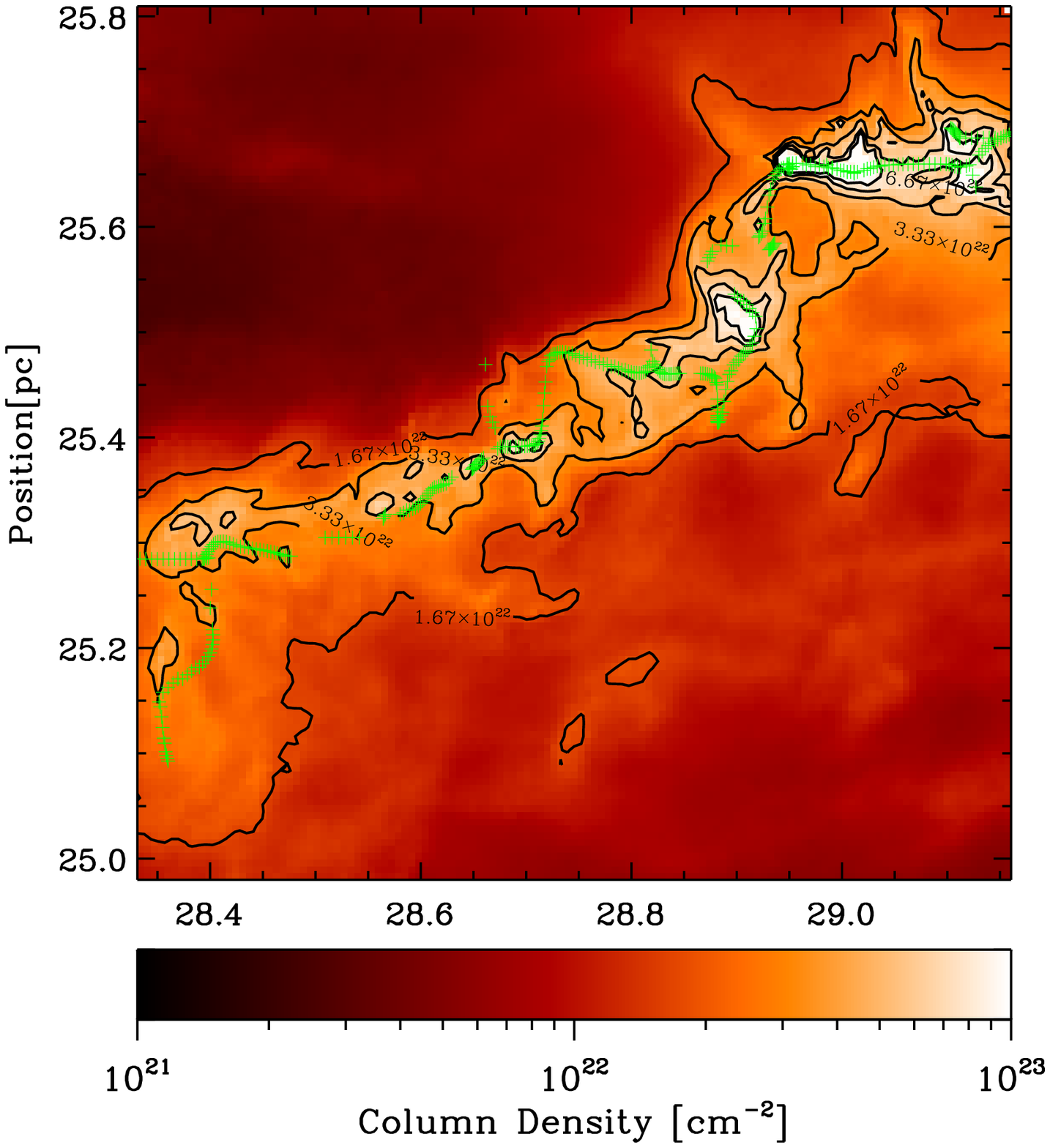}
\put (35,90) {\makebox(0,0){{\color{white} yz plane}}}
\end{overpic}
\end{tabular}
\caption{Filament 1 from S1 (\textit{top}) and Filament 1 from S2 (\textit{bottom}). The panels on the left show our standard $x$-$y$ projection view in which we identified the filaments using the column densities. The middle panels shows the same gas projected in the $y$-$z$ plane. The green points in these panels are at the same $x$-$y$ position as on the left but the $z$ co-ordinate is taken from the cell that has the maximum density at that $x$-$y$ position. The filaments are clearly not one continuous maximum density structure. The panels on the right show the $y$-$z$ projection, finding the $z$ co-ordinate by finding a new filament in $y$-$z$ column density.}
\label{2D3D}
\end{center}
\end{figure*}

In Figure \ref{2D3D} we compare the standard $x$-$y$ plane projection of the simulations used in Section \ref{results-col} with the same gas when projected in the $y$-$z$ plane. In order to do this, the $z$ co-ordinate of the filament must be found. In the middle panel, we indicate, for each $x$-$y$ position, the point in the $z$ direction that corresponds to the maximum density along that line of sight.  When identified in this manner, the green filament points trace the general outline of the structure seen in the column density map, but do not form one single continuous structure. In the right-hand panels, we instead identify the position of the filament in the $z$ direction by performing a second filament identification in the $y$-$z$ plane using \disperse. The clear difference between the two sets of green points shows that the gas corresponding to the 2D spine of the filament is not necessarily the densest gas along the line of sight.

\begin{figure*}
\begin{center}
\begin{tabular}{c c c}
\includegraphics[width=2.4in]{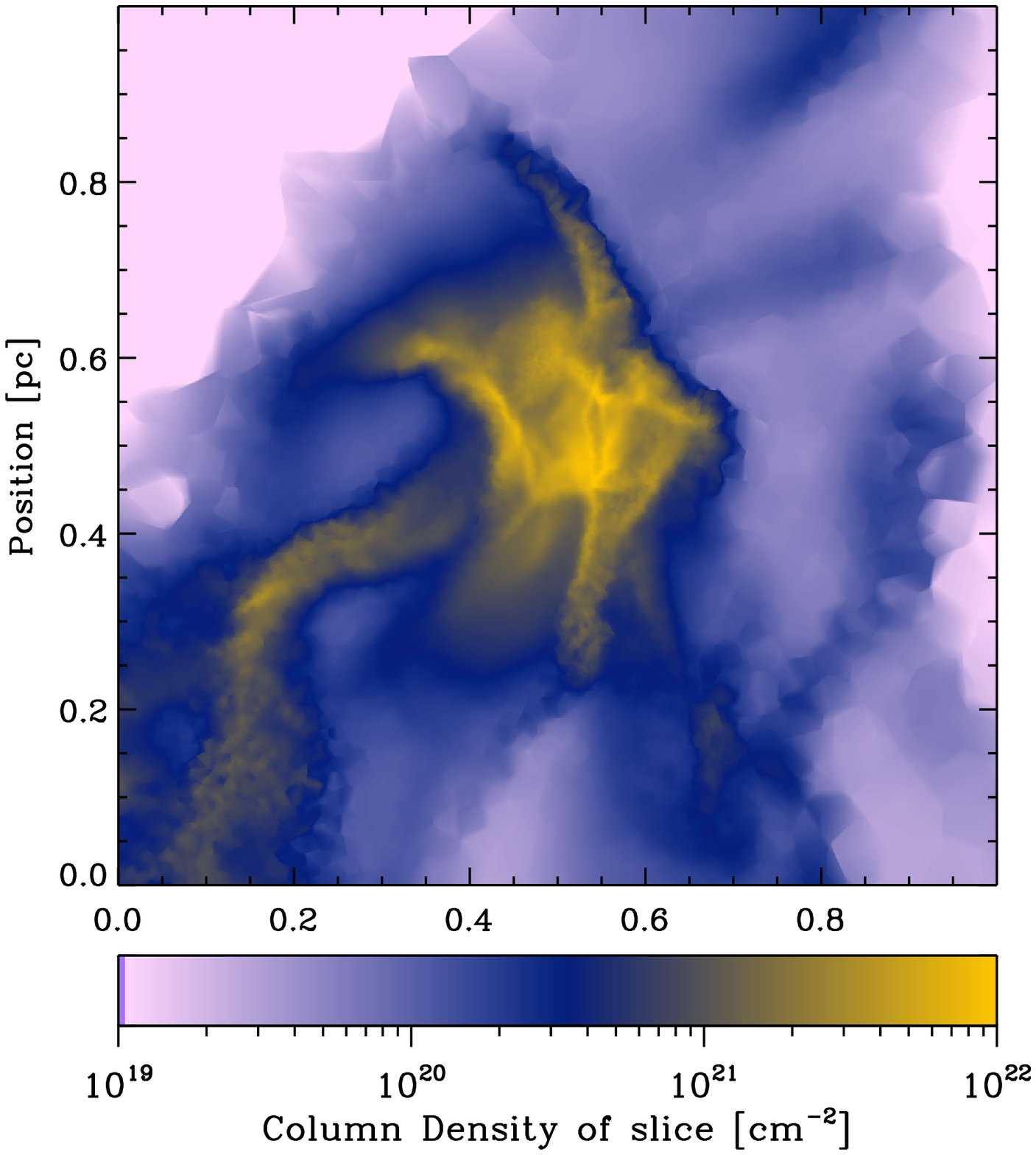}
\includegraphics[width=2.4in]{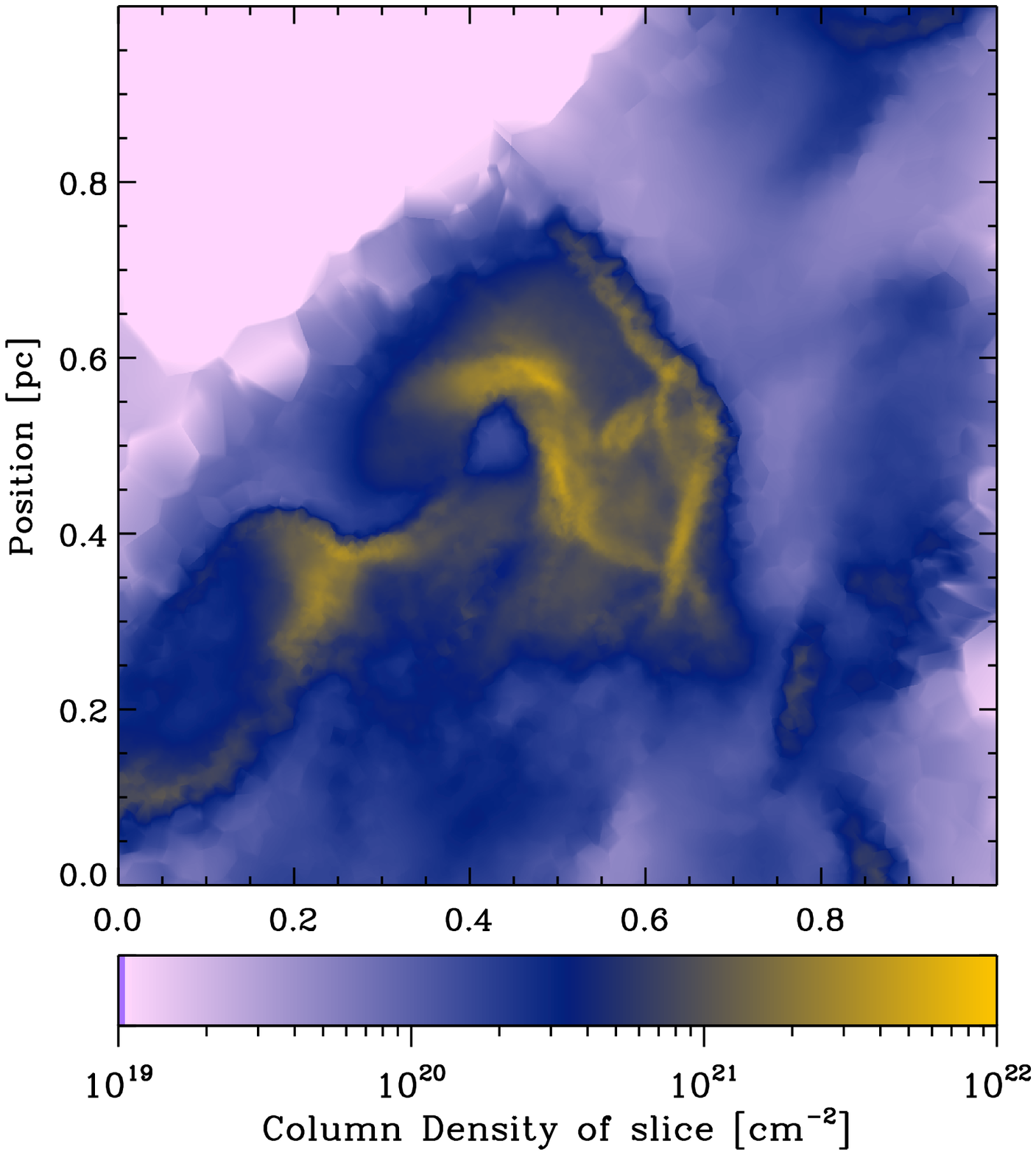}
\includegraphics[width=2.4in]{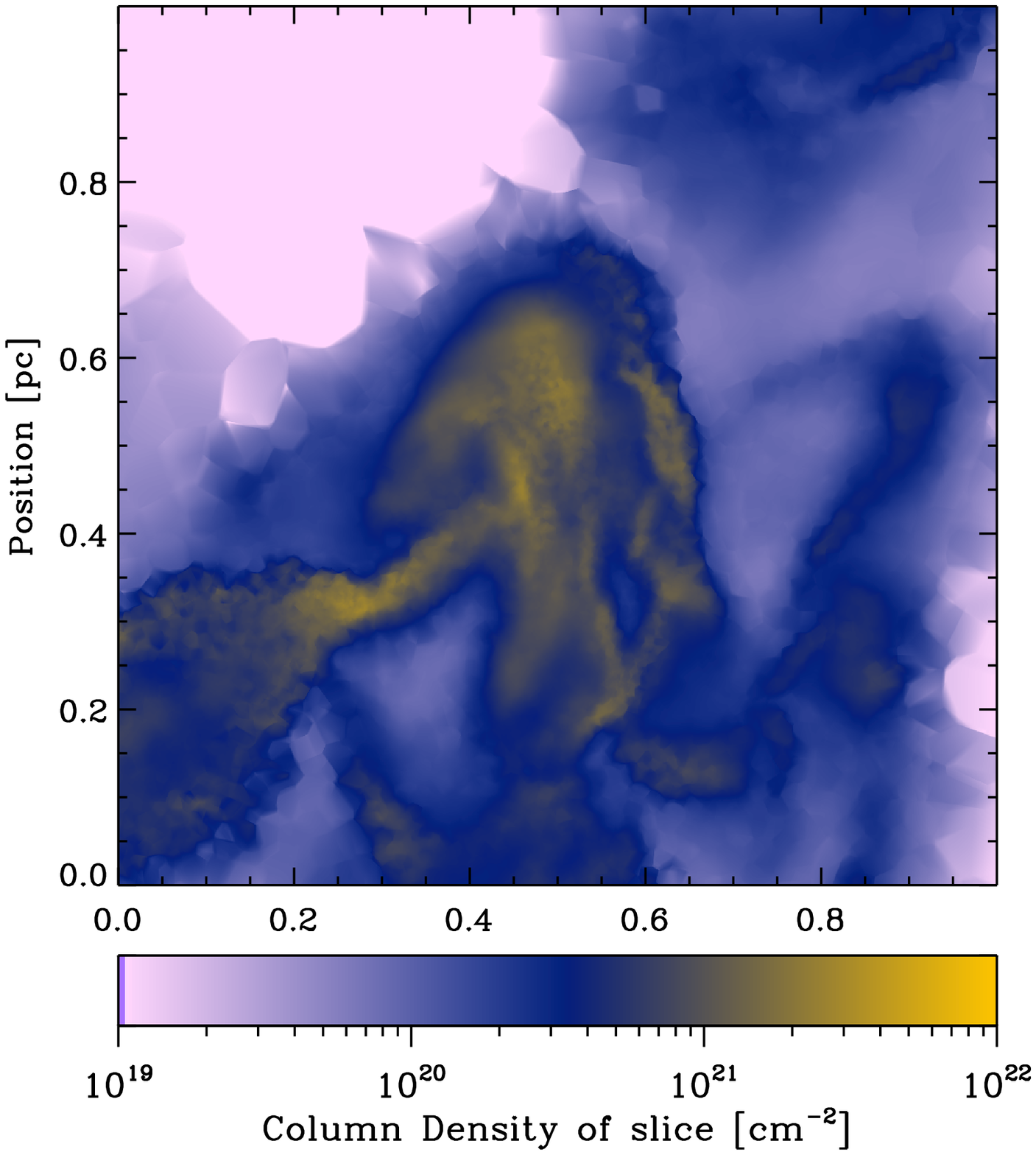}\\
\includegraphics[width=2.4in]{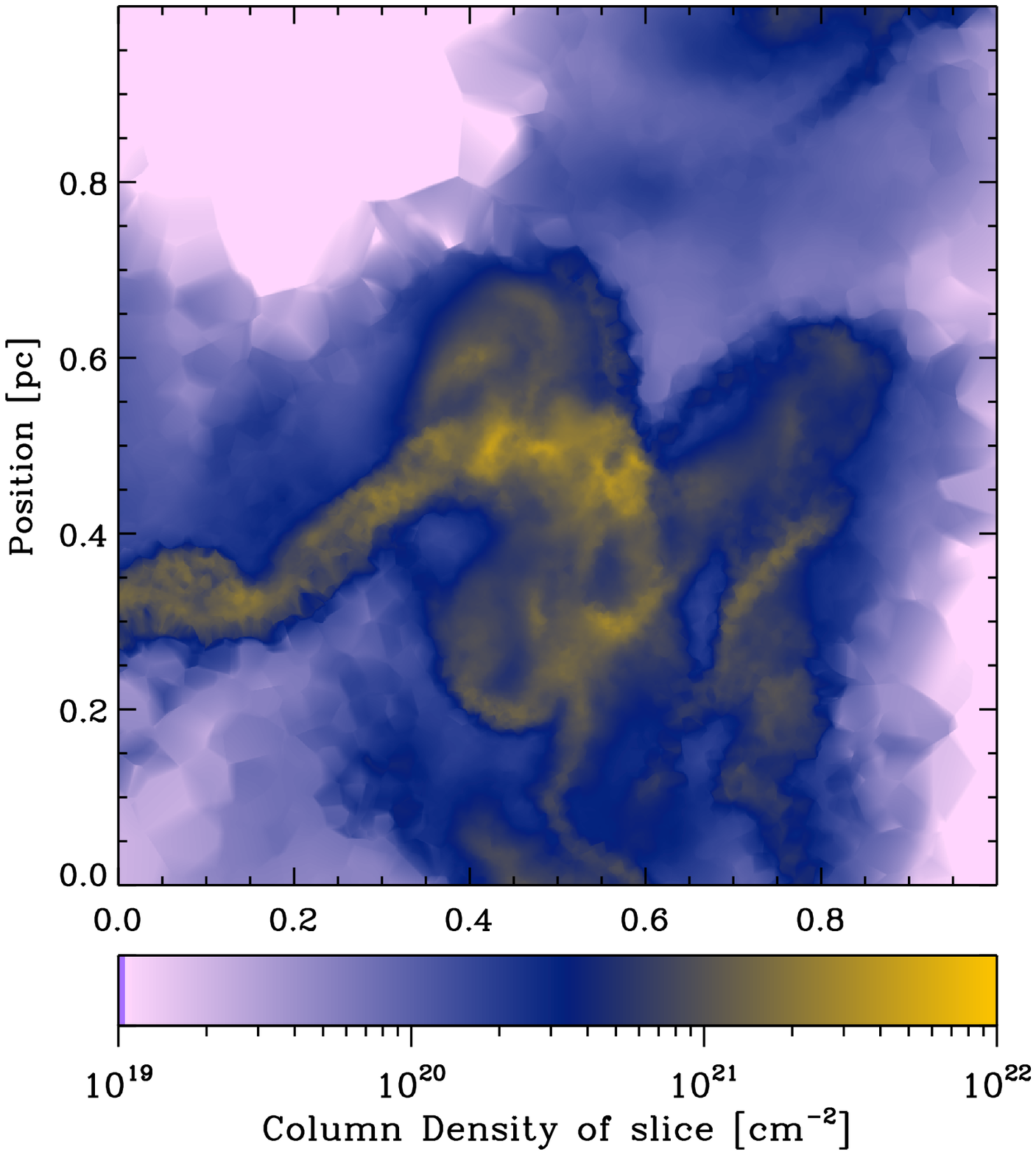}
\includegraphics[width=2.4in]{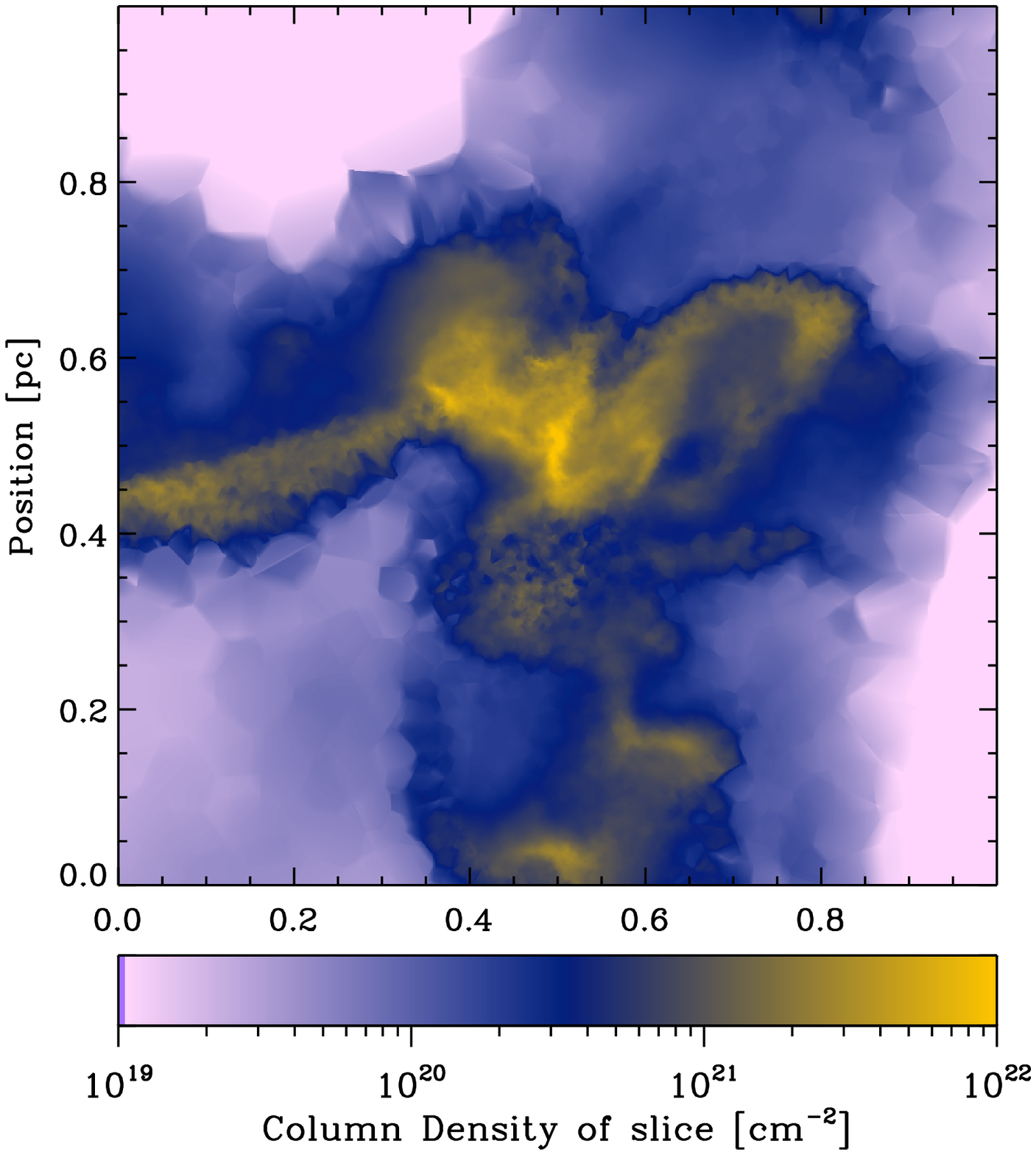}
\includegraphics[width=2.4in]{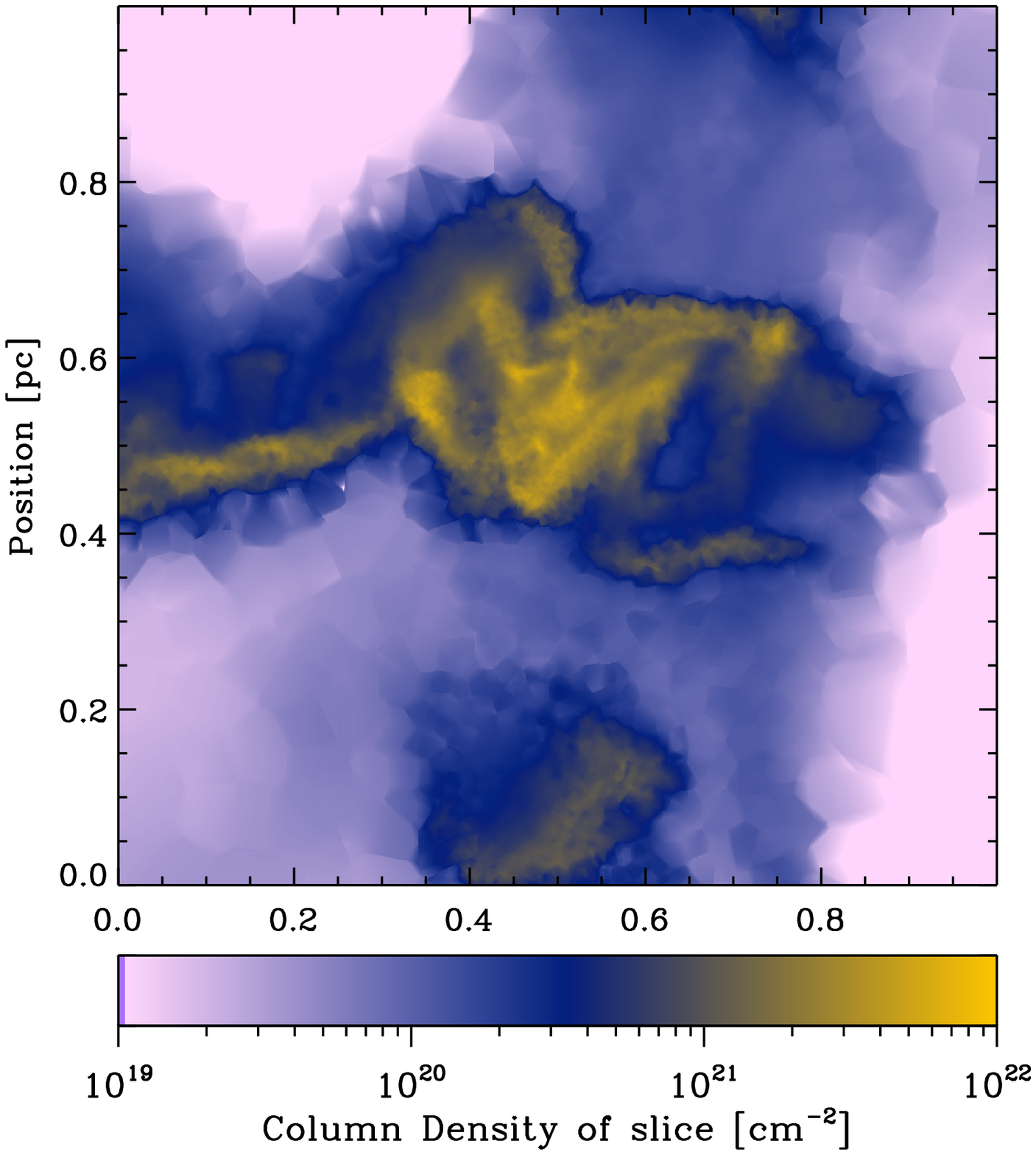}\\
\includegraphics[width=2.4in]{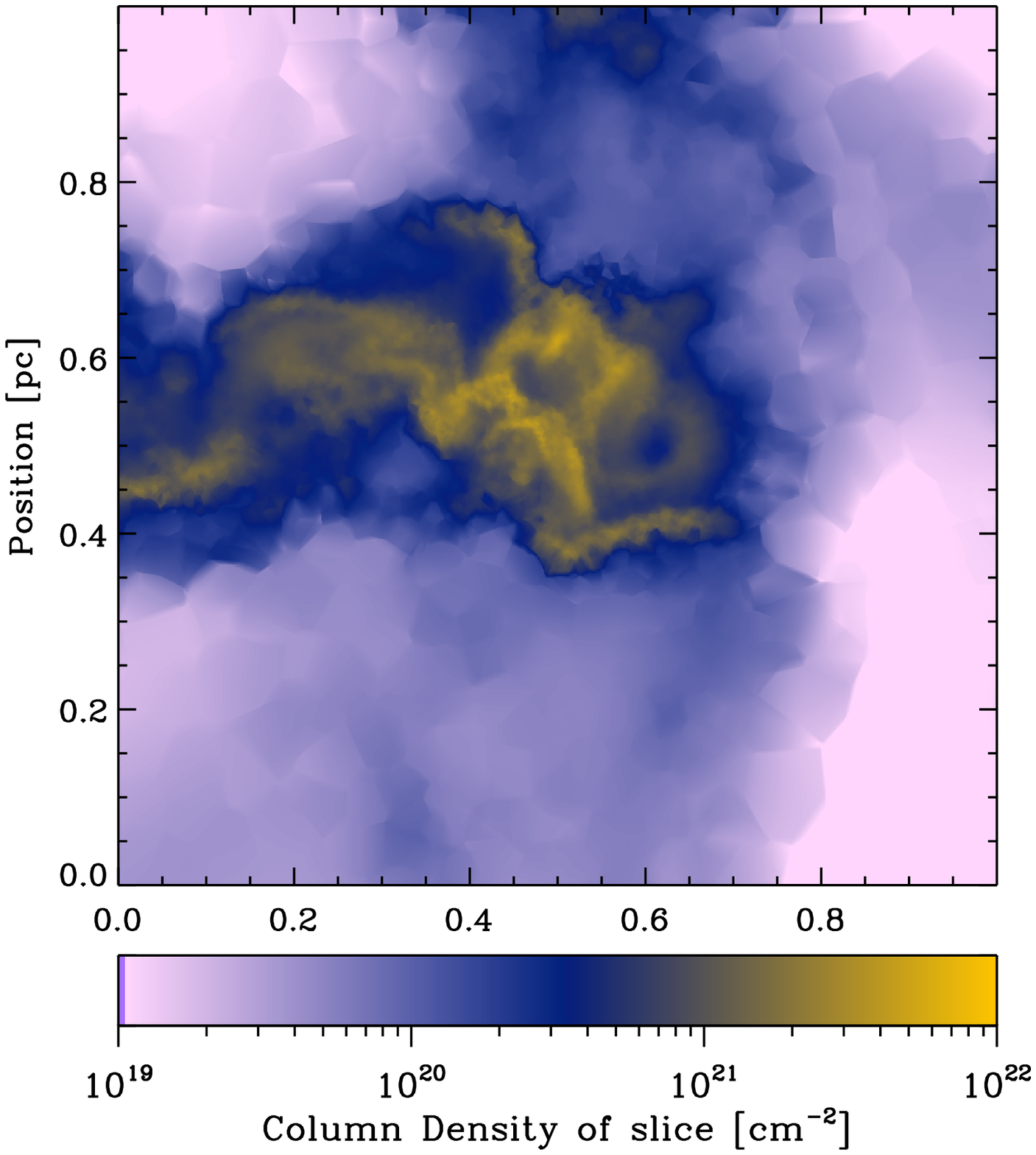}
\includegraphics[width=2.4in]{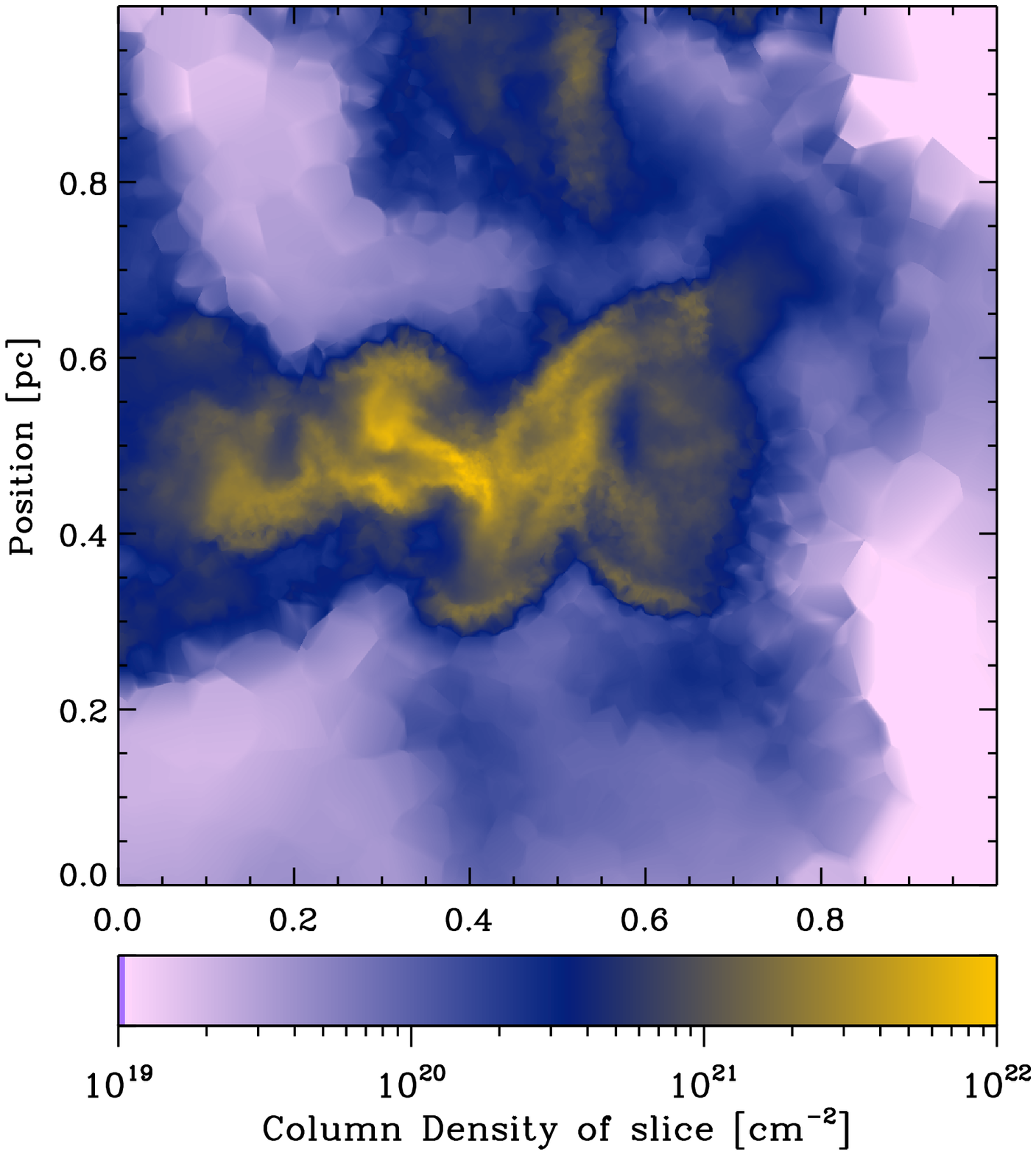}
\includegraphics[width=2.4in]{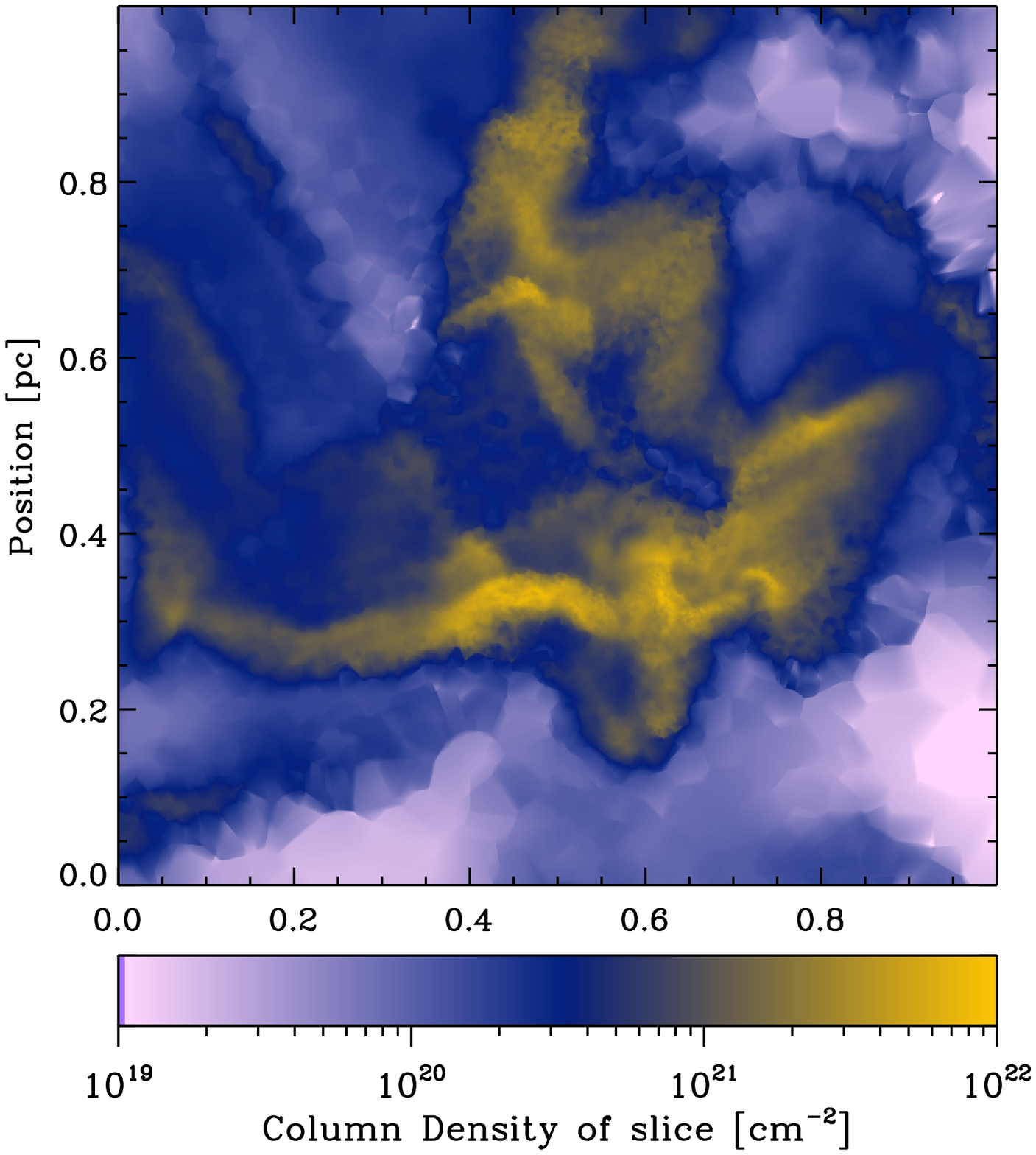}\\
\end{tabular}
\caption{Slices of 0.05 pc thickness along the simulation $y$-axis centred on the centre of filament S1-F1-T130.  A movie showing slices of the filament section is available online at http://www.ita.uni-heidelberg.de/$\sim$rowan/Filaments. The filament sections are not smooth but have considerable substructure and the peak column density does not necessarily correspond to the filament centre.}
\label{S1F1_walk}
\end{center}
\end{figure*}

To understand this behaviour, we examine the density distribution along the filament length using the original simulation data, not the column density projections. Figure \ref{S1F1_walk} shows cross-sections of the density as we move along the long axis of filament S1-F1-T130 using the raw simulation data. The positions are normalised such that the centre corresponds to the filament 2D spine centre identified using \disperse on two different column density projections to obtain the $x$, $y$ and $z$ co-ordinates. Each section shows the column density in a 0.05 pc thick slice.\footnote{A movie showing the slices along the full filament length is available online at http://www.ita.uni-heidelberg.de/$\sim$rowan/Filaments}

The filament cross-sections are not circular, as would be expected for a cylindrical filament. Instead, the cross sections themselves exhibit filamentary behaviour. For example, in the first panel of Figure \ref{S1F1_walk} there is a sub-filament oriented vertically, but as we walk along the long axis of the
filament,  the density distribution shifts and changes. New sub-filaments appear at different inclinations and then break up and disperse. What appears as a single structure when seen in column density projection is actually revealed to be made of several smaller structures when seen in 3D (see also \citealt{Moeckel14}). The distribution can be thought of as being made out of many short ribbons of gas which twist along the 2D spine of the filament traced in column density. 

\subsection{3D Filaments found using \disperse}

Up to this point, we have concentrated our analysis on filaments found by applying \disperse to column density projections taken from the simulations, as this best mimics what is done observationally when identifying filaments. In the previous section we investigated the true 3D morphology of gas along segments of the 2D spine and found that it was not smooth and contained multiple sub-filaments (often called fibres; see e.g.\ \citealt{Hacar13}). We now investigate where the spine points of filaments are found when we apply \disperse to a 3D grid of the density distribution from the simulation to get a 3D spine. Once this 3D spine is found, we can use it to determine the filament centre and then use the true simulation densities in 3D to investigate the density profile of the filament.

\begin{figure*}
\begin{center}
\begin{tabular}{c c c}
\includegraphics[width=1.5in]{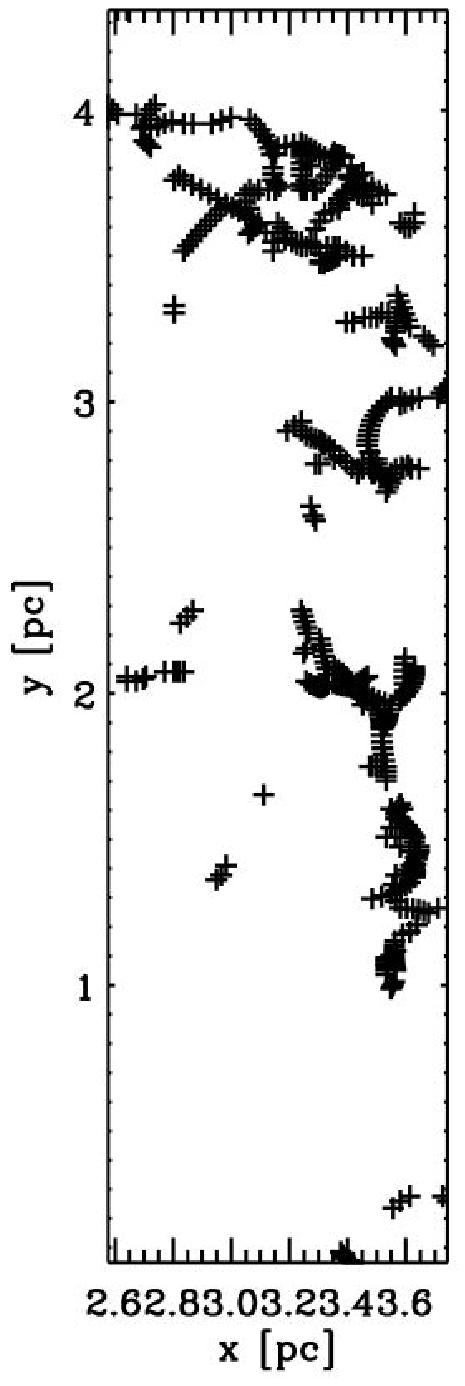}
\includegraphics[width=3.0in]{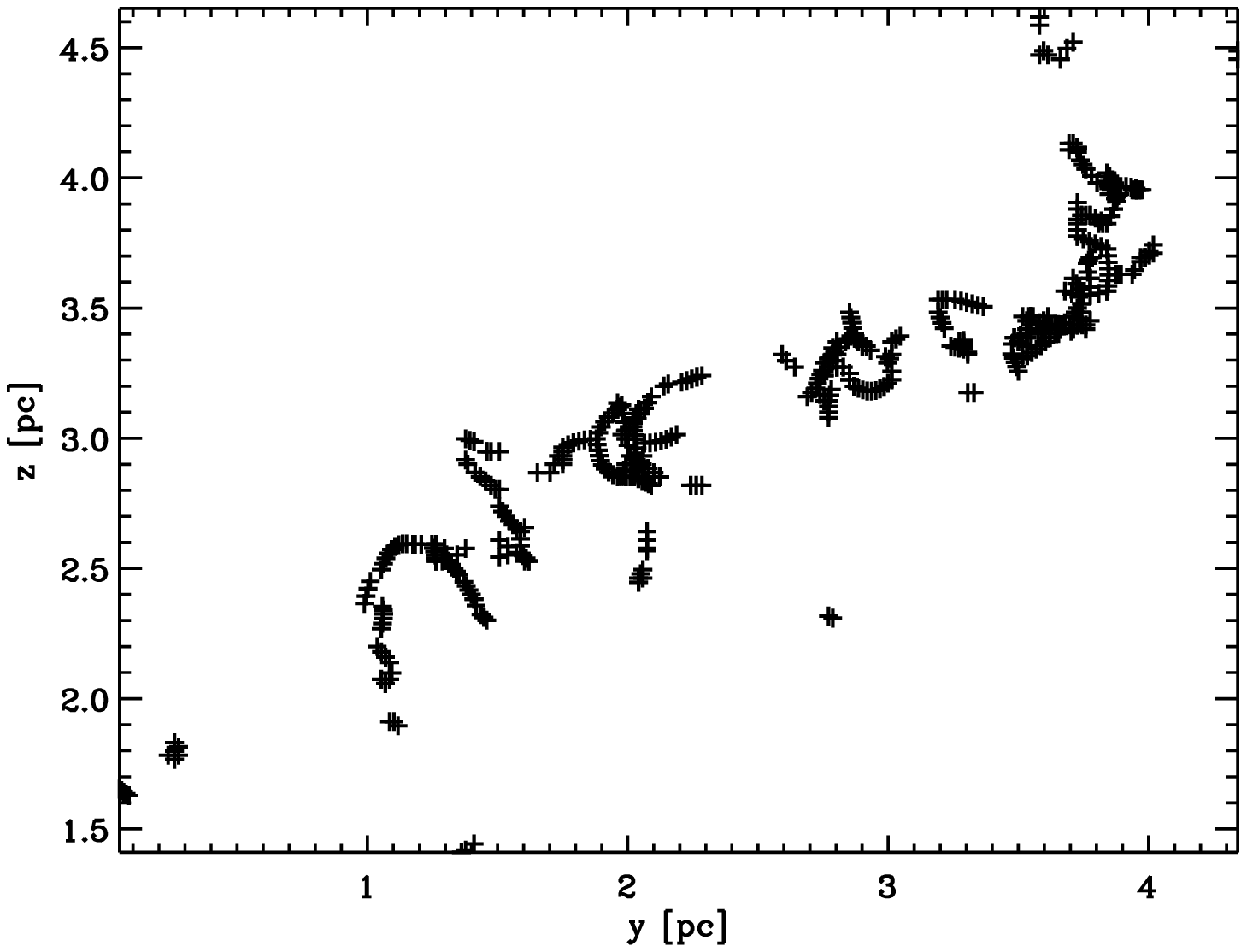}
\includegraphics[width=1.5in]{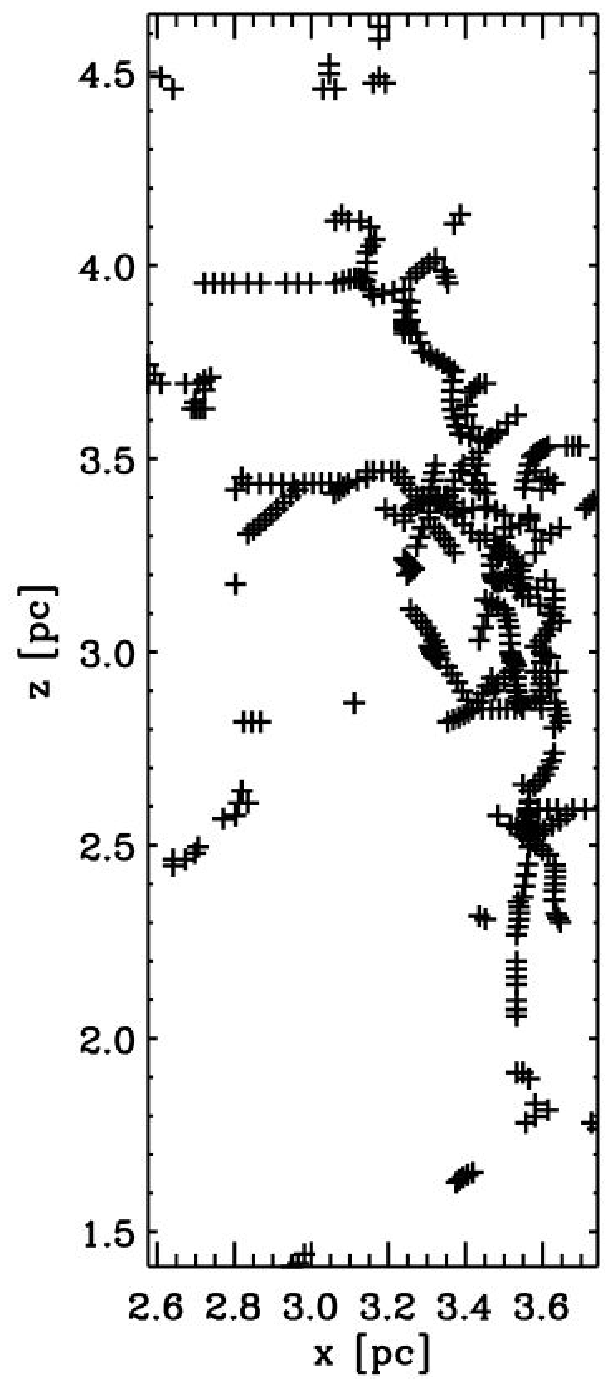}\\
\end{tabular}
\caption{The 3D spine points found using \disperse applied to 3D densities in the vicinity of S1-F1-T130. The filament seen in column density is made up of many smaller sub-filaments.}
\label{mse_pos}
\end{center}
\end{figure*}

Figure \ref{mse_pos} shows the 3D spine points for the 3D filaments found in the vicinity of S1-F1-T130. In the online material, we provide a movie showing a zoom around this distribution that more clearly demonstrates the morphology of the points relative to one another. As in the previous section, we find that what appears to be a single filament when observed in projection is actually made up of a number of smaller sub-filaments. These sub-filaments are not parallel to each other but branch out at a variety of angles from one another. They also frequently cross and intersect each other. However, at the same time, the 3D spine points have a global distribution which is elongated along the $y$-axis, as shown in Figure \ref{mse_pos} and online. 

\begin{figure}
\begin{center}
\includegraphics[width=2.5in]{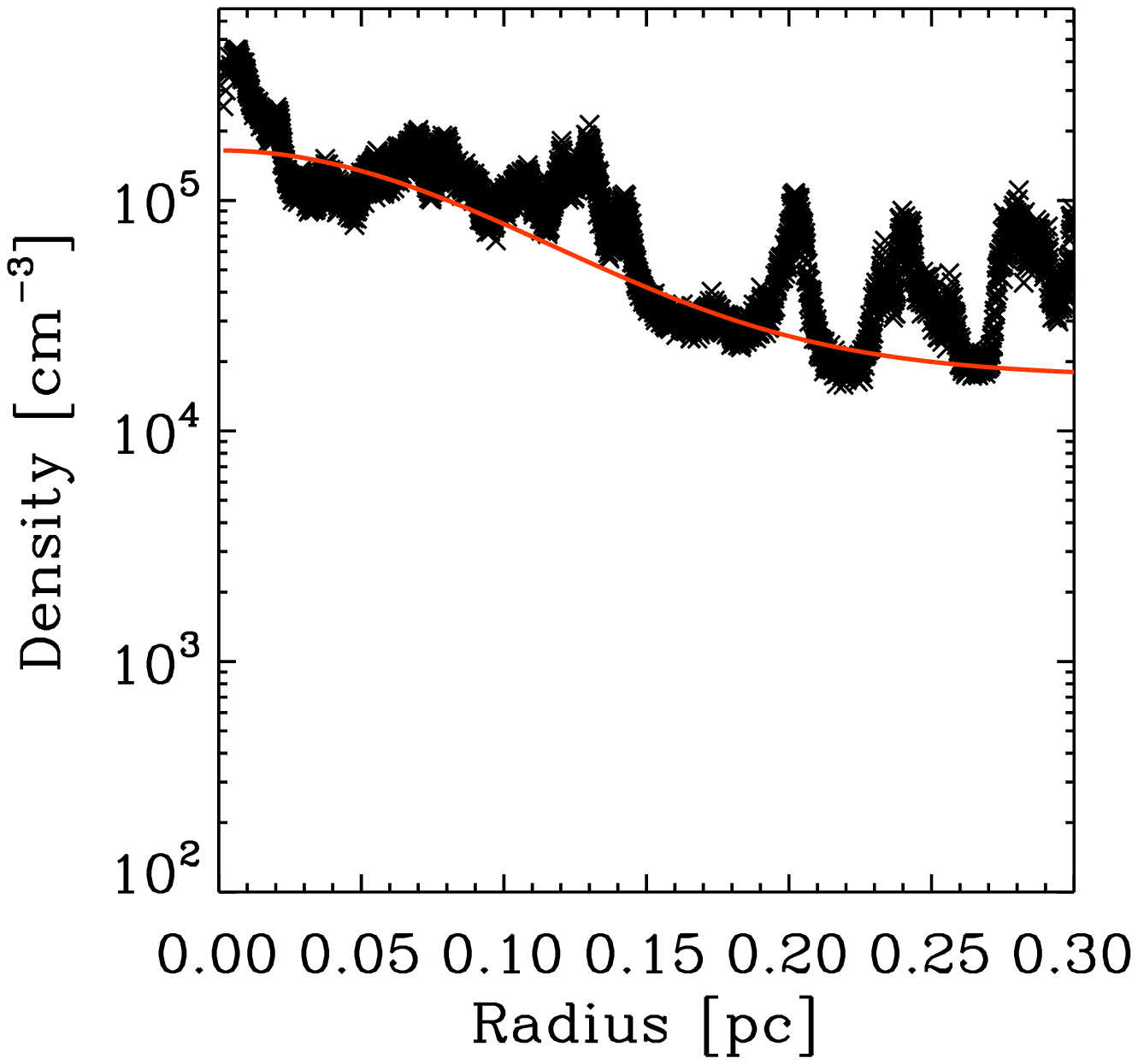}
\caption{The density profile found applying \disperse to the 3D data and calculating an average profile in a similar manner to that done for the column densities. The average 3D density profile is not smooth but has many peaks due to sub-filaments within the gas distribution.}
\label{3Dprofile_avg}
\end{center}
\end{figure}

Ideally, we would like to find the 3D density profile by fitting the average density profile perpendicular to the filament spine as in Section \ref{results-col}. However, the averaged 3D density distribution is extremely noisy. In Figure \ref{3Dprofile_avg} we show the average of the density profiles parallel to the filament 3D spine vectors for S1-F1-T140 shown in Figure \ref{mse_pos}. The profile contains multiple peaks due to the closely overlapping sub-filaments. While it is possible to fit the profile with Equation~\ref{Plum-dens}, the fit is poor. This can be easily understood by inspecting Figure \ref{S1F1_walk} which shows that the density distribution within the filaments is never radially symmetric.

\begin{figure*}
\begin{center}
\begin{tabular}{c c}
\includegraphics[width=2.5in]{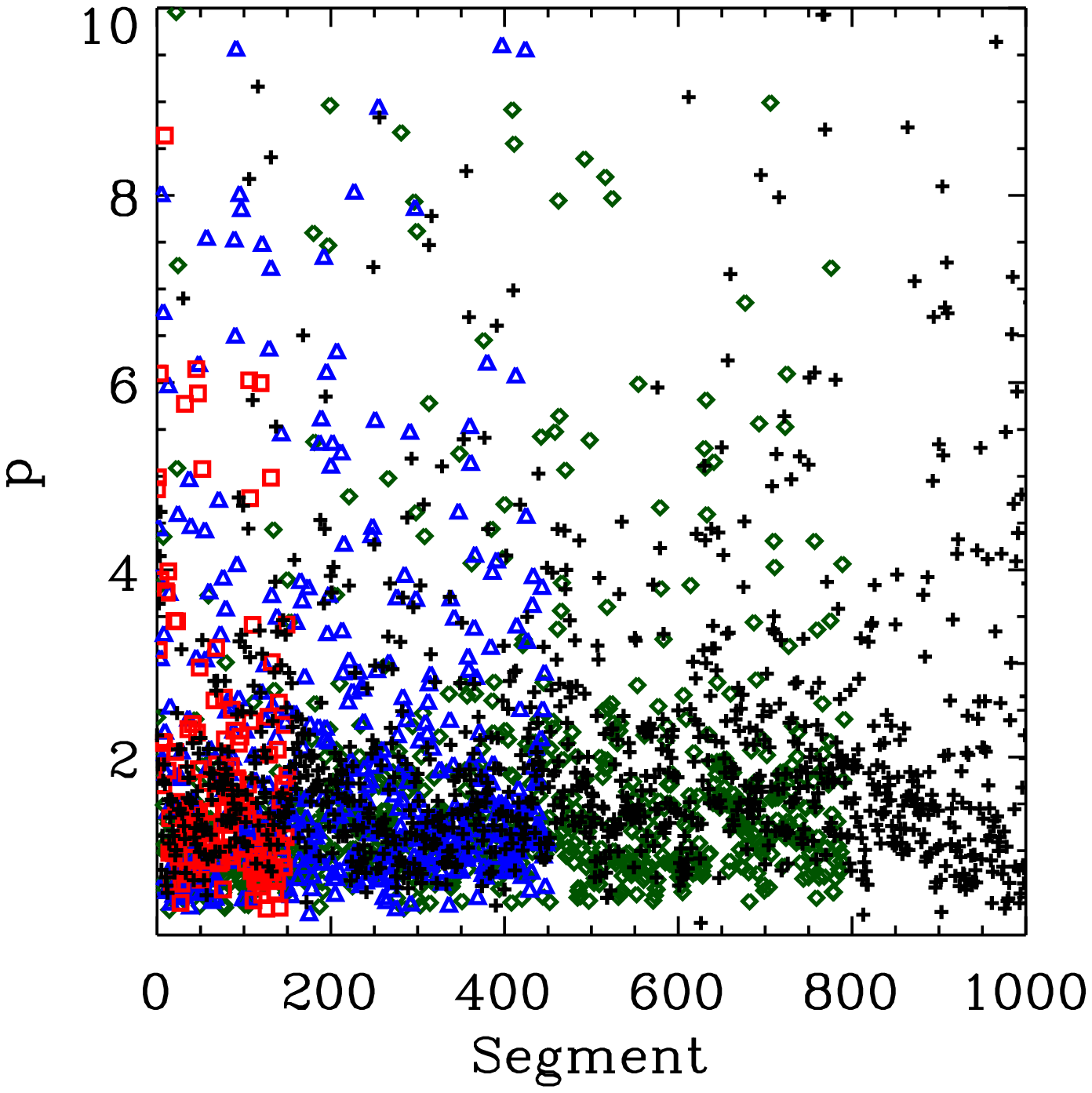}
\includegraphics[width=2.5in]{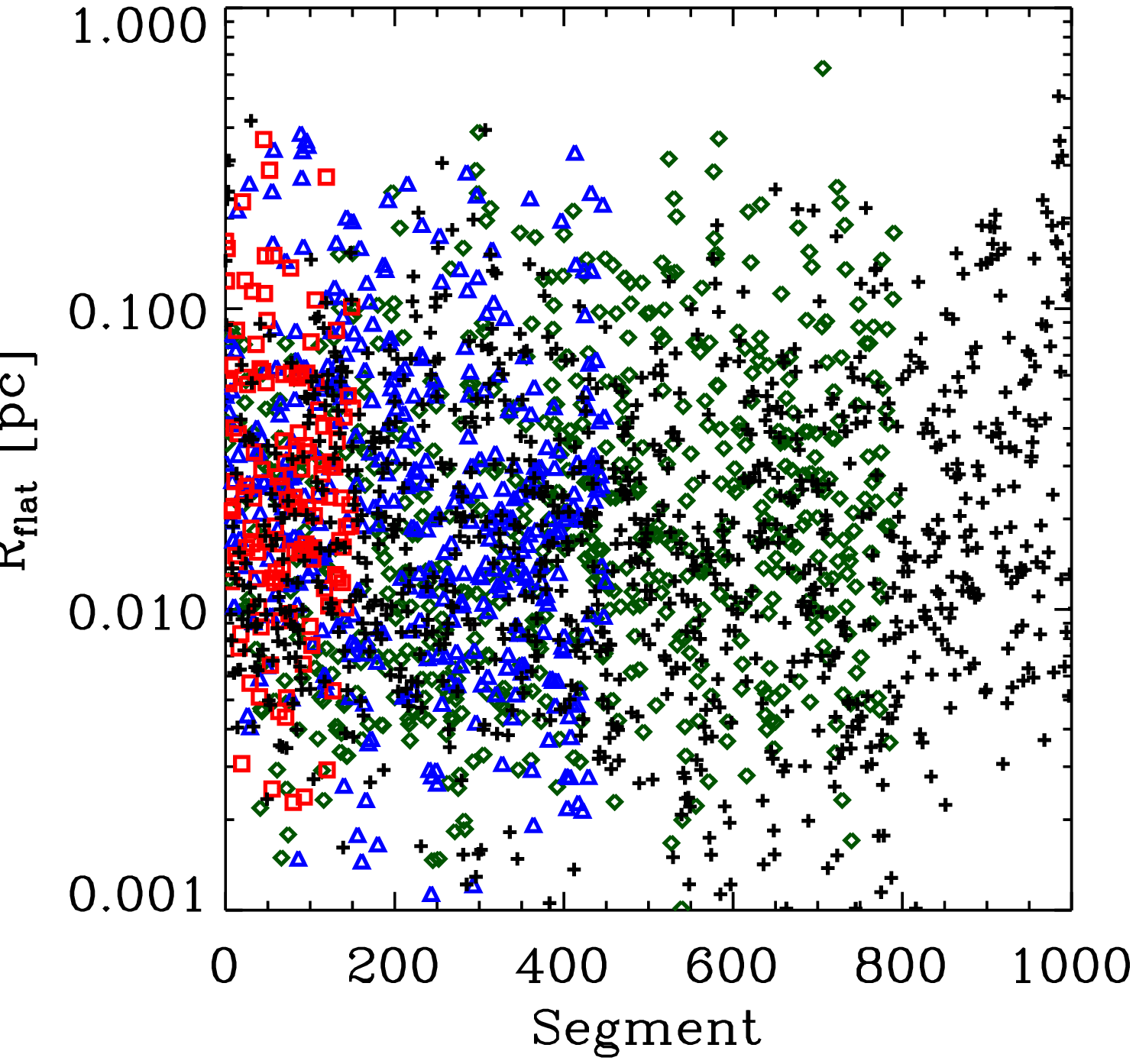}\\
\end{tabular}
\caption{The distribution of $p$ and $R_{\rm flat}$ found from fitting the segments of the 3D spine found using \disperse in the vicinity of filaments S1-F1-T140 \textit{(green diamonds)}, S2-F1-T180 \textit{(black crosses)}, S3-F1-T170 \textit{(blue triangles)}, and S4-F1-T030 \textit{(red squares)}.}
\label{3Dfits}
\end{center}
\end{figure*}

Instead, we fit each segment of all the 3D spine vectors identified from \disperse individually. For each segment we centre on the densest gas, rotate the simulation gas cells about the spine vector, and calculate the profile of the gas in that segment. As the 3D spine is made up of a variety of sub-filaments it is not a continuous structure but contains a mixture of segments from different sub-filaments. In Figure \ref{3Dfits} we show the distribution of the best fit values for the 3D Plummer-profile (Equation \ref{Plum-dens}) for the segments of the 3D spine found for the filaments in each of the simulations. Table \ref{3Dtable} summarises the average and median values of the best fits for the filaments in each simulation. In Table \ref{3Dtable} we do not include the FWHM as we found that it generally did not fit the filament well in 3D and was purely determined by the shape of the background gas. 

\begin{table}
	\centering
		\begin{tabular}{l c c c}
   	         \hline
	         \hline
		Simulation & $n_{\rm c}$ & $R_{\rm flat}$ & $p$\\
		 & [\cmc] & [pc] & \\
	         \hline
	         \textbf{Mean} & \\
	         \hline
	          S1 &  $8.15\E^5$ & 0.0316 & 1.56\\ 
		  S2 &  $2.66\E^6$ & 0.0328 & 2.08\\
		  S3 & $8.79\E^5$ & 0.0387 & 2.03\\
		  S4 & $4.81\E^5$ & 0.0382 & 1.84\\
		 \hline       
		 \textbf{Median} & \\
		 \hline
	          S1 &  $1.24\E^5$ & 0.0130 & 1.17\\ 
		  S2 &  $2.28\E^5$ & 0.0150 & 1.65\\
		  S3 & $9.01\E^4$ & 0.0185 & 1.43\\
		  S4 & $6.55\E^4$ & 0.0213 & 1.38\\
		 \hline  
		 \hline
		\end{tabular}
		\caption{The mean and median best fits values for the 3D Plummer-like profile (Equation \ref{Plum-dens}) when fitted to segments  in the vicinity of filaments S1-F1-T140, S2-F1-T180, S3-F1-T170, and S4-F1-T030.}
	\label{3Dtable}
\end{table}

Figure \ref{3Dfits} shows that there is over an order of magnitude scatter in the best fit values found for the individual 3D spine filament segments. There are few visible differences between the distributions, and a K-S test shows that the distributions are consistent with the null hypothesis that the samples are drawn from the same distribution. The mean flattening radius for the simulations are very similar ($R_{\rm flat}\sim0.035$~pc) and are smaller than the flattening radius found in 2D at the same time in Table \ref{master}. The Jeans length for gas with a number density of $10^5$~\cmc~at $T = 12$ K (typical of the gas at the centre of the filaments) is 0.032 pc. The filament flattening radii seen in 3D are therefore consistent with those predicted by thermal support.
		 
Generally the best power-law fits cluster between values of 1 and 2 (although again there is a large scatter). The departure of the power-law fits from the $p=4$ Ostriker profile seen in Section \ref{results-col} is confirmed in 3D and hence cannot be due to projection effects. The mean central density $n_{\rm c}$ of the 3D spine segments is generally an order of magnitude higher than that derived from the column density projections.  We include the median as well as the average in Table \ref{3Dtable} as an inspection of Figure \ref{3Dfits} shows that the points are not normally distributed. The value of the median shows that most of the gas is at lower densities than the mean would imply.

\section{Discussion}

\subsection{Interpreting the filament profiles}

In this paper we have compared the profiles of filaments identified in column density projections of our simulation results with those derived from observations and found that they share many similarities. In particular, the averaged filament density profiles are well-described using a Plummer-like  fitting function (Equations \ref{Plum-dens} and \ref{Plum-surface}), and consistently have a power-law index $p \sim 2$ in these fits. This power-law index is significantly shallower than the value of $p = 4$ derived in \citet{Ostriker64} for infinite isothermal filaments. Magnetohydrodynamical simulations of filaments have typically found $p \sim 2$ \citep{Tilley03,Hennebelle03}, and so the fact that observed filaments in real molecular clouds follow the same profile has been taken as evidence that they are magnetically supported \citep[see e.g.][]{Contreras13}. Our results show that this need not be the case -- purely hydrodynamical filaments, formed in a turbulent, non-isothermal cloud, also adopt a $p \sim 2$ profile, and so observations of this profile in real filaments tell us nothing one way or the other about the strength of the magnetic field in the filaments.

The average value of $R_{\rm flat}$ derived for the Plummer-like fits (Equation \ref{Plum-surface}) from our simulated filaments is 0.074 pc when using the projected column densities. The value of $R_{\rm flat}$ found in our simulations is larger than that derived by A11, but in very good agreement with that found in J12a, as discussed in Section \ref{results-fits}. For comparison, the Jeans length of the gas at  a density of $10^{5} \: {\rm cm^{-3}}$ and temperature of 12~K (which corresponds approximately to the mean density and temperature of the gas at the centre of the filaments) is around 0.032~pc. However we also calculated $R_{\rm flat}$ for the sub-filaments found in 3D and obtained values of $\sim 0.035$~pc, in excellent agreement with the predicted Jeans radius. We shall discuss this in more detail in the next section.

In general, the values derived for $n_{\rm c}$, $p$ and $R_{\rm flat}$ by a least squares minimisation using Equation \ref{Plum-surface} are quite uncertain. The central density in particular is poorly constrained, with uncertainties of up to 77\%. However, the power-law profiles are better determined with uncertainties of only $10-20\%$. These large uncertainties are at least in part due to the intrinsic degeneracies between $n_{\rm c}$, $p$ and $R_{\rm flat}$ as discussed in Section \ref{robust}. However, despite these large uncertainties, there is still value in fitting the filament profiles, as it allows one to make quantitative comparisons. Specifically, we have been able to show that our simulated filaments have shallow profiles, a flattening radius of the same magnitude as the Jeans radius for the dense gas, and are comparable to observations.

In addition to the Plummer-like density profile which is fit to the entire profile, A11 characterised their filaments by fitting a Gaussian profile to the inner parts of the filament profile. There is a degree of subjectivity involved when choosing the regime in which the data should be fit. A11 chose to fit out to a radius of 0.3-0.4 pc from the filament spine when fitting the Gaussian profile (Arzoumanian, private communication). In Section \ref{width} we tried out a variety of criteria and found that the mean FWHM depended sensitively on the data range chosen. Fitting the column densities out to a distance of 1~pc from the filament centre yielded a value FWHM$_{\rm 1pc} = 0.348$~pc, while using only the data within 0.35~pc of the filament centre yielded a significantly smaller value, FWHM$_{\rm 0.35pc} = 0.202$~pc. In all cases the widths were larger than the 0.1 pc width found in A11 and were not constant. While the simulations are inconsistent with the results reported in A11, they are consistent with the observations of J12a and \citet{Hennemann12} who both find widths of around $0.3$~pc.

When a Gaussian is fitted to data within one parsec of the filament spine a broader mean width of FWHM$_{\rm all} = 0.348$~pc is obtained.  \citet{Heitsch13a} studied the analytical density profiles that would be expected for accreting filaments and found that the FWHM of such filaments  typically stays constant at a value of around 0.3 pc throughout most of their evolution. The filaments in this study are accreting from their environment and grow in mass and central density with time. Our simulation results are therefore in good agreement with the predictions of \citet{Heitsch13a}.

\subsection{Filament formation in molecular clouds}\label{form}

\begin{figure*}
\begin{center}
\begin{tabular}{c c}
\begin{overpic}[scale=.4]{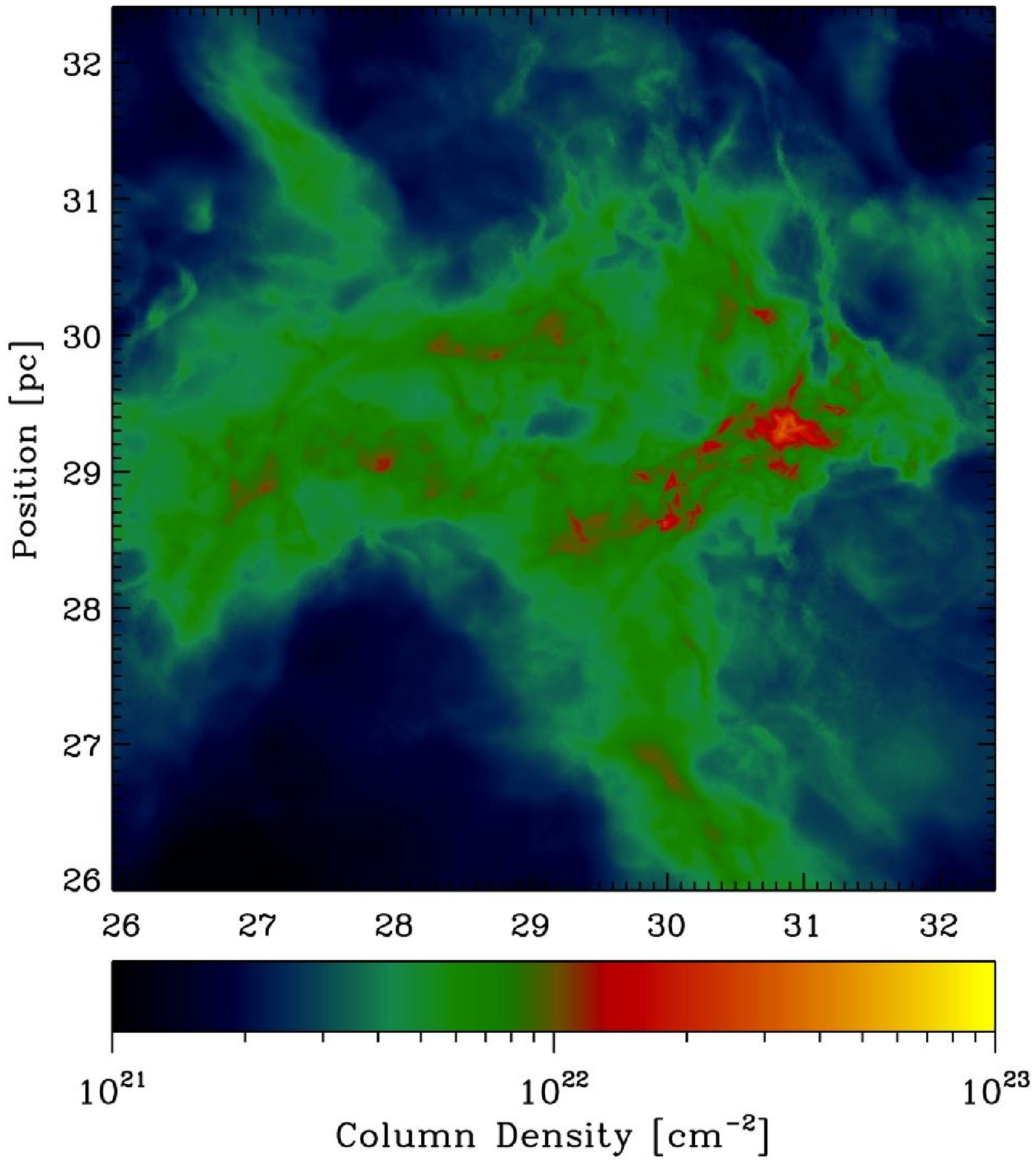}
\put (75,90) {\makebox(0,0){{\color{white}S2-T140}}}
\end{overpic}
\begin{overpic}[scale=.4]{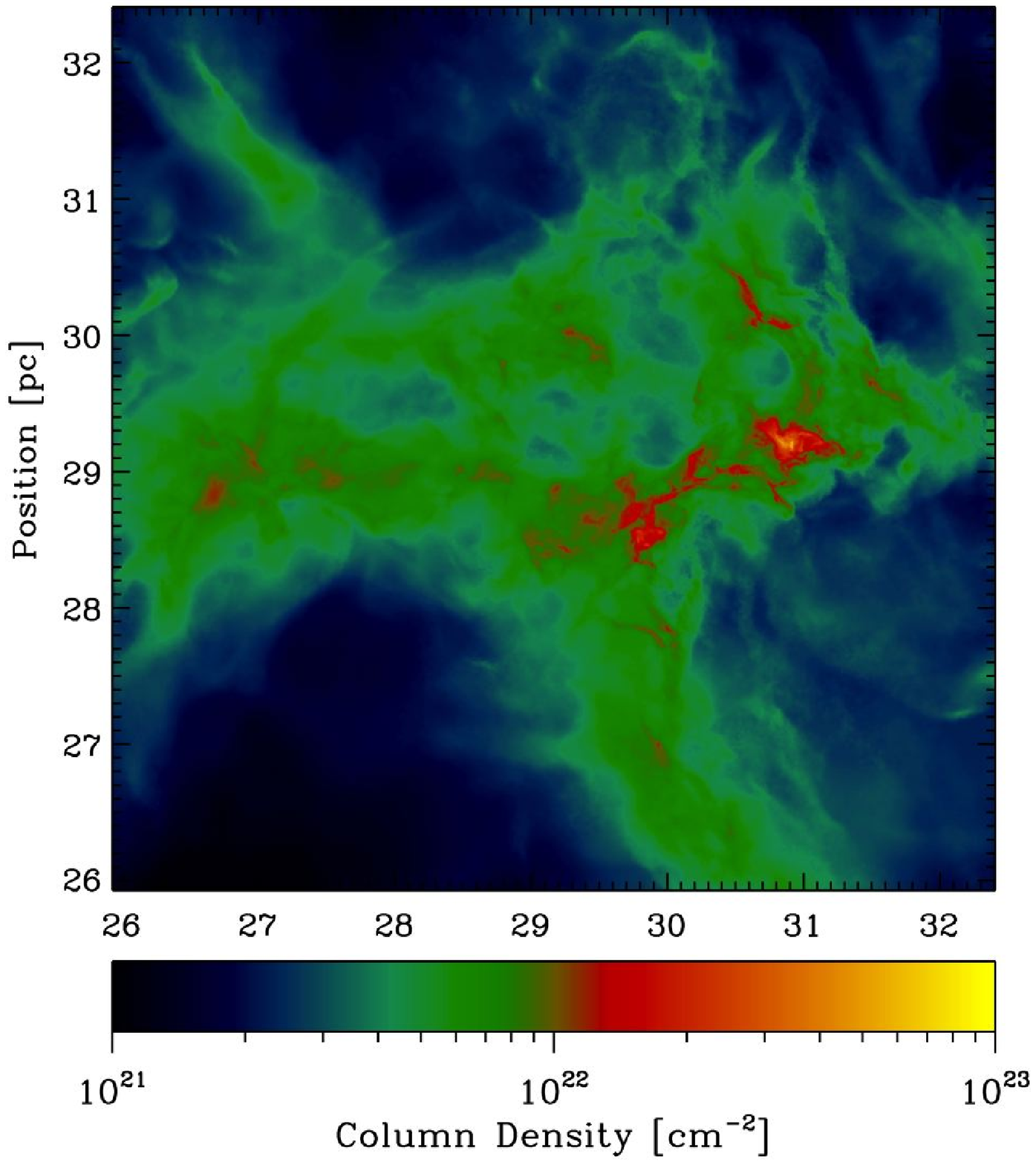}
\put (75,90) {\makebox(0,0){{\color{white}S2-T150}}}
\end{overpic}
\\
\begin{overpic}[scale=.4]{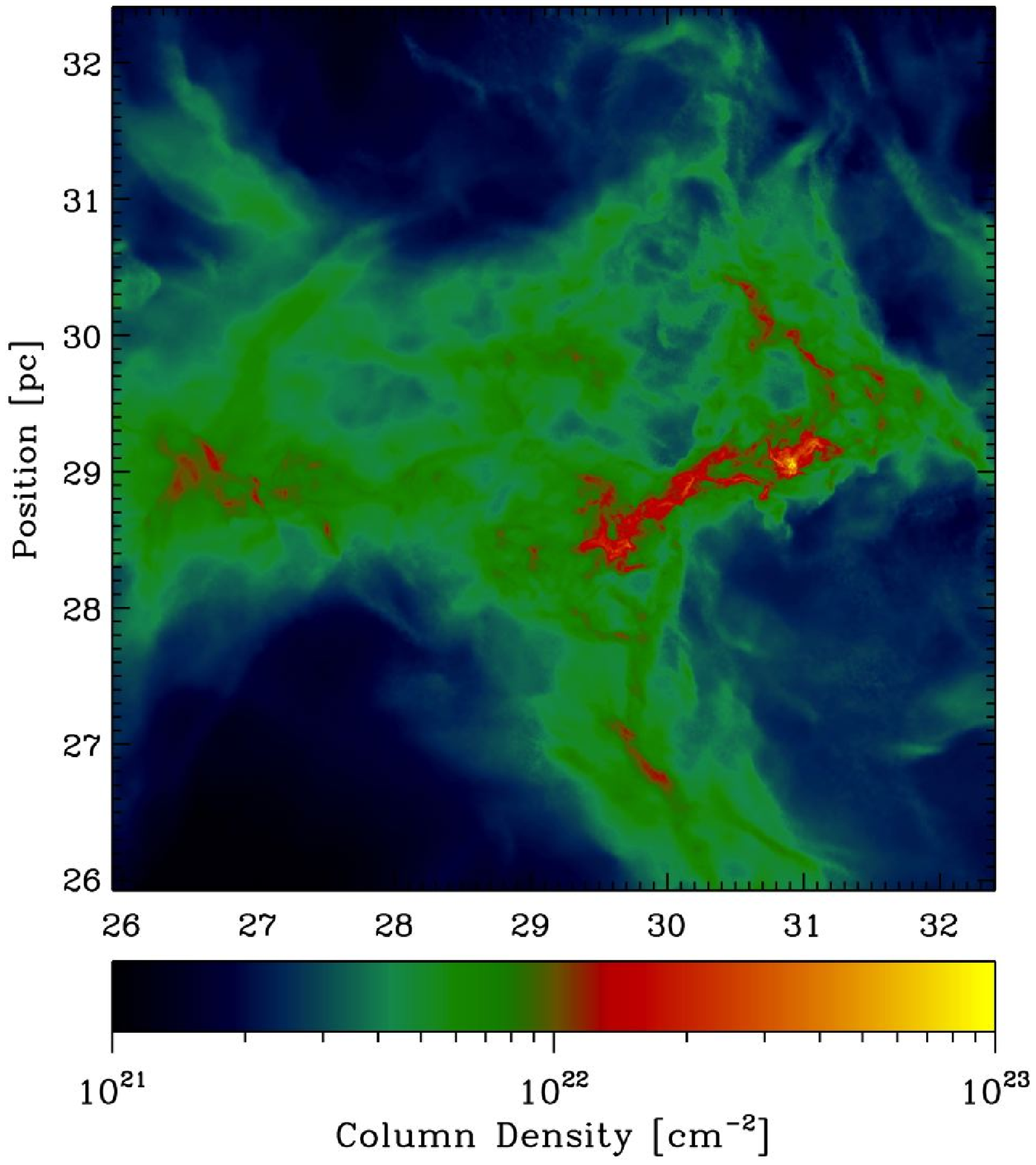}
\put (75,90) {\makebox(0,0){{\color{white}S2-T160}}}
\end{overpic}
\begin{overpic}[scale=.4]{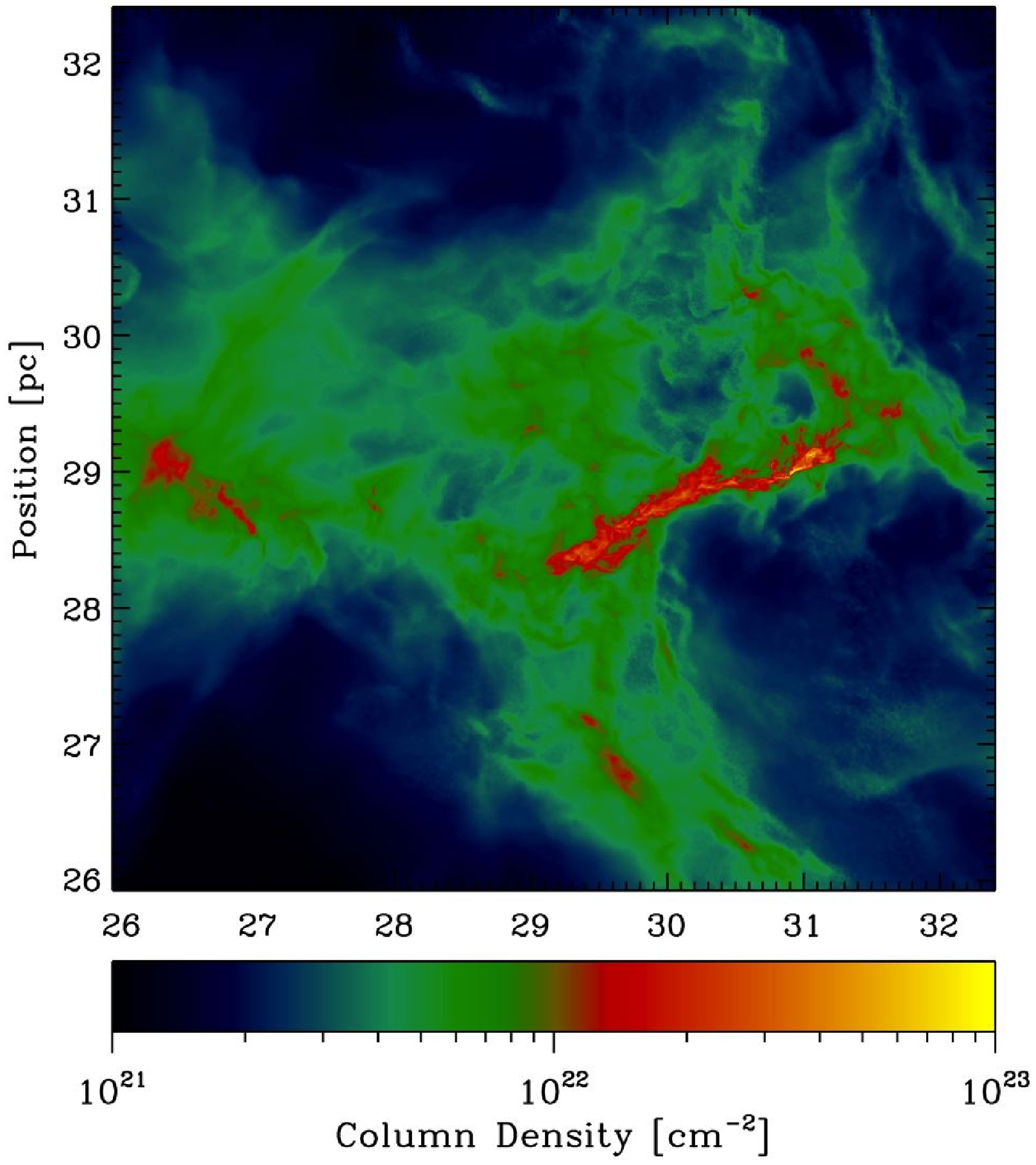}
\put (75,90) {\makebox(0,0){{\color{white}S2-T170}}}
\end{overpic}
\end{tabular}
\caption{A series of column density projections at $8.7\E^5$ yr intervals showing the formation of Filament S2-F1. The filament forms from the large scale collapse within the cloud sweeping up small filaments and clumps of gas into one large structure. Movies of the formation of filaments S1-F1, S2-F1, S3-F1 and S4-F1 can be found online at www.ita.uni-heidelberg.de/$\sim$rowan/Filaments.}
\label{movie}
\end{center}
\end{figure*}

In Section \ref{results-3D} we investigated how the filaments seen in column density projection correspond to the intrinsic 3D gas distribution. We found that the filaments seen in projection are not continuous density structures in 3D. Instead, they contain many ribbon-like sub-filaments that twist and branch off along the axis of the filament. When the sub-filament segments are fitted with Plummer-like profiles we obtain power-law indices similar to those found for the column density projections, but flattening radii that are smaller and densities that are higher. The column density filaments have a larger flattening radius which is consistent with them being made up of a superposition of smaller thermally supported sub-filaments.

The presence of short sub-filaments in the simulations is reminiscent of the observations of \citet{Hacar13}. These authors have shown that filaments in L1495/B213 in Taurus are made up of shorter $\sim0.5$ pc coherent sub-filaments. \citet{Hacar13} propose a scenario where filaments first fragment into sub-filaments, which then further fragment into cores. However, our simulations do not seem fully consistent with this scenario as in Figure \ref{mse_pos} we frequently see 3D spine sections that are perpendicular to the main filament axis. If the sub-filaments were formed by fragmentation we would expect the sub-filaments to align along the main filament axis. Furthermore, if an initially smooth filament were to fragment we would expect to see only one fragment at a given position corresponding to the maximum in density. On the contrary, we find that more than one sub-filament is often seen at the same point along the main filament axis, something which is also seen in \citet{Hacar13}.

To explore this further we made movies showing the formation of the filaments that we study in 3D. These can be found online at http://www.ita.uni-heidelberg.de/$\sim$rowan/Filaments, but we also present in Figure \ref{movie} a series of snapshots at $8.7\E^5$ yr intervals showing the formation of S2-F1. Clearly, the filaments' density structure changes over time and so this type of behaviour cannot be studied in hydrostatic models.

All four of the simulations show a common trend. Smaller filamentary clumps of gas exist within the cloud from an early stage (formed by turbulent compression and energy dissipation within the cloud). These smaller clumps are then swept up into a single column density feature by a large-scale flow of gas. In our simulations, this large-scale flow is due to the gravitational collapse of the molecular cloud. As the density distribution of the cloud is not spherical, the collapse is not spherical either and there are additional shearing motions which stretch the filament, further enhancing the filamentary morphology. The collapse also increases the mean density of the sub-filaments through compression and accretion, which will encourage star formation within them. Filament formation may therefore be an extremely important process in triggering star formation within pre-existing gas clumps.

Gravity is a natural mechanism for forming filaments by either the fragmentation of sheet-like clouds \citep{Miyama87} or by amplifying asymmetries and edges in the gas density distribution \citep{Hartmann07}. However, any large-scale converging motion \citep[e.g.][]{Gomez13} would have the same effect. Filaments are easily built by turbulence in gas with $\gamma < 1$ \citep{Larson85,Peters12b}. This may explain the similarities of filaments within diffuse clouds to those within more dense, gravitationally bound clouds \citep{Arzoumanian11}, such as those we study here. The difference between the turbulent initial conditions of our simulations allow us to explore this in more detail. Simulation S4 was initially seeded with purely compressible turbulence which is extremely efficient at forming filaments, and as a result there is less fragmentation and large scale collapse before the filament forms. Despite this, the filaments that form in this simulation are again made out of smaller pre-existing sub-filaments, although in this case they are longer and thinner compared to the other simulations, and are more strongly aligned with the main filament axis. However, it should be noted that S4 is quite an idealised example, as the filament forms close to the beginning of the simulation where the initial conditions are still quite apparent.


\section{Conclusions}
In this paper we used high resolution \arepo simulations to investigate the morphology of dense star-forming filaments. We created column density projections of turbulent molecular cloud simulations and used the \disperse algorithm to identify the spine of major filaments. We calculated the average column density profile perpendicular to the spine and fit it with a Plummer-like profile, as has previously been done for filaments identified in dust emission images of real molecular clouds. We investigated the widths of our simulated filaments by fitting them with a Gaussian. Finally, we compared the filaments observed in column density to the underlying 3D density structure to learn more about the nature of the filamentary structure. Our conclusions are as follows.
\begin{enumerate}
\item When the simulated filaments are fit with a Plummer-like density profile they have an average power-law index of $p=2.2$ without the presence of magnetic fields. This is shallower than the value of $p = 4$ expected for an isothermal filament in a vacuum \citep{Ostriker64}. Magnetic fields are often invoked to explain filament profiles that have $p < 4$, but our simulations show that filaments forming in turbulent clouds naturally have flatter profiles, without the need for magnetic support. Consequently, observations of shallow filament column density profiles cannot be used to infer anything about the magnetic field strength.
\item The best fit values obtained from our Plummer-like fits are in agreement with the observations of \cite{Arzoumanian11} and \citeauthor{Juvela12a}~(2012a). The observations of \citeauthor{Juvela12a}~(2012a) of filamentary structures in the \textit{Planck} Galactic cold cores are a particularly good match as they agree to within a few percent.
\item The parameters derived from the Plummer-like density profiles are quite uncertain due to a degeneracy between the fitted parameters.
\item The filament widths inferred from our Gaussian fits depend on the data range that is is fitted. When data out to 1~pc from the filament spine is fitted the filament FWHM is $\sim0.35$~pc, in agreement with predictions for accreting filaments \citep{Heitsch13a}. When only data within 0.35~pc of the spine is fitted (as is done in many observations), a FWHM of only $\sim0.2$~pc is found. 
\item The mean Gaussian FWHM of our simulated filaments is higher than the 0.1~pc found by \cite{Arzoumanian11} and is not constant. However while our simulated FWHM are inconsistent with the findings of \cite{Arzoumanian11}, they are in agreement with the studies of \citeauthor{Juvela12a}~(2012a) and \citet{Hennemann12}.
\item The filament profiles are not substantially affected by the nature of the initial turbulent velocity field in the simulations. They are similar regardless of whether solenoidal, compressive or a natural mix of turbulent modes is used. There is little systematic time evolution in the filaments over the studied period.
\item In 3D, the filaments seen in column density do not belong to a single structure. Instead, they are made up of a network of short ribbon-like sub-filaments reminiscent of those seen in Taurus by \citet{Hacar13}. The small sub-filaments do not always lie parallel to the main filament axis, but instead branch off from it and cross each other at a variety of angles.
\item The flattening radius of the 3D sub-filaments is consistent with the expected Jeans radius for the dense, cold gas at the filament centre. The sub-filaments may thus be thermally supported objects whose superposition forms the longer filament seen in column density. 
\item The small sub-filaments are pre-existing within the simulated clouds, and are not primarily formed through fragmentation of the larger filament seen in column density. Instead, small filamentary clumps are swept together into a single structure by the large-scale collapse of the cloud due to gravity. This increases the density of the sub-filaments and may induce future star formation within them.
\end{enumerate}

\section*{Acknowledgements}
We gratefully acknowledge Doris Arzoumanian whose assistance was invaluable in determining how best to compare our simulations to the \herschel data. We would also like to thank Paul Clark, Fabian Heitsch, Patrick Hennebelle, Helen Kirk, and Jaime Pineda for helpful discussions about how best to describe filaments. R.J.S.\ acknowledges support from the German Science Foundation (DFG) priority program 1573 {\em Physics of the Interstellar Medium} via project SM 321/1-1. R.J.S.\ also acknowledges the Aspen Centre for Physics for their hospitality while some of the basic analysis tools used in this work were being developed. R.S.K.\ acknowledges support from the European Research Council under the European Community's Seventh Framework Programme (FP7/2007-2013) via the ERC Advanced Grant {\em STARLIGHT} (project number 339177). The research leading to these results has also been supported by the DFG via the Collaborative Research Centre SFB 881 {\em The Milky Way System} (sub-projects B1, B2, B5 and B8).

\bibliography{./Bibliography}

\appendix
\section{Best fit plummer-like model when $R_{\rm flat}$ is held constant.}

The best fit parameters for the Plummer-like profile shown in Equation \ref{Plum-surface} when the flattening radius is held constant at a value of $R_{\rm flat}=0.04$ pc. The uncertainty in the best fit parameters drops significantly and the central density and power are in closer agreement with the values found by A11.

\begin{table}
	\centering
		\begin{tabular}{l c c}
   	         \hline
	         \hline
		ID  & peak & p \\
		 & [$10^4$ \cmc] & \\
	         \hline                  
	         S1-F1-T130 &  $14.6 \pm 14\%$ & $1.33 \pm 8\%$\\
		 S1-F1-T140 &  $23.2 \pm 7\% $ & $1.73 \pm 5\%$\\
		 S1-F1-T150 &  $29.8 \pm 8\% $ & $1.81 \pm 6\%$\\
		\hline
		S1-F2-T130 & $14.0 \pm 6\%$ & $1.42 \pm 5\%$\\
		S1-F2-T140 & $17.7 \pm 5\%$ & $1.59 \pm 6\%$\\
		S1-F2-T150 & $24.2 \pm 6\%$ & $1.49 \pm 6\%$\\	
	         \hline
		S2-F1-T170 & $15.2 \pm 6\%$ & $2.02 \pm 4\%$\\
		S2-F1-T180 & $19.7 \pm 8\%$ & $2.51 \pm 4\%$\\
		S2-F1-T190 & $15.8 \pm 8\%$ & $2.33 \pm 5\%$\\	
	        \hline
	         S3-F1-T160 & $10.6 \pm 3\%$ & $1.56 \pm 4\%$\\
		S3-F1-T170 & $18.8 \pm 4\%$ & $1.57 \pm 1\%$\\
		S3-F1-T180 & $25.7 \pm 5\%$ & $1.71 \pm 3\%$\\
	         \hline
	         S3-F2-T160 & $9.00 \pm 2\%$ & $1.32 \pm 6\%$\\
		S3-F2-T170 & $11.1 \pm 5\%$ & $1.50 \pm 6\%$\\
		S3-F2-T180 & $16.5 \pm 9\%$ & $1.84 \pm 6\%$\\		
	         \hline
	        S4-F1-T020 & $8.52 \pm 3\%$ & $1.30 \pm 1\%$\\
		S4-F1-T030 & $12.6 \pm 3\%$ & $1.62 \pm 1\%$\\
		S4-F1-T040 & $16.8 \pm 3\%$ & $1.84 \pm 1\%$\\	
	         \hline
	         S4-F2-T020 & $6.47 \pm 4\%$ & $1.63 \pm 5\%$ \\
		 S4-F2-T030 & $13.2 \pm 5\%$ & $1.58 \pm 6\%$\\
		 S4-F2-T040 & $23.5 \pm 10\%$ & $1.75 \pm 6\%$\\	
		 \hline
		 \hline
		 Mean & 16.5 & 1.67\\
		 St. Deviation & 5.99 & 0.30 \\
	         \hline
	         \hline
		\end{tabular}
	\label{rfixed}
	\caption{Peak and power when R$_{\rm flat}$ is held constant at 0.04pc. }
\end{table}


\label{lastpage}

\end{document}